\documentclass[twocolumn,showpacs,preprintnumbers,amsmath,amssymb,superscriptaddress,pra]{revtex4}
\usepackage[latin1]{inputenc}
\usepackage{dcolumn}
\usepackage{bm}
\usepackage{graphicx}
\usepackage[english]{babel}

\DeclareMathOperator{\Tr}{Tr}
\DeclareMathOperator{\Rea}{Re}
\DeclareMathOperator{\Ima}{Im}
\def\bbm[#1]{\mbox{\boldmath $#1$}}
\newcommand{\ket}[1]{\displaystyle{|#1\rangle}}
\newcommand{\bra}[1]{\displaystyle{\langle #1|}}
\newcommand{\TE}{\text{TE}}
\newcommand{\TM}{\text{TM}}

\begin{document}

\title{Scattering-matrix approach to Casimir-Lifshitz force and heat transfer\\out of thermal equilibrium between arbitrary bodies}

\author{Riccardo Messina}
\affiliation{LNE-SYRTE, Observatoire de Paris, CNRS UMR8630,
UPMC\\61 avenue de l'Observatoire, 75014 Paris, France}

\author{Mauro Antezza}
\affiliation{Universit\'e  Montpellier 2, Laboratoire Charles Coulomb UMR 5221, F-34095, Montpellier, France}
\affiliation{CNRS, Laboratoire Charles Coulomb UMR 5221, F-34095, Montpellier, France}

\date{\today}

\begin{abstract}
We study the radiative heat transfer and the Casimir-Lifshitz force occurring between two bodies in a system out of thermal equilibrium. We consider bodies of arbitrary shape and dielectric properties, held
at two different temperatures, and immersed in a environmental radiation at a third different temperature. We derive explicit closed-form analytic expressions for the correlations of the electromagnetic field, and
for the heat transfer and Casimir-Lifshitz force, in terms of the bodies scattering matrices. We then consider some particular cases which we investigate in detail: the atom-surface and the slab-slab configurations.
\end{abstract}


\maketitle

\section{Introduction}

The electromagnetic field interacting with bodies gives rise to
several effects, among which the radiative heat transfer
\cite{PolderPRB71} and the Casimir-Lifshitz force
\cite{CasimirProcKNedAkadWet48,CasimirPhysRev48,DzyaloshinskiiAdvPhys61}.
These two effects are intimately connected, and can be described by
a common formalism. If a definite temperature is assigned to
bodies and radiation, two main situations are possible. Radiation
can be at thermal equilibrium with matter everywhere: in this case
the heat transfer is identically zero, whereas the
Casimir-Lifshitz force assumes its equilibrium value. If, instead,
radiation is not at thermal equilibrium with matter, both the
appearance of heat transfer between the bodies and a variation of
the forces acting on them occur.

In considering the phenomena of heat transfer and Casimir-Lifshitz
forces, the bodies shapes, geometric configurations, and
dielectric functions are main issues, together with the thermal
configuration of the system. A general framework used in solving
such kind of problems is that of macroscopic electrodynamics
\cite{Rytov89,DzyaloshinskiiAdvPhys61,JoulainSurfSciRep05}.
Casimir-Lifshitz interaction for systems at thermal equilibrium
(at $T\geq0$) has been largely studied in the last 60 years reaching a
consistent and almost complete theoretical formulation, as well as experimental observation \cite{CasimirDalvit}. In particular, much more recently, the thermal component of the force at thermal equilibrium
has been measured \cite{SushkovNaturePhys11}. On the contrary, systems out of
thermal equilibrium have been much less explored, and mainly
simple configurations and idealized cases (such as infinite bodies) have been considered.
Nonetheless, systems out of thermal equilibrium showed remarkable
features, already object of theoretical and experimental
investigations both concerning the force
\cite{AntezzaPRL05,AntezzaPRL06,AntezzaJPhysA06,ObrechtPRL07,AntezzaPRA08,BuhmannPRL08,SherkunovPRA09,BimontePRA09,BehuninPRA10,BehuninJPhysA10,BehuninArXiv11}
and the heat transfer \cite{VolokitinRevModPhys07,RousseauNaturePhoton09,ShenNanoLetters09,KralikRevSciInstrum11,OttensPRL11,BenAbdallahJApplPhys09,BenAbdallahPRB10,BiehsArXiv11}. This is the case of the atom-surface force, where the absence of peculiar cancellations among
the different components of the radiation out of thermal equilibrium leads to large quantitative and qualitative modifications such as new asymptotic behaviors and possibility of repulsive interactions \cite{AntezzaPRL05}. These
new features have allowed the first experimental observation of thermal effects \cite{ObrechtPRL07}. Recently,
motivated by the necessity to develop a more complete theory for
systems out of thermal equilibrium, several studies have been
developed \cite{MessinaArXiv11,KrugerPRL11,KrugerEurophysLett11,BimonteArXiv11,RodriguezArXiv11,McCauleyArXiv11}. In
particular, a general closed-form analytic expression for both
heat transfer and Casimir-Lifshitz interactions has been derived,
valid for arbitrary bodies shapes, dielectric functions, and
arbitrary temperatures of the bodies and of the environment
\cite{MessinaArXiv11}. To this purpose a scattering-matrix
approach has been used. This approach has been already successfully employed to calculate Casimir-Lifshitz interactions \cite{LambrechtNewJPhys06,RahiPRD09,BimontePRA09,MessinaArXiv11}: its main advantage with respect to the
standard Green-function formulation, where the electromagnetic problem needs to be solved for the complete system, consists in requiring the solution for each body composing the system independently.

In this paper we provide a systematic derivation for the heat
transfer and for the Casimir-Lifshitz force out of thermal
equilibrium between two bodies at two different temperatures,
immersed in and external environmental radiation characterized by
another temperature \cite{MessinaArXiv11}. Particular attention is
devoted to the expression of the average values of the
electromagnetic field in terms of the scattering operators. The
general expression is finally analytically and numerically applied to several simple but
already interesting configurations, by calculating the heat
transfer and the force for slab-slab and atom-slab systems in an
out of equilibrium scenario. The role of finite-size effects and of the environmental temperature are shown to be qualitatively and quantitatively relevant.

The organization of the paper is the following. In Sec. \ref{SecPhysSyst} we present the physical system, the hypotheses
and the general theoretical framework. In Sec. \ref{SecT} and \ref{SecS} we evaluate the analytic expression and the flux of
the two main ingredients of our calculation: the Maxwell stress tensor and the Poynting vector. Section \ref{SRT} is dedicated to
the definition of the reflection and transmission operators associated to each body. Then, in Sec. \ref{FieldCorr} the
correlators of the fields emitted by the bodies and by the environment are expressed as a function of the scattering
operators. This allows us to calculate, in Sec. \ref{FinalFH}, the fluxes of the stress tensor and the Poynting vector in any region and then the final expressions of the Casimir-Lifshitz force and the heat transfer.
This expression is first applied to the case of a single body alone out of thermal equilibrium in Sec. \ref{BodyAlone}. Then, in Sec. \ref{NumAppl}, we discuss the case of the force acting on a neutral atom in front
of a planar slab, as well as both the force and the heat transfer in a two-slab configuration. We finally give some conclusive remarks.

\section{Physical system and electromagnetic field}\label{SecPhysSyst}

\begin{figure}[htb]
\includegraphics[height=4.5cm]{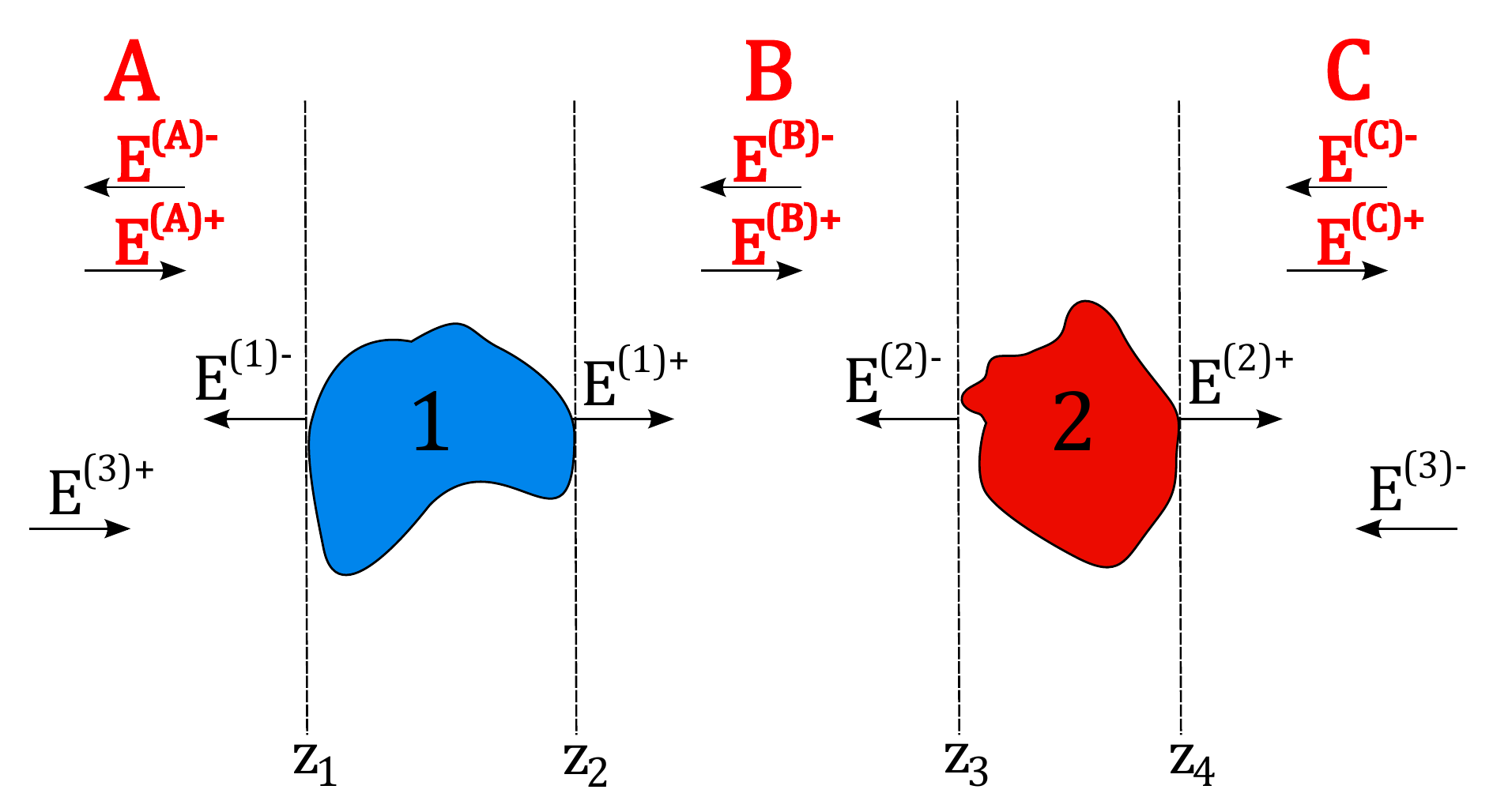}
\caption{The geometry of the system. The bodies are separated by
the strip $z_2\leq z\leq z_3$. This defines the three regions A, B
and C.}\label{Fig1}\end{figure}

Let us start by describing the geometrical configuration of our
physical system. We are going to deal with two bodies, labeled
with 1 and 2. From a geometrical point of view we will assume that
the two bodies are separated by a planar surface. This hypothesis
is not strictly necessary in a scattering-matrix approach but it
is nonetheless verified in all the relevant experimental
configurations such as for example two parallel planes, a sphere
or a cylinder in front of a plane, two cylinders, as well as an
atom in proximity of a planar surface. At the same time, this
assumption allows us to choose a convenient plane-wave
decomposition for the electric and magnetic fields, leading in
this way to quite simple expressions of the Casimir-Lifshitz force
and the heat transfer. To be more specific, the geometry of our
system is depicted in figure \ref{Fig1}: the bodies 1 and 2 are
respectively enclosed in the strips $z_1<z<z_2$ and $z_3<z<z_4$,
where $z_2<z_3$. As a consequence, any plane $z=\bar{z}$ with
$z_2<\bar{z}<z_3$ separates the two bodies, and three regions A, B
and C are defined.
 Our geometrical description
coherently includes as a limiting case the possibility of bodies
having infinite thickness, taking for example for body 1 (body 2)
the limit $z_1\to-\infty$ ($z_4\to+\infty$).

As far as the thermodynamical description of our system is
concerned, we will assume that we are able to define for body 1
(2) a temperature $T_1$ ($T_2$) and that it is in local thermal
equilibrium, i.e. the temperature of each body is assumed to be
constant. Moreover, we assume that the two bodies are immersed in
a vacuum environment (having $\varepsilon=1$) characterized by a
third temperature $T_3$. We will make the further important
assumption that the composite system is in a stationary regime,
which means that the three temperatures remain constant in time.

Let us now first describe the general framework of the
calculation. We are interested in calculating the force
$\mathbf{F}$ acting on any of the two bodies, as well as the heat
transfer $H$ on it, defined as the energy it absorbs per unit of
time. Focusing for example on body 1, these
two quantities can be expressed under the form of surface
integrals through a closed surface $\Sigma$ enclosing the body 1
\begin{equation}\label{FH}\begin{split}\mathbf{F}&=\int_\Sigma\langle\mathbb{T}(\mathbf{R},t)\rangle_\text{sym}\,\cdot d\bbm[\Sigma]\\
H&=-\int_{\Sigma}\langle\mathbf{S}(\mathbf{R},t)\rangle_\text{sym}\,\cdot
d\bbm[\Sigma]\end{split}\end{equation} of the quantum symmetrized
average of the Maxwell stress tensor $\mathbb{T}$ (having
cartesian components $T_{ij}$, with $i,j=x,y,z$) and the Poynting
vector $\mathbf{S}$. These quantities are classically defined in
SI units as
\begin{equation}\label{TijS}\begin{split}T_{ij}(\mathbf{R},t)&=\epsilon_0\Bigl[E_i(\mathbf{R},t)E_j(\mathbf{R},t)+c^2B_i(\mathbf{R},t)B_j(\mathbf{R},t)\\
&\,-\frac{1}{2}\Bigl(E^2(\mathbf{R},t)+c^2B^2(\mathbf{R},t)\Bigr)\delta_{ij}\Bigr],\\
\mathbf{S}(\mathbf{R},t)&=\epsilon_0c^2\mathbf{E}(\mathbf{R},t)\times\mathbf{B}(\mathbf{R},t).\end{split}\end{equation}
and the quantum symmetrized average value $\langle AB\rangle_\text{sym}$ is defined as
\begin{equation}\langle AB\rangle_\text{sym}=\frac{1}{2}\Bigl(\langle AB\rangle+\langle BA\rangle\Bigr)\end{equation}
being $\langle A\rangle$ an ordinary quantum average value. Before
working on eq. \eqref{FH} it is evident that a complete description
of the electric and magnetic fields in any region is mandatory.

In our coordinate system the $z$ axis clearly represents a
privileged direction, being the axis perpendicular to the plane
separating the two bodies. Inspired by this property, for the mode
decomposition of our electromagnetic field in any region we
replace the common plane-wave representation in which a mode of
the field is represented by the three-dimensional wavevector
$\mathbf{K}=(k_x,k_y,k_z)$ by a description in terms of the
transverse wavevector $\mathbf{k}=(k_x,k_y)$ and the frequency
$\omega$. The result of this choice is that in this approach the
$z$ component $k_z$ of the wavevector becomes a dependent
variable, defined by the relation
$k_z^2=\frac{\omega^2}{c^2}-\mathbf{k}^2$. Since this relation is
quadratic in $k_z$, we are obliged to introduce explicitly a
variable $\phi$ taking the values $\phi=\pm1$ (with shorthand notation $\phi=\pm$ in the expression of the polarization vectors
and field amplitudes) corresponding to the
sign of $k_z$. As a consequence, the complete wavevector will be
noted with
\begin{equation}\begin{split}\mathbf{K}^\phi&=(\mathbf{k},\phi k_z)=(k_x,k_y,\phi k_z)\\
k_z&=\sqrt{\frac{\omega^2}{c^2}-\mathbf{k}^2}.\end{split}\end{equation}
For $k\leq\frac{\omega}{c}$, this relations gives a real value of
$k_z$, and then a propagative wave, for which the $\phi$
represents the direction of propagation along the $z$ axis. On the
contrary, for $k>\frac{\omega}{c}$, $k_z$ becomes imaginary and we
get an evanescent wave whose amplitude depends on $z$: in this
case $\phi$ is the direction along with the amplitude of the
evanescent wave decays. Moreover, we will as usual need an index
$p$ associated to the polarization, taking the values $p=1,2$
corresponding to TE and TM modes respectively. Finally, in our
approach a mode of the field is identified by the set of variables
$(\omega,\mathbf{k},p,\phi)$.

The expression of the electric field in any region can be first
given under the form a frequency decomposition
\begin{equation}\mathbf{E}(\mathbf{R},t)=2\Rea\Biggl[\int_0^{+\infty}\frac{d\omega}{2\pi}\exp(-i\omega t)\mathbf{E}(\mathbf{R},\omega)\Biggr]\end{equation}
where a single-frequency component has the following mode
decomposition
\begin{equation}\label{DefE}\mathbf{E}(\mathbf{R},\omega)=\sum_{\phi,p}\int\frac{d^2\mathbf{k}}{(2\pi)^2}
\exp(i\mathbf{K}^\phi\cdot\mathbf{R})\hat{\bbm[\epsilon]}_p^\phi(\mathbf{k},\omega)E_p^\phi(\mathbf{k},\omega)\end{equation}
where from now on the sum on $\phi$ runs over the values
$\{+,-\}$, the sum on $p$ over the values $\{1,2\}$. The quantity
$E_p^\phi(\mathbf{k},\omega)$ represents the amplitude of the
electric field associated to a given mode
$(\omega,\mathbf{k},p,\phi)$. For the polarization vectors
$\hat{\bbm[\epsilon]}_p^\phi(\mathbf{k},\omega)$ appearing in eq.
\eqref{DefE} we adopt the following standard definitions
\begin{equation}\label{PolVect}\begin{split}\hat{\bbm[\epsilon]}_\TE^\phi(\mathbf{k},\omega)&=\hat{\mathbf{z}}\times\hat{\mathbf{k}}=\frac{1}{k}(-k_y\hat{\mathbf{x}}+k_x\hat{\mathbf{y}})\\
\hat{\bbm[\epsilon]}_\TM^\phi(\mathbf{k},\omega)&=\frac{c}{\omega}\hat{\bbm[\epsilon]}_\TE^\phi(\mathbf{k},\omega)\times\mathbf{K}^\phi=\frac{c}{\omega}(-k\hat{\mathbf{z}}+\phi
k_z\hat{\mathbf{k}})\\\end{split}\end{equation} where
$\hat{\mathbf{x}}$, $\hat{\mathbf{y}}$ and $\hat{\mathbf{z}}$ are
the unit vectors along the directions $x$, $y$ and $z$
respectively and $\hat{\mathbf{k}}=\mathbf{k}/k$. The unit vectors
defined in \eqref{PolVect} obey the following useful properties
\begin{equation}\label{PropEps}\begin{split}\hat{\bbm[\epsilon]}_\TE^\phi(-\mathbf{k},\omega)&=-\hat{\bbm[\epsilon]}_\TE^\phi(\mathbf{k},\omega)\qquad\hat{\bbm[\epsilon]}_\TE^{-\phi}(\mathbf{k},\omega)=\hat{\bbm[\epsilon]}_\TE^\phi(\mathbf{k},\omega)\\
\Bigl(\hat{\bbm[\epsilon]}_\TE^\phi(\mathbf{k},\omega)\Bigr)^*&=\hat{\bbm[\epsilon]}_\TE^\phi(\mathbf{k},\omega)\qquad\hat{\bbm[\epsilon]}_\TM^\phi(-\mathbf{k},\omega)=\hat{\bbm[\epsilon]}_\TM^{-\phi}(\mathbf{k},\omega)\\
\Bigl(\hat{\bbm[\epsilon]}_\TM^\phi(\mathbf{k},\omega)\Bigr)^*&=\begin{cases}\hat{\bbm[\epsilon]}_\TM^\phi(\mathbf{k},\omega) & k_z\in\mathbb{R}\\
\hat{\bbm[\epsilon]}_\TM^{-\phi}(\mathbf{k},\omega) & k_z\notin\mathbb{R}\\\end{cases}\\
\end{split}\end{equation}

The expression of the magnetic field can be easily deduced from
Maxwell's equations, and is given by
\begin{equation}\label{DefB}\mathbf{B}(\mathbf{R},\omega)=\frac{1}{c}\sum_{\phi,p}\int\frac{d^2\mathbf{k}}{(2\pi)^2}
\exp(i\mathbf{K}^\phi\cdot\mathbf{R})\hat{\bbm[\beta]}_p^\phi(\mathbf{k},\omega)E_p^\phi(\mathbf{k},\omega)\end{equation}
where
\begin{equation}\hat{\bbm[\beta]}_p^\phi(\mathbf{k},\omega)=(-1)^p\hat{\bbm[\epsilon]}_{S(p)}^\phi(\mathbf{k},\omega)\end{equation}
being $S(p)$ the function which switches between the two
polarization, acting as $S(1)=2$ and $S(2)=1$. We are going to
gather the expressions \eqref{DefE} and \eqref{DefB} of the
electric and magnetic fields at a given frequency $\omega$ in a
column vector and write
\begin{equation}\label{Vector}\begin{split}\begin{pmatrix}\mathbf{E}(\mathbf{R},\omega)\\c\mathbf{B}(\mathbf{R},\omega)\end{pmatrix}&=\sum_{\phi,p}\int\frac{d^2\mathbf{k}}{(2\pi)^2}
\exp(i\mathbf{K}^\phi\cdot\mathbf{R})\\
&\,\times\hat{\bbm[\epsilon]}_{A(p)}^\phi(\mathbf{k},\omega)\begin{pmatrix}1
\\ (-1)^p\end{pmatrix}E_p^\phi(\mathbf{k},\omega)\end{split}\end{equation}
where in the right-hand side the 1 and the $(-1)^p$ correspond to
the electric and magnetic field respectively and we have
introduced the vectorial index
\begin{equation}A(p)=\begin{pmatrix}p\\S(p)\end{pmatrix}\end{equation}
where again the upper (lower) index is associated to the electric
(magnetic) field.

Now that the expressions of the electric and magnetic fields are
explicitly given in terms of a set of field amplitudes
$E_p^\phi(\mathbf{k},\omega)$ we are ready to work, in the next
sections, on the explicit expressions of the Maxwell stress tensor
and Poynting vector.

\section{The Maxwell stress tensor}\label{SecT}

\subsection{General expression of the tensor}

From the definition of the stress tensor \eqref{TijS} it is clear
that we need to calculate explicitly the quantities $E_iE_j$ and
$B_iB_j$ for $i,j=x,y,z$ and as a further step their symmetrized
quantum average. Using the compact vectorial notation introduced
in eq. \eqref{Vector}, we have
\begin{widetext}
\begin{equation}\label{EiEj}\begin{split}\begin{pmatrix}E_iE_j\\c^2B_iB_j\end{pmatrix}&=\sum_{\phi\phi'}\sum_{pp'}
\int\frac{d^2\mathbf{k}}{(2\pi)^2}\int\frac{d^2\mathbf{k}'}{(2\pi)^2}\int_0^{+\infty}\frac{d\omega}{2\pi}\int_0^{+\infty}\frac{d\omega'}{2\pi}\begin{pmatrix}1 \\ (-1)^{p+p'}\end{pmatrix}\exp[i(\mathbf{k}-\mathbf{k}')\cdot\mathbf{r}]\exp[i(\phi k_z-\phi'k_z^{'*})z]\\
&\,\times\exp[-i(\omega-\omega')t]\Bigl\{E^\phi_p(\mathbf{k},\omega)E^{\phi'\dag}_{p'}(\mathbf{k}',\omega')\Bigl(\hat{\mathbf{\epsilon}}^\phi_{A(p)}(\mathbf{k},\omega)\Bigr)_i\Bigl(\hat{\mathbf{\epsilon}}^{\phi'*}_{A(p')}(\mathbf{k}',\omega')\Bigr)_j\\
&\,+ E^{\phi'\dag}_{p'}(\mathbf{k}',\omega')E^\phi_p(\mathbf{k},\omega)\Bigl(\hat{\mathbf{\epsilon}}^\phi_{A(p)}(\mathbf{k},\omega)\Bigr)_j\Bigl(\hat{\mathbf{\epsilon}}^{\phi'*}_{A(p')}(\mathbf{k}',\omega')\Bigr)_i\Bigr\}+\mathcal{Z}.\end{split}\end{equation}
\end{widetext}
where the fields implicitly depend on $(\mathbf{R},t)$ and
$\mathcal{Z}$ generically gathers all the terms proportional to
$EE$ or $E^\dag E^\dag$ whose average quantum value is zero.
Having in mind the transition to the quantum symmetrized average
values, we now give the following definition for the commutators
$C^{\phi\phi'}$ of the field amplitudes
\begin{equation}\label{CommC}\begin{split}\langle E^\phi_p(\mathbf{k},\omega)E^{\phi'\dag}_{p'}(\mathbf{k}',\omega')\rangle_\text{sym}&=\frac{1}{2}\langle E^\phi_p(\mathbf{k},\omega)E^{\phi'\dag}_{p'}(\mathbf{k}',\omega')\\
&\,+E^{\phi'\dag}_{p'}(\mathbf{k}',\omega')E^\phi_p(\mathbf{k},\omega)\rangle\\
&=2\pi\delta(\omega-\omega')\bra{p,\mathbf{k}}C^{\phi\phi'}\ket{p',\mathbf{k}'}\end{split}\end{equation}
where we stress the fact that in general two modes of the field
propagating in opposite directions do not necessarily commute.
Moreover we have explicitly inserted the conservation of
frequency: since in our system no dynamics is considered, the
field amplitudes at different frequencies necessarily commute as a
consequence of time invariance. In virtue of this conservation any correlation function of the electromagnetic field analogous to \eqref{CommC} is function of a single frequency $\omega$ (and not of both $\omega$ and
$\omega'$), which will not be explicitly
written from now on. Moreover, in eq. \eqref{CommC} we have
expressed the correlators as matrix elements of a matrix
$C^{\phi\phi'}$: in our notation, the matrices are defined on the
space $(p,\mathbf{k})$ being $p=1,2$ and
$\mathbf{k}\in\mathbb{R}^2$ and thus the product of two matrices
$A$ and $B$ is given by
\begin{equation}\begin{split}\bra{p,\mathbf{k}}\mathcal{A}\mathcal{B}\ket{p',\mathbf{k}'}&=\sum_{p''}\int\frac{d^2\mathbf{k}''}{(2\pi)^2}\bra{p,\mathbf{k}}\mathcal{A}\ket{p'',\mathbf{k}''}\\
&\,\times\bra{p'',\mathbf{k}''}\mathcal{B}\ket{p',\mathbf{k}'}\end{split}\end{equation}
where the matrices $\mathcal{A}$ and $\mathcal{B}$ are both calculated at the same fixed frequency $\omega$.

Using eq. \eqref{EiEj} and the definition \eqref{CommC} we have
\begin{equation}\label{EiEj2}\begin{split}&\langle\begin{pmatrix}E_iE_j\\c^2B_iB_j\end{pmatrix}\rangle_\text{sym}=\sum_{\phi\phi'}\sum_{pp'}
\int\frac{d^2\mathbf{k}}{(2\pi)^2}\int\frac{d^2\mathbf{k}'}{(2\pi)^2}\int_0^{+\infty}\frac{d\omega}{2\pi}\\
&\,\times\begin{pmatrix}1 \\ (-1)^{p+p'}\end{pmatrix}\exp[i(\mathbf{k}-\mathbf{k}')\cdot\mathbf{r}]\exp[i(\phi k_z-\phi'k_z^*)z]\\
&\,\times\bra{p,\mathbf{k}}C^{\phi\phi'}\ket{p',\mathbf{k}'}\Bigl\{\Bigl(\hat{\mathbf{\epsilon}}^\phi_{A(p)}(\mathbf{k},\omega)\Bigr)_i\Bigl(\hat{\mathbf{\epsilon}}^{\phi'*}_{A(p')}(\mathbf{k}',\omega)\Bigr)_j\\
&\,+\Bigl(\hat{\mathbf{\epsilon}}^\phi_{A(p)}(\mathbf{k},\omega)\Bigr)_j\Bigl(\hat{\mathbf{\epsilon}}^{\phi'*}_{A(p')}(\mathbf{k}',\omega)\Bigr)_i\Bigr\}.\end{split}\end{equation}
It is important to note that the correlator $C^{\phi\phi'}$ defined in \eqref{CommC} and appearing in \eqref{EiEj2} depends on the region (A, B or C) in which the average is calculated: this information is contained in the
$z$ dependence of the fields in the left-hand side of eq. \eqref{EiEj2}, as well as in the $z$ coordinate explicitly present in its right-hand side. As we will see in section \ref{SubFlux}, we only need to evaluate
the symmetrized average value of the flux of the $iz$ components
($i=x,y,z$) of the electromagnetic stress tensor. These quantities
can be calculated using eqs. \eqref{TijS} and \eqref{EiEj2}. In
the case of $T_{zz}$ we have
\begin{widetext}
\begin{equation}\label{Tzz}\begin{split}&\langle T_{zz}\rangle_\text{sym}=\epsilon_0\sum_{\phi\phi'}\sum_{pp'}
\int\frac{d^2\mathbf{k}}{(2\pi)^2}\int\frac{d^2\mathbf{k}'}{(2\pi)^2}\int_0^{+\infty}\frac{d\omega}{2\pi}\exp[i(\mathbf{k}-\mathbf{k}')\cdot\mathbf{r}]\exp[i(\phi k_z-\phi'k_z^{'*})z]\bra{p,\mathbf{k}}C^{\phi\phi'}\ket{p',\mathbf{k}'}\\
&\,\times \Bigl\{\Bigl(\hat{\mathbf{\epsilon}}^\phi_p(\mathbf{k},\omega)\Bigr)_z\Bigl(\hat{\mathbf{\epsilon}}^{\phi'*}_{p'}(\mathbf{k}',\omega)\Bigr)_z-\Bigl(\hat{\mathbf{\epsilon}}^\phi_p(\mathbf{k},\omega)\Bigr)_x\Bigl(\hat{\mathbf{\epsilon}}^{\phi'*}_{p'}(\mathbf{k}',\omega)\Bigr)_x-\Bigl(\hat{\mathbf{\epsilon}}^\phi_p(\mathbf{k},\omega)\Bigr)_y\Bigl(\hat{\mathbf{\epsilon}}^{\phi'*}_{p'}(\mathbf{k}',\omega)\Bigr)_y\\
&\,+(-1)^{p+p'}\Bigl[\Bigl(\hat{\mathbf{\epsilon}}^\phi_{S(p)}(\mathbf{k},\omega)\Bigr)_z\Bigl(\hat{\mathbf{\epsilon}}^{\phi'*}_{S(p')}(\mathbf{k}',\omega)\Bigr)_z-\Bigl(\hat{\mathbf{\epsilon}}^\phi_{S(p)}(\mathbf{k},\omega)\Bigr)_x\Bigl(\hat{\mathbf{\epsilon}}^{\phi'*}_{S(p')}(\mathbf{k}',\omega)\Bigr)_x-\Bigl(\hat{\mathbf{\epsilon}}^\phi_{S(p)}(\mathbf{k},\omega)\Bigr)_y\Bigl(\hat{\mathbf{\epsilon}}^{\phi'*}_{S(p')}(\mathbf{k}',\omega)\Bigr)_y\Bigr]\Bigr\}.\end{split}\end{equation}
whereas for $T_{mz}$ (being $m=x,y$) a straightforward calculation
gives
\begin{equation}\label{Tmz}\begin{split}&\langle T_{mz}\rangle_\text{sym}=\epsilon_0\sum_{\phi\phi'}\sum_{pp'}
\int\frac{d^2\mathbf{k}}{(2\pi)^2}\int\frac{d^2\mathbf{k}'}{(2\pi)^2}\int_0^{+\infty}\frac{d\omega}{2\pi}\exp[i(\mathbf{k}-\mathbf{k}')\cdot\mathbf{r}]\exp[i(\phi k_z-\phi'k_z^{'*})z]\bra{p,\mathbf{k}}C^{\phi\phi'}\ket{p',\mathbf{k}'}\\
&\,\times\Bigl\{\Bigl[\Bigl(\hat{\mathbf{\epsilon}}^\phi_p(\mathbf{k},\omega)\Bigr)_m\Bigl(\hat{\mathbf{\epsilon}}^{\phi'*}_{p'}(\mathbf{k}',\omega)\Bigr)_z+(-1)^{p+p'}\Bigl(\hat{\mathbf{\epsilon}}^\phi_{S(p)}(\mathbf{k},\omega)\Bigr)_m\Bigl(\hat{\mathbf{\epsilon}}^{\phi'*}_{S(p')}(\mathbf{k}',\omega)\Bigr)_z\Bigr]+[m\rightleftarrows
z]\Bigr\}.\end{split}\end{equation}
\end{widetext}
where the symbol $m\rightleftarrows z$ represents two more terms
obtained by interchanging in the first two $m$ with $z$. Let us
make some comments on these expressions. First we note that, even
if the stress tensor is in general a function of time, this is not
true for its quantum average value. Besides, this is a function of the
correlators of the electric-field amplitudes (propagating either
in the same or in opposite directions) and of the components of
the polarization unit vectors. As we will see in the next section,
the components \eqref{Tzz} and \eqref{Tmz} are the only ones we need to
calculate all the components of the force acting on the two
bodies.

\subsection{Flux of the stress tensor in terms of field correlators}\label{SubFlux}

\begin{figure}[htb]\centering
\includegraphics[height=5.5cm]{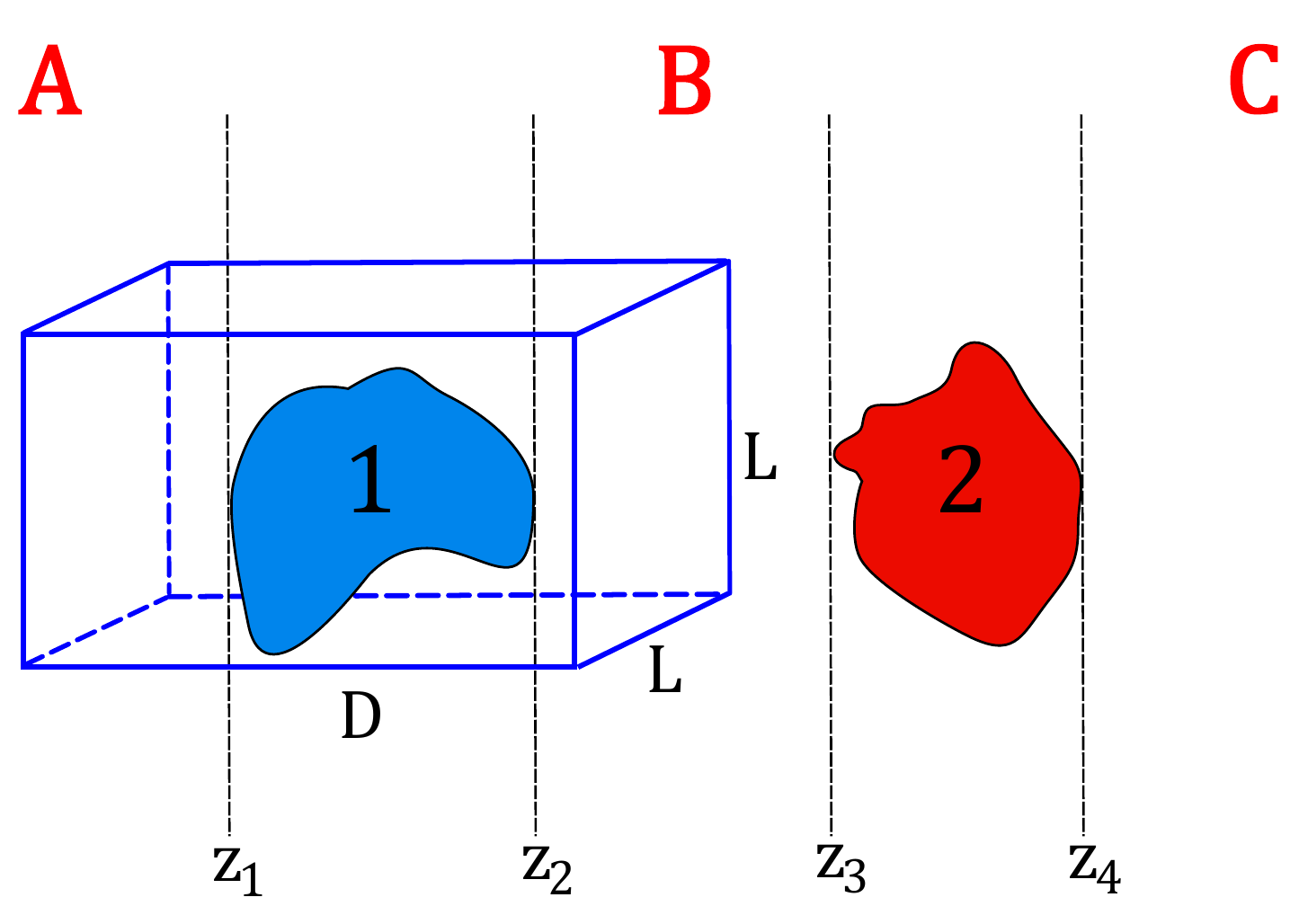}
\caption{The \emph{box} chosen for the calculation of the force
acting on body 1.}\label{Box}\end{figure}

As anticipated, the electromagnetic force acting on a given body inside a volume region
can be calculated by taking the flux of the stress tensor
through a closed surface enclosing this volume. As a consequence,
the $m$ component of the force on a given body is given by the
flux
\begin{equation}F_m=\int_\Sigma\langle T_{mj}\rangle_\text{sym}\,d\Sigma_j\end{equation}
where the summation over repeated indices is assumed and $\Sigma$
represents any closed \emph{box} entirely enclosing the body. Let us
choose for example the box depicted in fig. \ref{Box}, i.e. a
parallelepiped having one side of length $D$ and as a base
orthogonal to the $z$ axis a square of side $L$.
 According to the
definition of $T_{ij}$, the $m$ component of the force ($m=x,y,z$)
is given in this case by the flux of $T_{mz}$ through the two
surfaces orthogonal to the $z$ axis, plus the fluxes of $T_{mx}$
and $T_{my}$ through the surfaces of the parallelepiped orthogonal
to the $x$ and $y$ axes respectively. Taking now the limit
$L\to+\infty$ we see that the surface of the two bases orthogonal
to the $z$ axis diverges more rapidly (like $L^2$) than the other
four surfaces (like $L$). As a consequence, we deduce that in
order to calculate $F_m$ one simply needs to calculate the flux of
$T_{mz}$ on the surface (which has now become a plane) in region A
and subtract this result from the flux of $T_{mz}$ through the
plane in region B. Moreover, due to the arbitrariness of the box,
these two fluxes must not depend on the $z$ coordinates of the
respective planes, even if in general the stress tensor depends on
$z$.

From this discussion we conclude that we need the flux of $T_{mz}$
through a plane $z=\bar{z}$. We will then have $\bar{z}<z_1$ for
region A, $z_2<\bar{z}<z_3$ for region B and $\bar{z}>z_4$ for
region C. Integrating eq. \eqref{Tzz} on the plane $z=\bar{z}$ and
noticing that this gives a Dirac delta
$(2\pi)^2\delta(\mathbf{k}-\mathbf{k}')$, we get the flux of
$T_{zz}$ expressed as a function of the field correlators
\begin{equation}\begin{split}&\Phi_z(\bar{z})=\int_{z=\bar{z}}d^2\mathbf{r}\,\langle T_{zz}\rangle_\text{sym}\\
&\,=\epsilon_0\sum_{\phi\phi'}\sum_{pp'}
\int\frac{d^2\mathbf{k}}{(2\pi)^2}\int_0^{+\infty}\frac{d\omega}{2\pi}\exp[i(\phi k_z-\phi'k_z^*)\bar{z}]\\
&\,\times\Bigl\{\Bigl(\hat{\mathbf{\epsilon}}^\phi_p\Bigr)_z\Bigl(\hat{\mathbf{\epsilon}}^{\phi'*}_{p'}\Bigr)_z-\Bigl(\hat{\mathbf{\epsilon}}^\phi_p\Bigr)_x\Bigl(\hat{\mathbf{\epsilon}}^{\phi'*}_{p'}\Bigr)_x-\Bigl(\hat{\mathbf{\epsilon}}^\phi_p\Bigr)_y\Bigl(\hat{\mathbf{\epsilon}}^{\phi'*}_{p'}\Bigr)_y\\
&\,+(-1)^{p+p'}\Bigl[\Bigl(\hat{\mathbf{\epsilon}}^\phi_{S(p)}\Bigr)_z\Bigl(\hat{\mathbf{\epsilon}}^{\phi'*}_{S(p')}\Bigr)_z-\Bigl(\hat{\mathbf{\epsilon}}^\phi_{S(p)}\Bigr)_x\Bigl(\hat{\mathbf{\epsilon}}^{\phi'*}_{S(p')}\Bigr)_x\\
&\,-\Bigl(\hat{\mathbf{\epsilon}}^\phi_{S(p)}\Bigr)_y\Bigl(\hat{\mathbf{\epsilon}}^{\phi'*}_{S(p')}\Bigr)_y\Bigr]\Bigr\}\bra{p,\mathbf{k}}C^{\phi\phi'}\ket{p',\mathbf{k}}\end{split}\end{equation}
where all the polarization unit vectors are calculated in
$(\mathbf{k},\omega)$. For the other two components of the stress
tensor ($m=x,y$) we have, integrating eq. \eqref{Tmz},
\begin{equation}\begin{split}&\Phi_m(\bar{z})=\int_{z=\bar{z}}d^2\mathbf{r}\,\langle T_{mz}\rangle_\text{sym}\\
&\,=\epsilon_0\sum_{\phi\phi'}\sum_{pp'}
\int\frac{d^2\mathbf{k}}{(2\pi)^2}\int_0^{+\infty}\frac{d\omega}{2\pi}\exp[i(\phi k_z-\phi'k_z^*)\bar{z}]\\
&\,\times\Bigl\{\Bigl[\Bigl(\hat{\mathbf{\epsilon}}^\phi_p\Bigr)_m\Bigl(\hat{\mathbf{\epsilon}}^{\phi'*}_{p'}\Bigr)_z+(-1)^{p+p'}\Bigl(\hat{\mathbf{\epsilon}}^\phi_{S(p)}\Bigr)_m\Bigl(\hat{\mathbf{\epsilon}}^{\phi'*}_{S(p')}\Bigr)_z\Bigr]\\
&\,+[m\rightleftarrows
z]\Bigr\}\bra{p,\mathbf{k}}C^{\phi\phi'}\ket{p',\mathbf{k}}.\end{split}\end{equation}
From the definitions \eqref{PolVect} of the polarization unit
vectors we deduce that
\begin{equation}\begin{split}&\Bigl(\hat{\mathbf{\epsilon}}^\phi_p\Bigr)_z\Bigl(\hat{\mathbf{\epsilon}}^{\phi'*}_{p'}\Bigr)_z-\Bigl(\hat{\mathbf{\epsilon}}^\phi_p\Bigr)_x\Bigl(\hat{\mathbf{\epsilon}}^{\phi'*}_{p'}\Bigr)_x-\Bigl(\hat{\mathbf{\epsilon}}^\phi_p\Bigr)_y\Bigl(\hat{\mathbf{\epsilon}}^{\phi'*}_{p'}\Bigr)_y\\
&\,+(-1)^{p+p'}\Bigl[\Bigl(\hat{\mathbf{\epsilon}}^\phi_{S(p)}\Bigr)_z\Bigl(\hat{\mathbf{\epsilon}}^{\phi'*}_{S(p')}\Bigr)_z-\Bigl(\hat{\mathbf{\epsilon}}^\phi_{S(p)}\Bigr)_x\Bigl(\hat{\mathbf{\epsilon}}^{\phi'*}_{S(p')}\Bigr)_x\\
&\,-\Bigl(\hat{\mathbf{\epsilon}}^\phi_{S(p)}\Bigr)_y\Bigl(\hat{\mathbf{\epsilon}}^{\phi'*}_{S(p')}\Bigr)_y\Bigr]=-\delta_{pp'}\frac{c^2}{\omega^2}(k_z^2+\phi\phi'|k_z|^2)\end{split}\end{equation}
and
\begin{equation}\begin{split}&\Bigl[\Bigl(\hat{\mathbf{\epsilon}}^\phi_p\Bigr)_m\Bigl(\hat{\mathbf{\epsilon}}^{\phi'*}_{p'}\Bigr)_z+(-1)^{p+p'}\Bigl(\hat{\mathbf{\epsilon}}^\phi_{S(p)}\Bigr)_m\Bigl(\hat{\mathbf{\epsilon}}^{\phi'*}_{S(p')}\Bigr)_z\Bigr]+\Bigl[m\rightleftarrows z\Bigr]\\
&\,=-\delta_{pp'}\frac{c^2k_m}{\omega^2}(\phi k_z+\phi'k_z^*).\end{split}\end{equation}
We now observe that
\begin{equation}\begin{split}k_z^2+\phi\phi'|k_z|^2=\begin{cases}k_z^2+|k_z|^2 & \phi=\phi'\\
k_z^2-|k_z|^2 & \phi\neq\phi'\\\end{cases}\\
\phi k_z+\phi'k_z^*=\begin{cases}2\phi\Rea k_z & \phi=\phi'\\
2i\phi\Ima k_z &
\phi\neq\phi'\\\end{cases}\end{split}\end{equation} For both
expressions, in the former case $(\phi=\phi')$ only the
contribution coming from propagative waves plays a role, whilst in
the latter only evanescent waves are relevant. At the same time,
for both expressions, and both for $\phi=\phi'$ and
$\phi\neq\phi'$, the exponential term containing $\bar{z}$
disappears, as expected: whereas the stress tensor depends on $z$,
this is not the case for its flux calculated on a plane having an
arbitrary position $z=\bar{z}$, provided that $z=\bar{z}$ remains in a given region (A, B or C). Finally, the flux of $T_{mz}$ on
the plane $z=\bar{z}$ can be cast for any $m=x,y,z$ in the
form
\begin{equation}\label{FluxTmz}\begin{split}&\Phi_m(\bar{z})=-\sum_p
\int\frac{d^2\mathbf{k}}{(2\pi)^2}\Biggl(\sum_{\phi=\phi'}\int_{ck}^{+\infty}\frac{d\omega}{2\pi}+\sum_{\phi\neq\phi'}\int_0^{ck}\frac{d\omega}{2\pi}\Biggr)\\
&\,\times\frac{2\epsilon_0c^2k_z}{\omega^2}\bra{p,\mathbf{k}}C^{\phi\phi'}\ket{p,\mathbf{k}}\times\begin{cases}\phi
k_m & m=x,y\\k_z & m=z\end{cases}\end{split}\end{equation}
This equation represents the main result of this section, describing the flux of the component $T_{mz}$ of the stress tensor as a function of the field correlators. As we have shown before, this quantity, together with the explicit
knowledge of the correlators in any region of our system which will be discussed in Sec. \ref{FieldCorr}, is sufficient to deduce any component of the force acting on the two bodies. From eq. \eqref{FluxTmz} we deduce that the flux is written as the
sum of two separate contributions, coming from the propagative and evanescent sectors respectively: the former depends on the
correlators between the field propagating in a direction $\phi$ and itself, whilst the latter implies the correlators of
counterpropagating fields. We also remark that the quantity $C^{\phi\phi'}$ is the only term in eq. \eqref{FluxTmz} depending
on the region in which the flux is calculated, which means on the position of $\bar{z}$. Finally we observe that the result deduced
for the flux of $T_{zz}$ ($m=z$ in eq. \eqref{FluxTmz}) coincides with the expression obtained in \cite{BimontePRA09}.

\section{The Poynting vector}\label{SecS}

\subsection{General expression of the vector}

Let us now focus our attention on the Poynting vector defined in
eq. \eqref{TijS}. In order to evaluate its quantum symmetrized
average we first need to work out the generic field product $E_iB_j$.
Using the same conventions of the last section we obtain
\begin{widetext}
\begin{equation}\begin{split}&\begin{pmatrix}cE_iB_j\\cB_jE_i\end{pmatrix}=\sum_{\phi\phi'}\sum_{pp'}
\int\frac{d^2\mathbf{k}}{(2\pi)^2}\int\frac{d^2\mathbf{k}'}{(2\pi)^2}\int_0^{+\infty}\frac{d\omega}{2\pi}\int_0^{+\infty}\frac{d\omega'}{2\pi}\\
&\qquad\times\Bigl\{\exp[i(\mathbf{K}^\phi-\mathbf{K}^{'\phi'*})\cdot\mathbf{R}]\exp[-i(\omega-\omega')t](-1)^{p'}\begin{pmatrix}E^\phi_p(\mathbf{k},\omega)E^{\phi'\dag}_{p'}(\mathbf{k}',\omega')\\E^{\phi'\dag}_{p'}(\mathbf{k}',\omega')E^\phi_p(\mathbf{k},\omega)\end{pmatrix}\Bigl(\hat{\mathbf{\epsilon}}^\phi_p(\mathbf{k},\omega)\Bigr)_i\Bigl(\hat{\mathbf{\epsilon}}^{\phi'*}_{S(p')}(\mathbf{k}',\omega')\Bigr)_j\\
&\qquad+\exp[i(-\mathbf{K}^{\phi*}+\mathbf{K}^{'\phi'})\cdot\mathbf{R}]\exp[i(\omega-\omega')t](-1)^{p'}\begin{pmatrix}E^{\phi\dag}_p(\mathbf{k},\omega)E^{\phi'}_{p'}(\mathbf{k}',\omega')\\E^{\phi'}_{p'}(\mathbf{k}',\omega')E^{\phi\dag}_p(\mathbf{k},\omega)\end{pmatrix}\Bigl(\hat{\mathbf{\epsilon}}^{\phi*}_p(\mathbf{k},\omega)\Bigr)_i\Bigl(\hat{\mathbf{\epsilon}}^{\phi'}_{S(p')}(\mathbf{k}',\omega')\Bigr)_j\Bigr\}+\mathcal{Z}\end{split}\end{equation}
\end{widetext}
from which we immediately get
\begin{equation}\label{EiBj}\begin{split}&\langle cE_iB_j\rangle_\text{sym}=\sum_{\phi\phi'}\sum_{pp'}
\int\frac{d^2\mathbf{k}}{(2\pi)^2}\int\frac{d^2\mathbf{k}'}{(2\pi)^2}\int_0^{+\infty}\frac{d\omega}{2\pi}\\
&\,\times\exp[i(\mathbf{k}-\mathbf{k}')\cdot\mathbf{r}]\exp[i(\phi k_z-\phi'k_z^*)z]\\
&\,\times\bra{p,\mathbf{k}}C^{\phi\phi'}\ket{p',\mathbf{k}'}\Bigl\{(-1)^{p'}\Bigl(\hat{\mathbf{\epsilon}}^\phi_p(\mathbf{k},\omega)\Bigr)_i\Bigl(\hat{\mathbf{\epsilon}}^{\phi'*}_{S(p')}(\mathbf{k}',\omega)\Bigr)_j\\
&\,+(-1)^p\Bigl(\hat{\mathbf{\epsilon}}^\phi_{S(p)}(\mathbf{k},\omega)\Bigr)_j\Bigl(\hat{\mathbf{\epsilon}}^{\phi'*}_{p'}(\mathbf{k}',\omega)\Bigr)_i\Bigr\}\end{split}\end{equation}
with the same conventions of eq. \eqref{EiEj}.

\subsection{Flux of the Poynting vector in terms of field correlators}

We now observe that, in virtue of the same discussion about the
closed surface $\Sigma$ given in the last section, we only need
the flux of the $z$ component of the Poynting vector in order to
evaluate the heat flux on one of the two bodies. For the flux of
$S_z$ on the plane $z=\bar{z}$ we have
\begin{equation}\begin{split}&\varphi(\bar{z})=\int_{z=\bar{z}}d^2\mathbf{r}\,\langle S_z\rangle_\text{sym}\\
&\,=\epsilon_0c\sum_{\phi\phi'}\sum_{pp'}
\int\frac{d^2\mathbf{k}}{(2\pi)^2}\int_0^{+\infty}\frac{d\omega}{2\pi}\exp[i(\phi k_z-\phi'k_z^*)\bar{z}]\\
&\,\times\Bigl\{\Bigl[(-1)^{p'}\Bigl(\hat{\mathbf{\epsilon}}^\phi_p\Bigr)_x\Bigl(\hat{\mathbf{\epsilon}}^{\phi'*}_{S(p')}\Bigr)_y+(-1)^p\Bigl(\hat{\mathbf{\epsilon}}^\phi_{S(p)}\Bigr)_y\Bigl(\hat{\mathbf{\epsilon}}^{\phi'*}_{p'}\Bigr)_x\Bigr]\\
&\,-[x\rightleftarrows
y]\Bigr\}\bra{p,\mathbf{k}}C^{\phi\phi'}\ket{p',\mathbf{k}}\end{split}\end{equation}
where we stress the fact that the two terms obtained by
interchanging $x$ and $y$ must in this case be changed in sign. We
then obtain
\begin{equation}\begin{split}&\Bigl[(-1)^{p'}\Bigl(\hat{\mathbf{\epsilon}}^\phi_p\Bigr)_x\Bigl(\hat{\mathbf{\epsilon}}^{\phi'*}_{S(p')}\Bigr)_y+(-1)^p\Bigl(\hat{\mathbf{\epsilon}}^\phi_{S(p)}\Bigr)_y\Bigl(\hat{\mathbf{\epsilon}}^{\phi'*}_{p'}\Bigr)_x\Bigr]-[x\rightleftarrows y]\\
&=\delta_{pp'}\frac{c}{\omega}(\phi k_z+\phi'k_z^*)
\end{split}\end{equation}
and finally cast the expression of the flux of the $z$ component of the Poynting vector
under the form
\begin{equation}\label{FluxSz}\begin{split}&\varphi(\bar{z})=\sum_p
\int\frac{d^2\mathbf{k}}{(2\pi)^2}\Biggl(\sum_{\phi=\phi'}\int_{ck}^{+\infty}\frac{d\omega}{2\pi}+\sum_{\phi\neq\phi'}\int_0^{ck}\frac{d\omega}{2\pi}\Biggr)\\
&\,\times\frac{2\epsilon_0c^2\phi
k_z}{\omega}\bra{p,\mathbf{k}}C^{\phi\phi'}\ket{p,\mathbf{k}}.\end{split}\end{equation}
This expression is the analogue for the Poynting vector of eq. \eqref{FluxTmz}. It constitutes, together with the expression of the matrix elements of $C^{\phi\phi'}$ at any frequency, the main ingredient in the calculation
of the heat transfer on the two bodies.

\section{Scattering formalism: reflection and transmission operators}\label{SRT}

\begin{figure}[htb]\centering
\includegraphics[height=4.5cm]{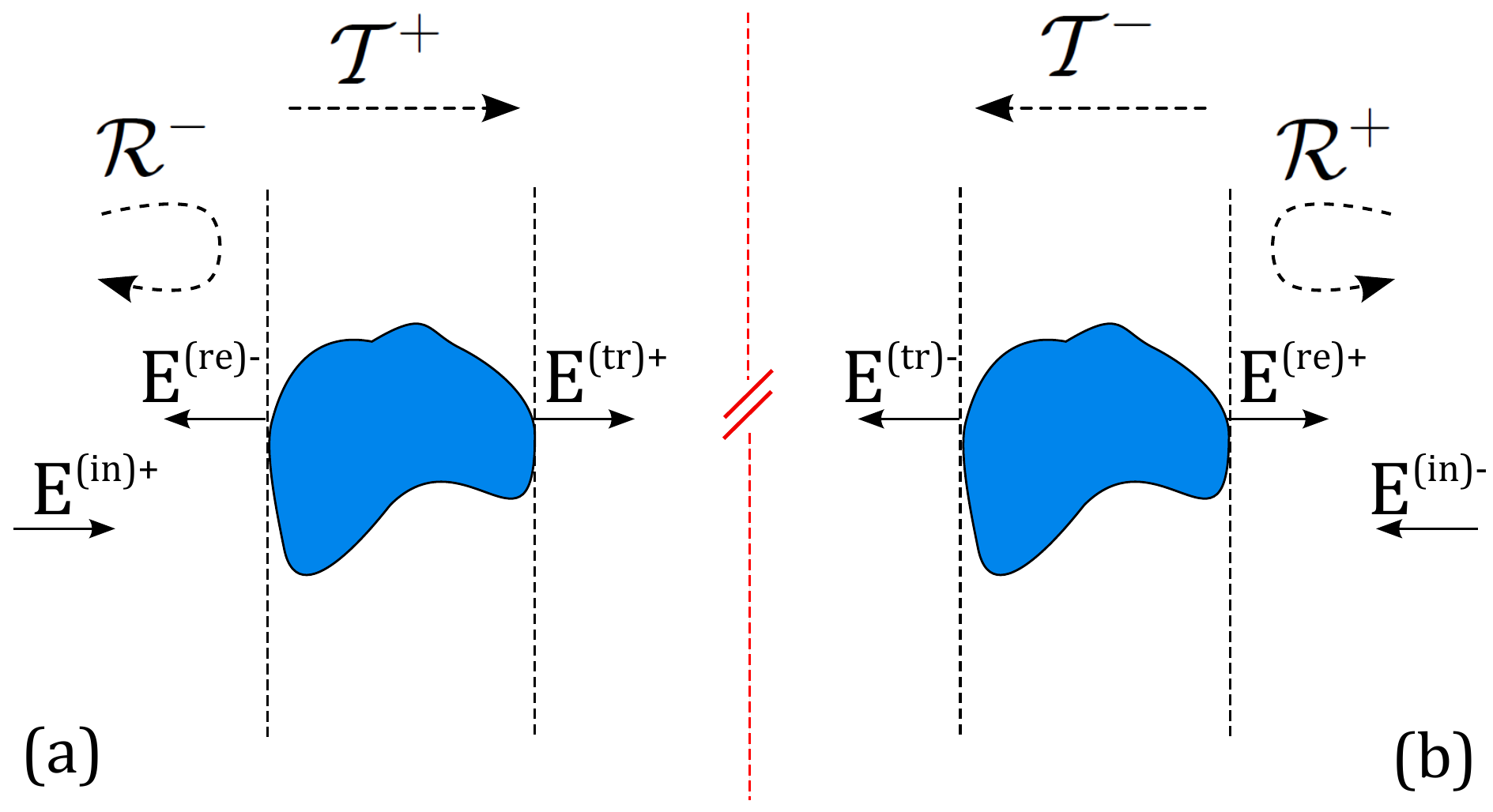}
\caption{The definition of the reflection and transmission
matrices.} \label{Fig3}\end{figure}

The quantities $C^{\phi\phi'}$ appearing in eqs. \eqref{FluxTmz}
and \eqref{FluxSz} and defined in eq. \eqref{CommC} are the
correlators of the total fields in each region. In our problem
these fields result from the ones emitted by the two bodies and
the environmental field, as well as from all the possible
scattering processes undergone in presence of the two bodies 1 and
2. We then need to introduce a set of operators describing the
scattering produced in presence of a single arbitrary body. Let us
suppose to have a body located in the region $z_1<z<z_2$.
Let us further assume that an external field is impinging on our
body, either from its left or from its right side. This field will
be scattered upon the body, producing in this way new components
of the field on both sides of the body. In particular, the field
coming from the left (right) will produce a reflected field
propagating toward the left (right) on the left (right) side, and
a transmitted one propagating toward the right (left) on the right
(left) side. The two possible configurations, depending on the
direction of propagation of the incoming field, are represented in
Figure \ref{Fig3}.
 The reflection and
transmission matrices $\mathcal{R}^\pm$ and $\mathcal{T}^\pm$ are
the operators linking each mode of the outgoing fields to the
incoming ones.

In particular, considering the case of a field coming from the
left side, the incoming field
\begin{equation}\label{Ein}\begin{split}\mathbf{E}^{\text{(in)+}}(\mathbf{R},t)&=2\Rea\Biggl[\sum_p\int_0^{+\infty}\frac{d\omega}{2\pi}\int\frac{d^2\mathbf{k}}{(2\pi)^2}\exp[i\mathbf{K}^+\cdot\mathbf{R}]\\
&\,\exp[-i\omega
t]\hat{\bbm[\epsilon]}_p^+(\mathbf{k},\omega)E_p^{\text{(in)+}}(\mathbf{k},\omega)\Biggr]\\\end{split}\end{equation}
will result in a reflected (on the left) and a transmitted (on the
right) field defined by
\begin{equation}\label{Ere}\begin{split}\mathbf{E}^{\text{(re)-}}(\mathbf{R},t)&=2\Rea\Biggl[\sum_p\int_0^{+\infty}\frac{d\omega}{2\pi}\int\frac{d^2\mathbf{k}}{(2\pi)^2}\exp[i\mathbf{K}^-\cdot\mathbf{R}]\\
&\,\exp[-i\omega
t]\hat{\bbm[\epsilon]}_p^-(\mathbf{k},\omega)E_p^{\text{(re)-}}(\mathbf{k},\omega)\Biggr]\\\end{split}\end{equation}
\begin{equation}\label{Etr}\begin{split}\mathbf{E}^{\text{(tr)+}}(\mathbf{R},t)&=2\Rea\Biggl[\sum_p\int_0^{+\infty}\frac{d\omega}{2\pi}\int\frac{d^2\mathbf{k}}{(2\pi)^2}\exp[i\mathbf{K}^+\cdot\mathbf{R}]\\
&\,\exp[-i\omega
t]\hat{\bbm[\epsilon]}_p^+(\mathbf{k},\omega)E_p^{\text{(tr)+}}(\mathbf{k},\omega)\Biggr].\\\end{split}\end{equation}
As remarked before, since the scattering process is stationary,
the frequency is conserved. We are thus able to define the
operators $\mathcal{R}^-$ and $\mathcal{T}^+$ through the
following relations involving the amplitudes defined in eqs.
\eqref{Ein}, \eqref{Ere} and \eqref{Etr}
\begin{equation}E_p^{\text{(re)-}}(\mathbf{k},\omega)=\sum_{p'}\int\frac{d^2\mathbf{k}'}{(2\pi)^2}\bra{p,\mathbf{k}}\mathcal{R}^-\ket{p',\mathbf{k}'}E_{p'}^{\text{(in)+}}(\mathbf{k}',\omega)\end{equation}
\begin{equation}E_p^{\text{(tr)+}}(\mathbf{k},\omega)=\sum_{p'}\int\frac{d^2\mathbf{k}'}{(2\pi)^2}\bra{p,\mathbf{k}}\mathcal{T}^+\ket{p',\mathbf{k}'}E_{p'}^{\text{(in)+}}(\mathbf{k}',\omega)\end{equation}
connecting each mode of the outgoing fields to all the modes of
the incoming one at the same frequency $\omega$. In analogy with the field correlators, we
are going to drop, for the sake of simplicity, the dependence on
the frequency $\omega$ in the reflection and transmission
operators. A perfectly analogous procedure leads to the definition
of the scattering operators $\mathcal{R}^-$ and $\mathcal{T}^+$.

Using the formalism we have just introduced, the case of the
absence of a given body is obtained, as far as its scattering
operators are concerned, by imposing
\begin{equation}\mathcal{R}^\phi=0\qquad\qquad\mathcal{T}^\phi=1.\end{equation}
As we will see later, it is convenient to introduce a modified
transmission operator which, in analogy with the reflection
operator, goes to zero as well in absence of the body. We thus
define
\begin{equation}\mathcal{T}^\phi=1+\tilde{\mathcal{T}}^\phi\end{equation}
writing the transmission operator as the sum of the identity,
describing the incoming field propagating unmodified on the other
side of the body, and of a new operator $\tilde{\mathcal{T}}^\phi$
accounting only for the scattered part of the field. In the limit
of the absence of the body, we have as desired
$\tilde{\mathcal{T}}^\phi=0$.

\section{Field correlators}\label{FieldCorr}

In order to proceed further and to calculate the force and heat
transfer on body 1, we need an expression for the correlators
$\bra{p,\mathbf{k}}C^{\phi\phi'}\ket{p',\mathbf{k}'}$, defined in
eq. \eqref{CommC}, in each region of our system. These correlators
will be expressed as a function of the correlators of the field
emitted by each body by means of the scattering operators
introduced in Sec. \ref{SRT}. In this section we will first
address the characterization of the single-body field correlators
(with this expression we refer both to the bodies 1 and 2 and to
the environment), and immediately after express the total-field
correlators as a function of these quantities through the
scattering operators. We will note with
$E^{(\gamma)\phi}_p(\mathbf{k},\omega)$ each mode of the
\emph{total} field propagating in direction $\phi$ in the region
$\gamma=A,B,C$, as shown in figure \ref{Fig1}. In order to
calculate the fluxes \eqref{FluxTmz} and \eqref{FluxSz} in the
three regions, we need to know the expression for the the
correlators $C^{\phi\phi'}_\gamma$ defined by
\begin{equation}\label{CorrRegion}\begin{split}\langle E^{(\gamma)\phi}_p(\mathbf{k},\omega)E^{(\gamma)\phi'\dag}_{p'}(\mathbf{k}',\omega')\rangle_\text{sym}&=2\pi\delta(\omega-\omega')\\
&\,\times\bra{p,\mathbf{k}}C^{\phi\phi'}_\gamma\ket{p',\mathbf{k}'}.\end{split}\end{equation}
For a system as the one in figure 1 at \emph{thermal equilibrium}
at temperature $T$ with the environment, the correlators of the
total electromagnetic field outside the body follow from the
fluctuation-dissipation theorem \cite{Landau63}
\begin{equation}\label{FluDiss}\begin{split}\langle E^\text{(tot)}_i(\mathbf{R},\omega)E_j^{\text{(tot)}\dag}(\mathbf{R}',\omega')\rangle_\text{sym}&=2\pi\delta(\omega-\omega')\frac{2}{\omega}N(\omega,T)\\
&\,\times\Ima G_{ij}(\mathbf{R},\mathbf{R}',\omega)\end{split}\end{equation}
being
\begin{equation}N(\omega,T)=\frac{\hbar\omega}{2}\coth\Bigl(\frac{\hbar\omega}{2k_BT}\Bigr)=\hbar\omega\Bigl[\frac{1}{2}+n(\omega,T)\Bigr]\end{equation}
with
\begin{equation}n(\omega,T)=\frac{1}{e^{\frac{\hbar\omega}{k_BT}}-1}.\end{equation}
In eq. \eqref{FluDiss} $i$ and $j$ refer to the cartesian components of the
field and $G_{ij}(\mathbf{R},\mathbf{R}',\omega)$ is the $ij$
component of the Green function of the system, solution of the
differential equation (see also appendix \ref{AppGreen})
\begin{equation}\Bigl[\nabla_\mathbf{R}\times\nabla_\mathbf{R}-\frac{\omega^2}{c^2}\epsilon(\omega,\mathbf{R})\Bigr]\mathbb{G}(\mathbf{R},\mathbf{R}',\omega)=\frac{\omega^2}{\epsilon_0c^2}\,\mathbb{I}\,\delta(\mathbf{R}-\mathbf{R}')\end{equation}
being $\mathbb{I}$ the identity dyad and
$\epsilon(\omega,\mathbf{R})$ the dielectric function of the
medium. The property \eqref{FluDiss} does not hold in the case of
a general nonequilibrium configuration. In our particular system
we have assumed that for each body a local temperature can be
defined, and remains constant in time. This assumption reasonably
leads to the hypothesis that the part of the total field
\emph{emitted} by each body is the same it would be if the body
was at thermal equilibrium with the environment at its own
temperature. In other words the emission process is not
considerably influenced by the modification of the external
radiation impinging on the body. This hypothesis implies that the
correlators of the field \emph{emitted} by each body can still be
deduced using the fluctuation-dissipation theorem eq.
\eqref{FluDiss} at its local temperature.  We note that the limits
of validity of this hypothesis, already used in
\cite{PolderPRB71,HenkelJOptA02,AntezzaPRL05,AntezzaPRA08,BimontePRA09}, require further
experimental and theoretical investigations.

We are now ready to write down the expressions of the correlators
of the environmental field as well as the fields emitted by the
bodies at local thermal equilibrium.

\subsection{Correlators of radiating bodies and environment}

\subsubsection{Environmental field}

The correlators of the environmental radiation in equilibrium at
temperature $T_3$ are well known: they are given, for
$\phi,\phi'\in\{+,-\}$, by
\begin{equation}\label{Co3}\begin{split}&\langle E_p^{(3)\phi}(\mathbf{k},\omega)E_{p'}^{(3)\phi'\dag}(\mathbf{k}',\omega')\rangle_\text{sym}\\
&=\delta_{\phi,\phi'}\frac{\omega}{2\epsilon_0c^2}N(\omega,T_3)\Rea\Bigl(\frac{1}{k_z}\Bigr)\delta_{pp'}(2\pi)^3\delta(\omega-\omega')\delta(\mathbf{k}-\mathbf{k}')\\
&=\delta_{\phi,\phi'}\frac{\omega}{2\epsilon_0c^2}N(\omega,T_3)2\pi\delta(\omega-\omega')\bra{p,\mathbf{k}}\mathcal{P}_{-1}^{\text{(pw)}}\ket{p',\mathbf{k}'}\\
&=\delta_{\phi,\phi'}2\pi\delta(\omega-\omega')\bra{p,\mathbf{k}}C^{(3)}\ket{p',\mathbf{k}'}.\end{split}\end{equation}
In this expression we have defined the matrix $C^{(3)}$ and
introduced the notation, valid for any integer $n$,
\begin{equation}\bra{p,\mathbf{k}}\mathcal{P}_n^\text{(pw/ew)}\ket{p',\mathbf{k}'}=k_z^n\bra{p,\mathbf{k}}\Pi^\text{(pw/ew)}\ket{p',\mathbf{k}'}\end{equation}
$\Pi^\text{(pw)}$ ($\Pi^\text{(ew)}$) being the projector on the
propagative (evanescent) sector. We remark that the operators
$\mathcal{P}_n^\text{(pw/ew)}$ and $\Pi^\text{(pw/ew)}$ depend
implicitly on the frequency $\omega$.

\subsubsection{Field emitted by each body}

We now calculate the correlators of the field emitted by body
$i$ ($i=1,2$) at temperature $T_i$ by ignoring the presence of the
other body and assuming thermal equilibrium at temperature $T_i$.
The main point of this derivation is the connection between the
Green function and the scattering operators: the details of this
calculation are presented in appendices \ref{AppGreen} and
\ref{AppCorrObj}. The result for the field correlators is, for two
modes of the field propagating in the same direction,
\begin{equation}\begin{split}\label{Cois}&\langle E_p^{(i)\phi}(\mathbf{k},\omega)E_{p'}^{(i)\phi\dag}(\mathbf{k}',\omega')\rangle_\text{sym}=\frac{\omega}{2\epsilon_0c^2}N(\omega,T_i)\\
&\,\times2\pi\delta(\omega-\omega')\bra{p,\mathbf{k}}\Bigl(\mathcal{P}_{-1}^\text{(pw)}-\mathcal{R}^{(i)\phi}\mathcal{P}_{-1}^\text{(pw)}\mathcal{R}^{(i)\phi\dag}\\
&\,+\mathcal{R}^{(i)\phi}\mathcal{P}_{-1}^\text{(ew)}-\mathcal{P}_{-1}^\text{(ew)}\mathcal{R}^{(i)\phi\dag}-\mathcal{T}^{(i)\phi}\mathcal{P}_{-1}^\text{(pw)}\mathcal{T}^{(i)\phi\dag}\Bigr)\ket{p',\mathbf{k}'}\end{split}\end{equation}
being $\mathcal{R}^{(i)\phi}$ and $\mathcal{T}^{(i)\phi}$
respectively the reflection and transmission operators associated
to the side $\phi$ of body $i$ defined in section \ref{SRT}. For
fields propagating in opposite directions ($\phi\neq\phi'$) we have
\begin{equation}\label{Coio}\begin{split}&\langle E_p^{(i)\phi}(\mathbf{k},\omega)E_{p'}^{(i)\phi'\dag}(\mathbf{k}',\omega')\rangle_\text{sym}=\frac{\omega}{2\epsilon_0c^2}N(\omega,T_i)\\
&\,\times2\pi\delta(\omega-\omega')\bra{p,\mathbf{k}}\Bigl(-\mathcal{R}^{(i)\phi}\mathcal{P}_{-1}^\text{(pw)}\mathcal{T}^{(i)\phi'\dag}\\
&\,-\mathcal{T}^{(i)\phi}\mathcal{P}_{-1}^\text{(pw)}\mathcal{R}^{(i)\phi'\dag}+\mathcal{T}^{(i)\phi}\mathcal{P}_{-1}^\text{(ew)}-\mathcal{P}_{-1}^\text{(ew)}\mathcal{T}^{(i)\phi'\dag}\Bigr)\ket{p',\mathbf{k}'}.\end{split}\end{equation}
In analogy with the previous definitions, the correlators of the
field produced by the body $i$ will be gathered in the matrix
$C^{(i)\phi\phi'}$, defined by the relation
\begin{equation}\begin{split}\langle E_p^{(i)\phi}(\mathbf{k},\omega)E_{p'}^{(i)\phi'\dag}&(\mathbf{k}',\omega')\rangle_\text{sym}=2\pi\delta(\omega-\omega')\\
&\times\bra{p,\mathbf{k}}C^{(i)\phi\phi'}\ket{p',\mathbf{k}'}.\end{split}\end{equation}

We are now ready to characterize the total field in each region by
means of the scattering operators, and then to deduce its
correlators using the results just obtained in this section.

\subsection{Correlators in region B}

In order to build up the field in the region B between the two bodies the ingredients we need
are the amplitudes $E_p^{(1)+}(\mathbf{k},\omega)$,
$E_p^{(2)-}(\mathbf{k},\omega)$ and
$E_p^{(3)\pm}(\mathbf{k},\omega)$. In the region B of figure
\ref{Fig1} the field propagates in both directions: its amplitudes
will be simply noted with
$E_p^{\text{(B)}\phi}(\mathbf{k},\omega)$. Gathering all the modes
$E_p^{\text{(B)}\phi}(\mathbf{k},\omega)$ in the symbol
$E^{\text{(B)}\phi}$, the amplitudes can be expressed as the
solutions of the system of equations
\begin{equation}\begin{cases}E^{\text{(B)}+}=E^{(1)+}+\mathcal{T}^{(1)+}E^{(3)+}+\mathcal{R}^{(1)+}E^{\text{(B)}-}\\
E^{\text{(B)}-}=E^{(2)-}+\mathcal{T}^{(2)-}E^{(3)-}+\mathcal{R}^{(2)-}E^{\text{(B)}+}\end{cases}\end{equation}
where all the operators and field amplitudes are calculated at a given frequency $\omega$, not
explicitly indicated, and the products between scattering
operators and fields are to be considered as matrix-vector
products. As an intermediate step, we have
\begin{equation}\begin{cases}E^{\text{(B)}+}=E^{(1)+}+\mathcal{T}^{(1)+}E^{(3)+}+\mathcal{R}^{(1)+}E^{\text{(B)}-}\\
E^{\text{(B)}-}=U^{(21)}\mathcal{R}^{(2)-}E^{(1)+}+U^{(21)}\mathcal{R}^{(2)-}\mathcal{T}^{(1)+}E^{(3)+}\\
\hspace{1.3cm}+U^{(21)}\mathcal{T}^{(2)-}E^{(3)-}+U^{(21)}E^{(2)-}\end{cases}\end{equation}
where we have introduced the operators
\begin{equation}\begin{split}U^{(12)}&=(1-\mathcal{R}^{(1)+}\mathcal{R}^{(2)-})^{-1}\\
U^{(21)}&=(1-\mathcal{R}^{(2)-}\mathcal{R}^{(1)+})^{-1}\end{split}\end{equation}
describing the series of \emph{intracavity} (between the two bodies) reflections produced by the
single-body operators $\mathcal{R}^{(1)+}$ and
$\mathcal{R}^{(2)-}$. From the definition
\begin{equation}U^{(12)}=(1-\mathcal{R}^{(1)+}\mathcal{R}^{(2)-})^{-1}=\sum_{n=0}^{+\infty}(\mathcal{R}^{(1)+}\mathcal{R}^{(2)-})^n\end{equation}
and its analogous counterpart for $U^{(21)}$ we easily deduce the
following useful properties
\begin{equation}\mathcal{R}^{(1)+}U^{(21)}=U^{(12)}\mathcal{R}^{(1)+}\qquad\mathcal{R}^{(2)-}U^{(12)}=U^{(21)}\mathcal{R}^{(2)-}\end{equation}
and
\begin{equation}\label{PropU12}\begin{split}\mathcal{R}^{(1)+}U^{(21)}\mathcal{R}^{(2)-}&=U^{(12)}-1\\
\mathcal{R}^{(2)-}U^{(12)}\mathcal{R}^{(1)+}&=U^{(21)}-1.\\\end{split}\end{equation}
These relations allow us to obtain the following final expression
of the field propagating in both directions in region B as a
function of the fields emitted by the bodies and the environment
\begin{equation}\label{EB}\begin{cases}E^{\text{(B)}+}=U^{(12)}E^{(1)+}+\mathcal{R}^{(1)+}U^{(21)}E^{(2)-}\\
\hspace{1.3cm}+\mathcal{R}^{(1)+}U^{(21)}\mathcal{T}^{(2)-}E^{(3)-}+U^{(12)}\mathcal{T}^{(1)+}E^{(3)+}\\
E^{\text{(B)}-}=\mathcal{R}^{(2)-}U^{(12)}E^{(1)+}+U^{(21)}E^{(2)-}\\
\hspace{1.3cm}+U^{(21)}\mathcal{T}^{(2)-}E^{(3)-}+\mathcal{R}^{(2)-}U^{(12)}\mathcal{T}^{(1)+}E^{(3)+}\end{cases}\end{equation}
We remark here that by taking $E^{(3)\pm}=0$ or
$\mathcal{T}^{(1)+}=\mathcal{T}^{(2)-}=0$ in eq. \eqref{EB} we go
back to eqs. (18) and (19) of \cite{BimontePRA09}, where for both
bodies infinite thickness was assumed. Since the fields $E^{(1)-}$ and $E^{(2)+}$ clearly do not participate in the expression of the total field between the two bodies, the expression of $E^{\text{(B)}\phi}$
(for $\phi=+,-$) can be cast without loss of generality in the form
\begin{equation}\label{Dec}E^{\text{(B)}\pm}=A^{\text{(B)}\pm}_1E^{(1)+}+A^{\text{(B)}\pm}_2E^{(2)-}+\sum_{\alpha=+,-}B^{\text{(B)}\pm}_\alpha E^{(3)\alpha}\end{equation}
which in this case gives, by comparison with eq. \eqref{EB},\\
\begin{center}
\begin{tabular}{lclclcl}
$A^{\text{(B)}+}_1=U^{(12)}$ & & $A^{\text{(B)}+}_2=\mathcal{R}^{(1)+}U^{(21)}$\\
$B^{\text{(B)}+}_-=\mathcal{R}^{(1)+}U^{(21)}\mathcal{T}^{(2)-}$ & & $B^{\text{(B)}+}_+=U^{(12)}\mathcal{T}^{(1)+}$\\
$A^{\text{(B)}-}_1=\mathcal{R}^{(2)-}U^{(12)}$ & & $A^{\text{(B)}-}_2=U^{(21)}$\\
$B^{\text{(B)}-}_-=U^{(21)}\mathcal{T}^{(2)-}$ & & $B^{\text{(B)}-}_+=\mathcal{R}^{(2)-}U^{(12)}\mathcal{T}^{(1)+}$.\\
\end{tabular}\end{center}
Using eq. \eqref{Dec} the correlators in region B can be expressed
as a function of the correlators $C^{(i)\phi\phi}$ (for $i=1,2$) and $C^{(3)}$
given by eqs. \eqref{Co3}, \eqref{Cois} and \eqref{Coio}. We
finally obtain the expression of the matrix
$C_\text{B}^{\phi\phi'}$ in terms of the scattering operators of
the two bodies
\begin{equation}\label{CorrB}\begin{split}&\langle E^{\text{(B)}\phi}_p(\mathbf{k},\omega)E^{\text{(B)}\phi'\dag}_{p'}(\mathbf{k}',\omega')\rangle_\text{sym}=2\pi\delta(\omega-\omega')\\
&\,\times\bra{p,\mathbf{k}}\Bigl[A^{\text{(B)}\phi}_1C^{(1)++}A^{\text{(B)}\phi'\dag}_1+A^{\text{(B)}\phi}_2C^{(2)--}A^{\text{(B)}\phi'\dag}_2\\
&\,+\sum_{\alpha=+,-}B^{\text{(B)}\phi}_\alpha C^{(3)}B^{\text{(B)}\phi'\dag}_\alpha\Bigr]\ket{p',\mathbf{k}'}\\
&=2\pi\delta(\omega-\omega')\bra{p,\mathbf{k}}C_\text{B}^{\phi\phi'}\ket{p',\mathbf{k}'}.\\\end{split}\end{equation}

\subsection{Correlators in regions A}

The complete knowledge of the properties of the field in region B
is not sufficient, in general, to deduce the force and the heat
transfer associated to any of the to bodies. Focusing our
attention on body 1, for example, we also need to characterize the
field in the region on its left side, namely region A. The field
$E^{\text{(A)}+}$ propagating toward the right in this region is
obviously only the environment field propagating in the same
direction $E^{(3)+}$. On the contrary, as far as the field
$E^{\text{(A)}-}$ is concerned, it will also include, assuming a
finite thickness for body 1, components from the fields produced
by bodies 1 and 2, as well as from the environment field
$E^{(3)-}$. The total field in region A is then entirely described
by the system of equations
\begin{equation}\begin{cases}E^{\text{(A)}+}=E^{(3)+}\\
E^{\text{(A)}-}=E^{\text{(1)}-}+\mathcal{R}^{(1)-}E^{(3)+}+\mathcal{T}^{(1)-}E^{\text{(B)}-}\end{cases}\end{equation}
which using the result \eqref{EB} for the intracavity field becomes
\begin{equation}\begin{cases}E^{\text{(A)}+}=E^{(3)+}\\
E^{\text{(A)}-}=E^{\text{(1)}-}+\mathcal{T}^{(1)-}\mathcal{R}^{(2)-}U^{(12)}E^{(1)+}\\
\hspace{1.3cm}+\mathcal{T}^{(1)-}U^{(21)}E^{(2)-}+\mathcal{T}^{(1)-}U^{(21)}\mathcal{T}^{(2)-}E^{(3)-}\\
\hspace{1.3cm}+\bigl(\mathcal{R}^{(1)-}+\mathcal{T}^{(1)-}\mathcal{R}^{(2)-}U^{(12)}\mathcal{T}^{(1)+}\bigr)E^{(3)+}\end{cases}\end{equation}
Using the general decomposition
\begin{equation}E^{\text{(A)}-}=\sum_{\alpha=+,-}A^{\text{(A)}}_{1\alpha}E^{(1)\alpha}+A^{\text{(A)}}_2E^{(2)-}+\sum_{\alpha=+,-}B^{\text{(A)}}_\alpha E^{(3)\alpha}\end{equation}
we have\\\begin{center}\begin{tabular}{lclcl}
$A^{\text{(A)}}_{1+}=\mathcal{T}^{(1)-}\mathcal{R}^{(2)-}U^{(12)}$ & & $A^{\text{(A)}}_{1-}=1$\\
$A^{\text{(A)}}_2=\mathcal{T}^{(1)-}U^{(21)}$\\
$B^{\text{(A)}}_+=\mathcal{R}^{(1)-}+\mathcal{T}^{(1)-}\mathcal{R}^{(2)-}U^{(12)}\mathcal{T}^{(1)+}$\\
$B^{\text{(A)}}_-=\mathcal{T}^{(1)-}U^{(21)}\mathcal{T}^{(2)-}$.\\
\end{tabular}\end{center}
We are now ready to give the final expression of the correlators
$C_A^{\phi\phi'}$ of the total field in region A in terms of the
scattering matrices, which reads
\begin{equation}\label{CorrA}\begin{split}C_A^{++}&=C^{(3)}\\
C_A^{+-}&=C^{(3)}B^{\text{(A)}\dag}_+\\
C_A^{-+}&=B^{\text{(A)}}_+C^{(3)}\\
C_A^{--}&=A^{\text{(A)}}_{1+}C^{(1)++}A^{\text{(A)}\dag}_{1+}+C^{(1)--}+A^{\text{(A)}}_{1+}C^{(1)+-}\\
&\,+C^{(1)-+}A^{\text{(A)}\dag}_{1+}+A^{\text{(A)}}_{2}C^{(2)--}A^{\text{(A)}\dag}_{2}+B^{\text{(A)}}_+C^{(3)}B^{\text{(A)}\dag}_+\\
&\,+B^{\text{(A)}}_-C^{(3)}B^{\text{(A)}\dag}_-.\\\end{split}\end{equation}

Due to the geometry of our system, the correlators of the field in
region C can be obtained from the ones given here for region A
performing the interchanges $\text{A}\rightleftarrows \text{C}$,
$1\rightleftarrows2$ and $+\rightleftarrows-$. This holds for all
the other quantities we are going to calculate in region A.

\section{Final expressions for the Casimir-Lifshitz force and heat transfer out of thermal equilibrium}\label{FinalFH}

\subsection{Casimir-Lifshitz force}

We are now going to calculate the flux of the stress tensor, in
order to deduce the expression of the force. For simplicity, we
will focus on the $z$ component of the force acting on body 1, using as a
consequence the expression \eqref{FluxTmz} with $m=z$ in regions B and A. The
calculation of the other components of the force follows the same
scheme we are going to present in the following. Let us now turn,
then, to the evaluation of the fluxes of $T_{zz}$ in regions B (Sec. \ref{FluxB}) and
A (Sec. \ref{FluxA}): we will assume that for both surfaces the vector orthogonal to
the surface is oriented toward the right, i.e. the positive
direction of $z$ axis. These two results will provide the final expression of the Casimir-Lifshitz force acting on body 1, deduced in Sec. \ref{CasLif1}.

\subsubsection{Flux in region B}\label{FluxB}

Using the relation \eqref{FluxTmz} (with $m=z$ and
$z_2<\bar{z}<z_3$) and \eqref{CorrB}, we are able to
express the flux in region B as a function of the correlators of
the field emitted by the bodies and the environment \eqref{Co3},
\eqref{Cois}, \eqref{Coio}
\begin{equation}\begin{split}&\Phi^{\text{(B)}}_z(T_1,T_2,T_3)=-2\epsilon_0\sum_p
\int\frac{d^2\mathbf{k}}{(2\pi)^2}\int_0^{+\infty}\frac{d\omega}{2\pi}\frac{c^2k_z^2}{\omega^2}\\
&\times\bra{p,\mathbf{k}}\Bigl\{\Pi^{\text{(pw)}}
\Bigl(A_1^{\text{(B)}+}C^{(1)++}A_1^{\text{(B)}+\dag}\\
&+A_1^{\text{(B)}-}C^{(1)++}A_1^{\text{(B)}-\dag}+A_2^{\text{(B)}+}C^{(2)--}A_2^{\text{(B)}+\dag}\\
&+A_2^{\text{(B)}-}C^{(2)--}A_2^{\text{(B)}-\dag}\Bigr)\\
&+\Pi^{\text{(ew)}}\Bigl(A_1^{\text{(B)}+}C^{(1)++}A_1^{\text{(B)}-\dag}+A_1^{\text{(B)}-}C^{(1)++}A_1^{\text{(B)}+\dag}\\
&+A_2^{\text{(B)}+}C^{(2)--}A_2^{\text{(B)}-\dag}+A_2^{\text{(B)}-}C^{(2)--}A_2^{\text{(B)}+\dag}\Bigr)\\
&+\sum_{\alpha=+,-}\Bigl[\Pi^{\text{(pw)}}
\Bigl(B_\alpha^{\text{(B)}+}C^{(3)}B_\alpha^{\text{(B)}+\dag}+B_\alpha^{\text{(B)}-}C^{(3)}B_\alpha^{\text{(B)}-\dag}\Bigr)\\
&+\Pi^{\text{(ew)}}\Bigl(B_\alpha^{\text{(B)}+}C^{(3)}B_\alpha^{\text{(B)}-\dag}+B_\alpha^{\text{(B)}-}C^{(3)}B_\alpha^{\text{(B)}+\dag}\Bigr)\Bigr]\Bigr\}\ket{p,\mathbf{k}}.\end{split}\end{equation}
Defining the trace operator for a frequency-dependent operator
$\mathcal{A}$ as
\begin{equation}\label{DefTrace}\Tr\mathcal{A}=\sum_p\int\frac{d^2\mathbf{k}}{(2\pi)^2}\int_0^{+\infty}\frac{d\omega}{2\pi}\bra{p,\mathbf{k}}\mathcal{A}\ket{p,\mathbf{k}}\end{equation}
we can write the flux in region B under the form
\begin{equation}\begin{split}&\Phi^{\text{(B)}}_z(T_1,T_2,T_3)=-2\epsilon_0c^2\Tr\Biggl\{\frac{1}{\omega^2}\Biggl[\mathcal{P}_2^{\text{(pw)}}\\
&\times\Bigl(A_1^{\text{(B)}+}C^{(1)++}A_1^{\text{(B)}+\dag}+A_1^{\text{(B)}-}C^{(1)++}A_1^{\text{(B)}-\dag}\\
&+A_2^{\text{(B)}+}C^{(2)--}A_2^{\text{(B)}+\dag}+A_2^{\text{(B)}-}C^{(2)--}A_2^{\text{(B)}-\dag}\Bigr)\\
&+\mathcal{P}_2^{\text{(ew)}}\Bigl(A_1^{\text{(B)}+}C^{(1)++}A_1^{\text{(B)}-\dag}+A_1^{\text{(B)}-}C^{(1)++}A_1^{\text{(B)}+\dag}\\
&+A_2^{\text{(B)}+}C^{(2)--}A_2^{\text{(B)}-\dag}+A_2^{\text{(B)}-}C^{(2)--}A_2^{\text{(B)}+\dag}\Bigr)\\
&+\sum_{\alpha=+,-}\Bigl[\mathcal{P}_2^{\text{(pw)}}
\Bigl(B_\alpha^{\text{(B)}+}C^{(3)}B_\alpha^{\text{(B)}+\dag}+B_\alpha^{\text{(B)}-}C^{(3)}B_\alpha^{\text{(B)}-\dag}\Bigr)\\
&+\mathcal{P}_2^{\text{(ew)}}\Bigl(B_\alpha^{\text{(B)}+}C^{(3)}B_\alpha^{\text{(B)}-\dag}+B_\alpha^{\text{(B)}-}C^{(3)}B_\alpha^{\text{(B)}+\dag}\Bigr)\Bigr]\Biggr]\Biggr\}.\end{split}\end{equation}
In both expressions it is clear that we have three separate
contributions associated to body 1, body 2 and environment
respectively. Using the fact that the trace is invariant under cyclic permutations, we have
\begin{equation}\label{Final1B}\begin{split}&\Phi^{\text{(B)}}_z(T_1,T_2,T_3)=-\Tr\Bigl\{\frac{1}{\omega}\\
&\times\Bigl[N(\omega,T_1)J(\mathcal{R}^{(1)+},\mathcal{R}^{(2)-})+N(\omega,T_2)J(\mathcal{R}^{(2)-},\mathcal{R}^{(1)+})\\
&+\Bigl(N(\omega,T_3)-N(\omega,T_1)\Bigr)H(\mathcal{R}^{(1)+},\mathcal{R}^{(2)-},\mathcal{T}^{(1)+})\\
&+\Bigl(N(\omega,T_3)-N(\omega,T_2)\Bigr)H(\mathcal{R}^{(2)-},\mathcal{R}^{(1)+},\mathcal{T}^{(2)-})\Bigr]\Bigr\}\end{split}\end{equation}
where
\begin{equation}\begin{split}&J(\mathcal{R}^{(1)+},\mathcal{R}^{(2)-})=U^{(12)}\Bigl(\mathcal{P}_{-1}^{\text{(pw)}}-\mathcal{R}^{(1)+}\mathcal{P}_{-1}^{\text{(pw)}}\mathcal{R}^{(1)+\dag}\\
&\qquad+\mathcal{R}^{(1)+}\mathcal{P}_{-1}^{\text{(ew)}}-\mathcal{P}_{-1}^{\text{(ew)}}\mathcal{R}^{(1)+\dag}\Bigr)U^{(12)^\dag}\\
&\qquad\times\Bigl(\mathcal{P}_2^{\text{(pw)}}+\mathcal{R}^{(2)-\dag}\mathcal{P}_2^{\text{(pw)}}\mathcal{R}^{(2)-}\\
&\qquad+\mathcal{R}^{(2)-\dag}\mathcal{P}_2^{\text{(ew)}}+\mathcal{P}_2^{\text{(ew)}}\mathcal{R}^{(2)-}\Bigr)\\
&H(\mathcal{R}^{(1)+},\mathcal{R}^{(2)-},\mathcal{T}^{(1)+})=U^{(12)}\mathcal{T}^{(1)+}\mathcal{P}_{-1}^{\text{pw}}\mathcal{T}^{(1)+\dag}U^{(12)\dag}\\
&\qquad\times\Bigl(\mathcal{P}_2^{\text{(pw)}}+\mathcal{R}^{(2)-\dag}\mathcal{P}_2^{\text{(pw)}}\mathcal{R}^{(2)-}\\
&\qquad+\mathcal{R}^{(2)-\dag}\mathcal{P}_2^{\text{(ew)}}+\mathcal{P}_2^{\text{(ew)}}\mathcal{R}^{(2)-}\Bigr).\\\end{split}\end{equation}
Note that when calculating
$J(\mathcal{R}^{(2)-},\mathcal{R}^{(1)+})$ one also need to change
$U^{(12)}$ into $U^{(21)}$.

Eq. \eqref{Final1B} can be cast in the form
\begin{equation}\label{FinalB}\begin{split}\Phi^{\text{(B)}}_z(T_1,T_2,T_3)&=\frac{\Phi_z^{\text{(B,eq)}}(T_1)+\Phi_z^{\text{(B,eq)}}(T_2)}{2}\\
&\,+\Delta\Phi_z^{\text{(B)}}(T_1,T_2,T_3)\end{split}\end{equation}
where
\begin{equation}\label{PhiBeq}\begin{split}&\Phi_z^{\text{(B,eq)}}(T)=-\Tr\Bigl\{\frac{1}{\omega}N(\omega,T)\Bigl[J(\mathcal{R}^{(1)+},\mathcal{R}^{(2)-})\\
&\qquad+J(\mathcal{R}^{(2)-},\mathcal{R}^{(1)+})\Bigr]\Bigr\}\\
&=-2\Rea\Tr\Bigl\{\frac{k_z}{\omega}N(\omega,T)\Bigl[U^{(12)}\mathcal{R}^{(1)+}\mathcal{R}^{(2)-}\\
&\qquad+U^{(21)}\mathcal{R}^{(2)-}\mathcal{R}^{(1)+}\Bigr]\Bigr\}-2\Tr\Bigl\{\frac{1}{\omega}N(\omega,T)\mathcal{P}_1^{\text{pw}}\Bigr\}\end{split}\end{equation}
and
\begin{equation}\label{DeltaB}\begin{split}&\Delta\Phi^{\text{(B)}}_z(T_1,T_2,T_3)=-\hbar\\
&\times\Tr\Bigl[\frac{n_{12}}{2}\Bigl(J(\mathcal{R}^{(1)+},\mathcal{R}^{(2)-})-J(\mathcal{R}^{(2)-},\mathcal{R}^{(1)+})\Bigr)\\
&\quad+n_{31}H(\mathcal{R}^{(1)+},\mathcal{R}^{(2)-},\mathcal{T}^{(1)+})\\
&\quad+n_{32}H(\mathcal{R}^{(2)-},\mathcal{R}^{(1)+},\mathcal{T}^{(2)-})\Bigr]\end{split}\end{equation}
where we have defined, for $i,j=1,2,3$,
\begin{equation}n_{ij}=n(\omega,T_i)-n(\omega,T_j).\end{equation}

At this point, a remark is important about the expression \eqref{PhiBeq}, giving the equilibrium part of the flux in region B of $T_{zz}$. We have to observe that,
since the operator $\mathcal{P}_{1}^{\text{(pw)}}$ is diagonal in the $(\mathbf{k},p)$ basis, its trace defined as in \eqref{DefTrace} is divergent. Moreover, this term
is independent of the bodies under scrutiny. Nevertheless, we have to keep in mind that the flux in region B does not have a direct physical meaning, since we still have to subtract from it
the flux in region A in order to obtain the value of the force. As we will see in Secs. \ref{CasLif1} and \ref{SecUnif}, the divergences present in the individual fluxes are completely regularized
when taking the difference $\Phi^{\text{(B)}}_z-\Phi^{\text{(A)}}_z$.

\subsubsection{Flux in region A}\label{FluxA}

In order to calculate the flux of $T_{zz}$ in A we have to use eq.
\eqref{FluxTmz} with $m=z$ and $\bar{z}<z_1$. Moreover, the
correlators in region A are given in eq. \eqref{CorrA}. After
algebraic manipulations analogous to the ones used in the last
section we have, for the flux in region A,
\begin{equation}\begin{split}&\Phi^{\text{(A)}}_z(T_1,T_2,T_3)=-2\epsilon_0c^2\Tr\Bigl[\frac{1}{\omega^2}\mathcal{P}_2^{\text{(pw)}}\Bigl(C^{(3)}\\
&+A^{\text{(A)}}_{1+}C^{(1)++}A^{\text{(A)}\dag}_{1+}+C^{(1)--}+A^{\text{(A)}}_{1+}C^{(1)+-}\\
&\,+C^{(1)-+}A^{\text{(A)}\dag}_{1+}+A^{\text{(A)}}_{2}C^{(2)--}A^{\text{(A)}\dag}_{2}+B^{\text{(A)}}_+C^{(3)}B^{\text{(A)}\dag}_+\\
&\,+B^{\text{(A)}}_-C^{(3)}B^{\text{(A)}\dag}_-\Bigr)\Bigr].\end{split}\end{equation}
This can be cast in the form
\begin{equation}\label{FinalA}\begin{split}\Phi^{\text{(A)}}_z(T_1,T_2,T_3)&=-\Tr\Bigl[\frac{1}{\omega}\Bigl(N(\omega,T_1)+N(\omega,T_3)\Bigr)\mathcal{P}_1^{\text{(pw)}}\Bigr]\\
&\,+\Delta\Phi^{\text{(A)}}_z(T_1,T_2,T_3)\end{split}\end{equation}
where
\begin{equation}\label{DeltaA}\begin{split}&\Delta\Phi^{\text{(A)}}_z(T_1,T_2,T_3)=-\hbar\Tr\Bigl[n_{31}\mathcal{P}_2^{\text{(pw)}}\mathcal{R}^{(1)-}\mathcal{P}_{-1}^\text{(pw)}\mathcal{R}^{(1)-\dag}\\
&+n_{32}\mathcal{P}_2^{\text{(pw)}}\mathcal{T}^{(1)-}U^{(21)}\mathcal{T}^{(2)-}\mathcal{P}_{-1}^{\text{(pw)}}\mathcal{T}^{(2)-\dag}U^{(21)\dag}\mathcal{T}^{(1)-\dag}\\
&+n_{21}\Bigl(\mathcal{P}_{-1}^\text{(pw)}+\mathcal{R}^{(2)-}\mathcal{P}_{-1}^\text{(ew)}-\mathcal{P}_{-1}^\text{(ew)}\mathcal{R}^{(2)-\dag}\\
&-\mathcal{R}^{(2)-}\mathcal{P}_{-1}^\text{(pw)}\mathcal{R}^{(2)-\dag}\Bigr)U^{(21)\dag}\mathcal{T}^{(1)-\dag}\mathcal{P}_2^\text{(pw)}\mathcal{T}^{(1)-}U^{(21)}\\
&+n_{31}\Bigl(\mathcal{P}_2^{\text{(pw)}}\mathcal{R}^{(1)-}\mathcal{P}_{-1}^{\text{(pw)}}\mathcal{T}^{(1)+\dag}U^{(12)\dag}\mathcal{R}^{(2)-\dag}\mathcal{T}^{(1)-\dag}\\
&+\mathcal{P}_2^{\text{(pw)}}\mathcal{T}^{(1)-}\mathcal{R}^{(2)-}U^{(12)}\mathcal{T}^{(1)+}\mathcal{P}_{-1}^{\text{(pw)}}\mathcal{R}^{(1)-\dag}\\
&+\mathcal{P}_2^{\text{(pw)}}\mathcal{T}^{(1)-}\mathcal{R}^{(2)-}U^{(12)}\mathcal{T}^{(1)+}\mathcal{P}_{-1}^{\text{(pw)}}\mathcal{T}^{(1)+\dag}\\
&\times
U^{(12)\dag}\mathcal{R}^{(2)-\dag}\mathcal{T}^{(1)-\dag}\Bigr)\Bigr].\end{split}\end{equation}
We repeat here that the flux in region C (necessary for the calculation of the force acting on body 2) can be obtained from eqs. \eqref{FinalA} and \eqref{DeltaA}
by performing the interchanges $A\rightleftarrows C$, $1\rightleftarrows2$ and $+\rightleftarrows-$.

\subsubsection{Casimir-Lifshitz force acting on body 1}\label{CasLif1}

We now have all the ingredients to give the $z$ component of the
force acting on body 1. From the definition of the stress tensor
we have
\begin{equation}F_{1z}=\Phi_z^{\text{(B)}}(T_1,T_2,T_3)-\Phi_z^{\text{(A)}}(T_1,T_2,T_3)\end{equation}
where the two fluxes are given by eqs. \eqref{FinalB} and
\eqref{FinalA}. Gathering all the results obtained in the previous
sections, the complete expression of the force reads
\begin{equation}\label{F1tot}\begin{split}F_{1z}(T_1,T_2,T_3)&=\frac{F_z^{\text{(eq)}}(T_1)+F_z^{\text{(eq)}}(T_2)}{2}\\
&\,+\Delta F_{1z}(T_1,T_2,T_3).\end{split}\end{equation} In this
expression the result is written as a sum of two terms. The
first contribution is the average, at the temperatures $T_1$ and
$T_2$ of the two bodies, of the equilibrium force
\begin{equation}\label{FeqSc}\begin{split}F_z^{\text{(eq)}}(T)&=-2\Rea\Tr\Bigl\{\frac{k_z}{\omega}N(\omega,T)\Bigl[U^{(12)}\mathcal{R}^{(1)+}\mathcal{R}^{(2)-}\\
&\,+U^{(21)}\mathcal{R}^{(2)-}\mathcal{R}^{(1)+}\Bigr]\Bigr\}\end{split}\end{equation}
which contains both the zero-temperature term and the thermal
correction. This result for the equilibrium force was already obtained by different authors in the framework of scattering-matrix theory \cite{LambrechtNewJPhys06,RahiPRD09}. As remarked in \cite{BimontePRA09}, the eq. \eqref{FeqSc}
gives a finite result for any choice of temperature and material properties for the two bodies. Moreover, the equilibrium force \eqref{FeqSc} shows the important property of depending only on the \emph{intracavity} reflection
operators $\mathcal{R}^{(1)+}$ and $\mathcal{R}^{(2)-}$, i.e. the operators describing the reflection produced by each body on the side of the other one.

The second term in \eqref{F1tot} is the non-equilibrium contribution, given by
\begin{equation}\label{Fnoneq}\begin{split}\Delta F_{1z}(T_1,T_2,T_3)&=\Delta\Phi^{\text{(B)}}_z(T_1,T_2,T_3)-\Delta\Phi^{\text{(A)}}_z(T_1,T_2,T_3)\\
&\,+\hbar\Tr\Bigl(n_{32}\mathcal{P}_1^{\text{(pw)}}\Bigr)\end{split}\end{equation}
where the two fluxes $\Delta\Phi^{\text{(B)}}_z$ and
$\Delta\Phi^{\text{(A)}}_z$ are explicitly given by eqs.
\eqref{DeltaB} and \eqref{DeltaA} respectively. The
non-equilibrium contribution manifestly satisfies the condition
\begin{equation}\Delta F_{1z}(T,T,T)=0.\end{equation}
Differently from the equilibrium force \eqref{FeqSc}, the nonequilibrium contribution \eqref{Fnoneq} still contains terms which are individually formally divergent. In Sec. \ref{SecUnif}, where a unified expression
for the Casimir-Lifshitz force and the heat transfer will be provided, we will see that this can be manipulated so that all these divergent terms disappear.

\subsection{Heat transfer}

In order to obtain the expression of the heat transfer on body 1
we have to follow the same steps we used in the case of the force.
We first need the fluxes of the Poynting vector in regions B and
A. Their difference will provide us the energy absorbed per unit
of time by body 1. The flux in region B can be obtained by
combining eq. \eqref{FluxSz} with the correlators in region B
given by eq. \eqref{CorrB}. The result can be cast under the form
\begin{equation}\label{FluxSB}\begin{split}&\varphi^{\text{(B)}}(T_1,T_2,T_3)=\hbar\Tr\Bigl\{\omega\Bigl[n_{31}\mathcal{L}(\mathcal{R}^{(1)+},\mathcal{R}^{(2)-},\mathcal{T}^{(1)+})\\
&+n_{23}\mathcal{L}(\mathcal{R}^{(2)-},\mathcal{R}^{(1)+},\mathcal{T}^{(2)-})+n_{12}\mathcal{J}(\mathcal{R}^{(1)+},\mathcal{R}^{(2)-})\Bigr]\Bigr\}\end{split}\end{equation}
where
\begin{equation}\begin{split}&\mathcal{J}(\mathcal{R}^{(1)+},\mathcal{R}^{(2)-})=U^{(12)}\Bigl(\mathcal{P}_{-1}^{\text{(pw)}}-\mathcal{R}^{(1)+}\mathcal{P}_{-1}^{\text{(pw)}}\mathcal{R}^{(1)+\dag}\\
&\qquad+\mathcal{R}^{(1)+}\mathcal{P}_{-1}^{\text{(ew)}}-\mathcal{P}_{-1}^{\text{(ew)}}\mathcal{R}^{(1)+\dag}\Bigr)U^{(12)^\dag}\\
&\qquad\times\Bigl(\mathcal{P}_1^{\text{(pw)}}-\mathcal{R}^{(2)-\dag}\mathcal{P}_1^{\text{(pw)}}\mathcal{R}^{(2)-}\\
&\qquad+\mathcal{R}^{(2)-\dag}\mathcal{P}_1^{\text{(ew)}}-\mathcal{P}_1^{\text{(ew)}}\mathcal{R}^{(2)-}\Bigr)\\
&\mathcal{L}(\mathcal{R}^{(1)+},\mathcal{R}^{(2)-},\mathcal{T}^{(1)+})=U^{(12)}\mathcal{T}^{(1)+}\mathcal{P}_{-1}^{\text{pw}}\mathcal{T}^{(1)+\dag}U^{(12)\dag}\\
&\qquad\times\Bigl(\mathcal{P}_1^{\text{(pw)}}-\mathcal{R}^{(2)-\dag}\mathcal{P}_1^{\text{(pw)}}\mathcal{R}^{(2)-}\\
&\qquad+\mathcal{R}^{(2)-\dag}\mathcal{P}_1^{\text{(ew)}}-\mathcal{P}_1^{\text{(ew)}}\mathcal{R}^{(2)-}\Bigr).\\\end{split}\end{equation}

An analogous calculation leads us to the following expression of
the flux in region A
\begin{equation}\label{FluxSA}\begin{split}&\varphi^{\text{(A)}}(T_1,T_2,T_3)=-\hbar\Tr\Bigl\{\omega\\
&\times\Bigl[n_{31}\Bigl(-\Pi^{\text{(pw)}}+\mathcal{P}_1^{\text{(pw)}}\mathcal{R}^{(1)-}\mathcal{P}_{-1}^\text{(pw)}\mathcal{R}^{(1)-\dag}\Bigr)\\
&+n_{32}\mathcal{P}_1^{\text{(pw)}}\mathcal{T}^{(1)-}U^{(21)}\mathcal{T}^{(2)-}\mathcal{P}_{-1}^{\text{(pw)}}\mathcal{T}^{(2)-\dag}U^{(21)\dag}\mathcal{T}^{(1)-\dag}\\
&+n_{21}\Bigl(\mathcal{P}_{-1}^\text{(pw)}+\mathcal{R}^{(2)-}\mathcal{P}_{-1}^\text{(ew)}-\mathcal{P}_{-1}^\text{(ew)}\mathcal{R}^{(2)-\dag}\\
&-\mathcal{R}^{(2)-}\mathcal{P}_{-1}^\text{(pw)}\mathcal{R}^{(2)-\dag}\Bigr)U^{(21)\dag}\mathcal{T}^{(1)-\dag}\mathcal{P}_1^\text{(pw)}\mathcal{T}^{(1)-}U^{(21)}\\
&+n_{31}\Bigl(\mathcal{P}_1^{\text{(pw)}}\mathcal{R}^{(1)-}\mathcal{P}_{-1}^{\text{(pw)}}\mathcal{T}^{(1)+\dag}U^{(12)\dag}\mathcal{R}^{(2)-\dag}\mathcal{T}^{(1)-\dag}\\
&+\mathcal{P}_1^{\text{(pw)}}\mathcal{T}^{(1)-}\mathcal{R}^{(2)-}U^{(12)}\mathcal{T}^{(1)+}\mathcal{P}_{-1}^{\text{(pw)}}\mathcal{R}^{(1)-\dag}\\
&+\mathcal{P}_1^{\text{(pw)}}\mathcal{T}^{(1)-}\mathcal{R}^{(2)-}U^{(12)}\mathcal{T}^{(1)+}\mathcal{P}_{-1}^{\text{(pw)}}\\
&\times\mathcal{T}^{(1)+\dag}U^{(12)\dag}\mathcal{R}^{(2)-\dag}\mathcal{T}^{(1)-\dag}\Bigr)\Bigr].\end{split}\end{equation}

The total heat flux on body 1 is finally given by the difference
of the two contributions
\begin{equation}\label{H1}H(T_1,T_2,T_3)=\varphi^{\text{(A)}}(T_1,T_2,T_3)-\varphi^{\text{(B)}}(T_1,T_2,T_3).\end{equation}
Since the fluxes \eqref{FluxSB} and \eqref{FluxSA} in regions B
and A respectively are zero for $T_1=T_2=T_3$, the heat flux \eqref{H1} on
body 1 satisfies the evident property
\begin{equation}H(T,T,T)=0\end{equation}
for any temperature $T\geq0$.

\subsection{Unified expression for force and heat transfer}\label{SecUnif}

We are now ready to give the main result of the paper, namely the analytic explicit expressions of the Casimir-Lifshitz force and heat transfer on the body 1. These expressions are valid for any choice of the shape
and dielectric properties of the two bodies. We are going to give the following definitions
\begin{equation}\label{Ffinal}F_{1z}(T_1,T_2,T_3)=\frac{F_z^{\text{(eq)}}(T_1)+F_z^{\text{(eq)}}(T_2)}{2}+\Delta_2(T_1,T_2,T_3)\end{equation}
\begin{equation}\label{Hfinal}H(T_1,T_2,T_3)=\Delta_1(T_1,T_2,T_3)\end{equation} we can collect
and to give a unified expression for the nonequilibrium contribution $\Delta_2$ to the force and the heat transfer $\Delta_1$, both relative to the body 1. Before providing the explicit analytic expression of $\Delta_m$
for $m=1,2$ we recall that the fluxes \eqref{FinalA} and \eqref{FinalB} of $T_{zz}$ in regions A and B respectively contain individual divergent terms. The same property holds for the heat transfer, as it is evident from example
from the flux \eqref{FluxSA} of $S_z$ in region A, containing as a first term the trace of $\Pi^{\text{(pw)}}$. We are going to show that the nonequilibrium force and the heat transfer are indeed convergent for any choice of
the two bodies. To this aim a fundamental intermediate step is the identification of the individual divergent terms in the expressions of the fluxes of $T_{zz}$ and $S_z$. We first observe that all the terms which do not contain
any reflection or transmission operator, as the ones we have already discussed, are indeed divergent. This in not the case, on the contrary, for the terms proportional to at least one reflection operator, since these ones tend
to zero in absence of the objects. As far as the transmission operators are concerned, we have then to express each $\mathcal{T}$ operator as $1+\tilde{\mathcal{T}}$: in analogy with the reflection operators, $\tilde{\mathcal{T}}$
tends to zero in absence of the bodies. Finally, considering the terms containing only the operators $U^{(12)}$, $U^{(21)}$ and projection operators, it is sufficient to use the relations \eqref{PropU12} in order to write each of
them as a sum of a divergent one, which is independent on the scattering operators, and another one proportional to the reflection matrices.

By following the procedure we have just described it can be shown
that all the divergent terms exactly cancel each other. All the
remaining terms are proportional to either a reflection or a
modified transmission operator, as explicitly shown in eqs.
\eqref{Noneq}-\eqref{ual} below. We are now ready to give the
final analytic unified expression for the Casimir-Lifshitz force
and heat transfer on body 1. This reads
\begin{widetext}
\begin{equation}\label{Noneq}\begin{split}&\Delta_m(T_1,T_2,T_3)=(-1)^{m+1}\hbar\Tr\Biggl[\omega^{2-m}\Bigl\{\frac{n_{12}}{2}\Bigl[\Bigl(U^{(21)\dag}\Bigl(2g_m(\mathcal{T}^{(1)-})-f_m(\mathcal{R}^{(1)+})\Bigr)U^{(21)}+u_m\Bigr)\Bigl(\mathcal{P}_{-1}^{\text{(pw)}}+f_{-1}(\mathcal{R}^{(2)-})\Bigr)\\
&\,+(-1)^m\Bigl(U^{(12)}f_{-1}(\mathcal{R}^{(1)+})U^{(12)\dag}+u_{-1}\Bigr)\Bigl(\mathcal{P}_m^{\text{(pw)}}+f_m(\mathcal{R}^{(2)-})\Bigr)\Bigr]\\
&+n_{13}\Bigl[-(-1)^m\Bigl(U^{(12)}g_{-1}(\mathcal{T}^{(1)+})U^{(12)\dag}+u_{-1}\Bigr)\Bigl(\mathcal{P}_m^{\text{(pw)}}+f_m(\mathcal{R}^{(2)-})\Bigr)+\mathcal{P}_{m}^{\text{(pw)}}\mathcal{R}^{(1)-}\mathcal{P}_{-1}^\text{(pw)}\mathcal{R}^{(1)-\dag}\\
&+\Bigl(U^{(21)\dag}g_m(\mathcal{T}^{(1)-})U^{(21)}+u_m\Bigr)\mathcal{R}^{(2)-}\mathcal{P}_{-1}^{\text{(pw)}}\mathcal{R}^{(2)-\dag}+U^{(21)\dag}\Bigl(\mathcal{P}_m^\text{(pw)}+g_m(\mathcal{T}^{(1)-})\Bigr)U^{(21)}\mathcal{R}^{(2)-}g_{-1}(\mathcal{T}^{(1)+})\mathcal{R}^{(2)-\dag}\\
&+\Bigl(\mathcal{P}_{m}^{\text{(pw)}}\mathcal{R}^{(1)-}\mathcal{P}_{-1}^{\text{(pw)}}\Bigl(1+\tilde{\mathcal{T}}^{(1)+\dag}\Bigr)U^{(12)\dag}\mathcal{R}^{(2)-\dag}\Bigl(1+\tilde{\mathcal{T}}^{(1)-\dag}\Bigr)+\text{h.c.}\Bigr)\Bigr]\\
&+n_{23}U^{(21)\dag}\Bigl(g_m(\mathcal{T}^{(1)-})-f_m(\mathcal{R}^{(1)+})\Bigr)U^{(21)}\Bigl(\mathcal{P}_{-1}^{\text{(pw)}}+g_{-1}(\mathcal{T}^{(2)-})\Bigr)\Bigr\}\Biggr]\\\end{split}\end{equation}
where we have defined the auxiliary functions
\begin{equation}\label{fal}f_\alpha(\mathcal{R})=\begin{cases}-\mathcal{R}\mathcal{P}_{-1}^{\text{(pw)}}\mathcal{R}^\dag+\mathcal{R}\mathcal{P}_{-1}^{\text{(ew)}}-\mathcal{P}_{-1}^{\text{(ew)}}\mathcal{R}^\dag & \alpha=-1\\
(-1)^m\mathcal{R}^\dag\mathcal{P}_m^{\text{(pw)}}\mathcal{R}+\mathcal{R}^\dag\mathcal{P}_m^{\text{(ew)}}+(-1)^m\mathcal{P}_m^{\text{(ew)}}\mathcal{R}
& \alpha=m\in\{1,2\}\end{cases}\end{equation}
\begin{equation}\label{gal}g_\alpha(\mathcal{T})=\begin{cases}\mathcal{T}\mathcal{P}_{-1}^{\text{(pw)}}\mathcal{T}^\dag-\mathcal{P}_{-1}^{\text{(pw)}}=\tilde{\mathcal{T}}\mathcal{P}_{-1}^{\text{(pw)}}\tilde{\mathcal{T}}^\dag+\mathcal{P}_{-1}^{\text{(pw)}}\tilde{\mathcal{T}}^\dag+\tilde{\mathcal{T}}\mathcal{P}_{-1}^{\text{(pw)}} & \alpha=-1\\
\mathcal{T}^\dag\mathcal{P}_m^{\text{(pw)}}\mathcal{T}-\mathcal{P}_m^{\text{(pw)}}=\tilde{\mathcal{T}}^\dag\mathcal{P}_m^{\text{(pw)}}\tilde{\mathcal{T}}+\mathcal{P}_m^{\text{(pw)}}\tilde{\mathcal{T}}+\tilde{\mathcal{T}}^\dag\mathcal{P}_m^{\text{(pw)}}
& \alpha=m\in\{1,2\}\end{cases}\end{equation}
\begin{equation}\label{ual}u_\alpha=\begin{cases}U^{(12)}\mathcal{P}_{-1}^{\text{(pw)}}U^{(12)\dag}-\mathcal{P}_{-1}^{\text{(pw)}}=\mathcal{P}_{-1}^{\text{(pw)}}U^{(12)\dag}\mathcal{R}^{(2)-\dag}\mathcal{R}^{(1)+\dag}\\
\vspace{.3cm}+\mathcal{R}^{(1)+}\mathcal{R}^{(2)-}U^{(12)}\mathcal{P}_{-1}^{\text{(pw)}}+\mathcal{R}^{(1)+}\mathcal{R}^{(2)-}U^{(12)}\mathcal{P}_{-1}^{\text{(pw)}}U^{(12)\dag}\mathcal{R}^{(2)-\dag}\mathcal{R}^{(1)+\dag} & \alpha=-1\\
U^{(21)\dag}\mathcal{P}_m^{\text{(pw)}}U^{(21)}-\mathcal{P}_m^{\text{(pw)}}=\mathcal{P}_m^{\text{(pw)}}\mathcal{R}^{(2)-}\mathcal{R}^{(1)+}U^{(21)}\\
+U^{(21)\dag}\mathcal{R}^{(1)+\dag}\mathcal{R}^{(2)-\dag}\mathcal{P}_m^{\text{(pw)}}+U^{(21)\dag}\mathcal{R}^{(1)+\dag}\mathcal{R}^{(2)-\dag}\mathcal{P}_m^{\text{(pw)}}\mathcal{R}^{(2)-}\mathcal{R}^{(1)+}U^{(21)} & \alpha=m\in\{1,2\}\end{cases}\end{equation}
\end{widetext}
Equations \eqref{Ffinal}-\eqref{ual} allows then to explicitly consider two bodies of arbitrary geometries and dielectric properties, in a system characterized by tree possibly different temperatures $T_1$, $T_2$ and $T_3$. In order 
to obtain the expression of the force and heat transfer on body 2, in Eq. \eqref{Noneq} the indexes 1 and 2 must be interchanged, as well as the indexes + and $-$. Moreover, in the case of the force the overall sign has to be changed. In 
what follows we analyze such expression for several cases.

\section{Force and heat transfer on a body alone out of thermal equilibrium}\label{BodyAlone}

Before discussing some numerical applications of eqs. \eqref{Ffinal} and \eqref{Hfinal} for particular choices of bodies 1 and 2, we will start by applying our formalism to a simpler configuration, providing an interesting
example of the role played by the reflection and transmission operators in the calculation of the force and the heat transfer. We are now going to consider the problem of a body (called body 1)
at temperature $T_1$ placed in absence of body 2 in the same
environment as before having temperature $T_3$. In order to obtain
the force and the heat transfer in this case we can exploit the
result \eqref{Noneq} and impose $\mathcal{R}^{(2)\pm}=0$,
$\tilde{\mathcal{T}}^{(2)\pm}=0$ as well as $T_2=T_3$. We remark that the
equilibrium contribution to the force in \eqref{Ffinal} goes to
zero in this limit. After straightforward manipulations, the
result reads
\begin{equation}\label{Alone}\begin{split}&\Delta_m(T_1,T_3)=(-1)^{m+1}\hbar\Tr\Bigl\{\omega^{2-m}n_{31}\mathcal{P}_m^{\text{(pw)}}\\
&\times\Bigl[(-1)^m\mathcal{R}^{(1)+}\mathcal{P}_{-1}^{\text{(pw)}}\mathcal{R}^{(1)+\dag}-\mathcal{R}^{(1)-}\mathcal{P}_{-1}^{\text{(pw)}}\mathcal{R}^{(1)-\dag}\\
&+(-1)^m\Bigl(\tilde{\mathcal{T}}^{(1)+}\mathcal{P}_{-1}^{\text{(pw)}}\tilde{\mathcal{T}}^{(1)+\dag}+\tilde{\mathcal{T}}^{(1)+}\mathcal{P}_{-1}^{\text{(pw)}}+\mathcal{P}_{-1}^{\text{(pw)}}\tilde{\mathcal{T}}^{(1)+\dag}\Bigr)\\
&-\Bigl(\tilde{\mathcal{T}}^{(1)-}\mathcal{P}_{-1}^{\text{(pw)}}\tilde{\mathcal{T}}^{(1)-\dag}+\tilde{\mathcal{T}}^{(1)-}\mathcal{P}_{-1}^{\text{(pw)}}+\mathcal{P}_{-1}^{\text{(pw)}}\tilde{\mathcal{T}}^{(1)-\dag}\Bigr)\Bigr]\Bigr\}.\end{split}\end{equation}
Clearly, this expression is in general different from zero.
Considering the particular case of thermal equilibrium $T_1=T_3$
we see that $\Delta_m$ goes to zero: as expected, at thermal
equilibrium no force is acting on a body alone independently from
its geometrical properties, and it does not exchange any heat with
the environment. On the contrary, if $T_1\neq T_3$, the force and
the heat transfer are linked to the different behavior of
reflection and transmission on the two sides of the body. In
particular, if the body is symmetric with respect to a plane
$z=z_0$, it is easy to show that the matrix elements of reflection
and transmission operators on the two sides cancel each other in
the case of the force ($m=1$). This is expected for evident reasons of symmetry. Nevertheless, even under this specific assumption,
the heat transfer ($m=2$ in eq. \eqref{Alone}) still remains
different from zero.

\section{Some applications}\label{NumAppl}

In this section we are going to perform some applications of eqs. \eqref{Ffinal} and \eqref{Hfinal}. In particular, we are going to discuss the force acting on a neutral atom in front of a planar slab
of finite thickness, as well as the force and the heat transfer in the case of two parallel slabs. To this aim we will provide the reflection and transmission operators associated to an atom and a planar
slab.

\subsection{Force between an atom and a slab}\label{SecAtomSlab}

\begin{figure}[htb]
\includegraphics[height=4.5cm]{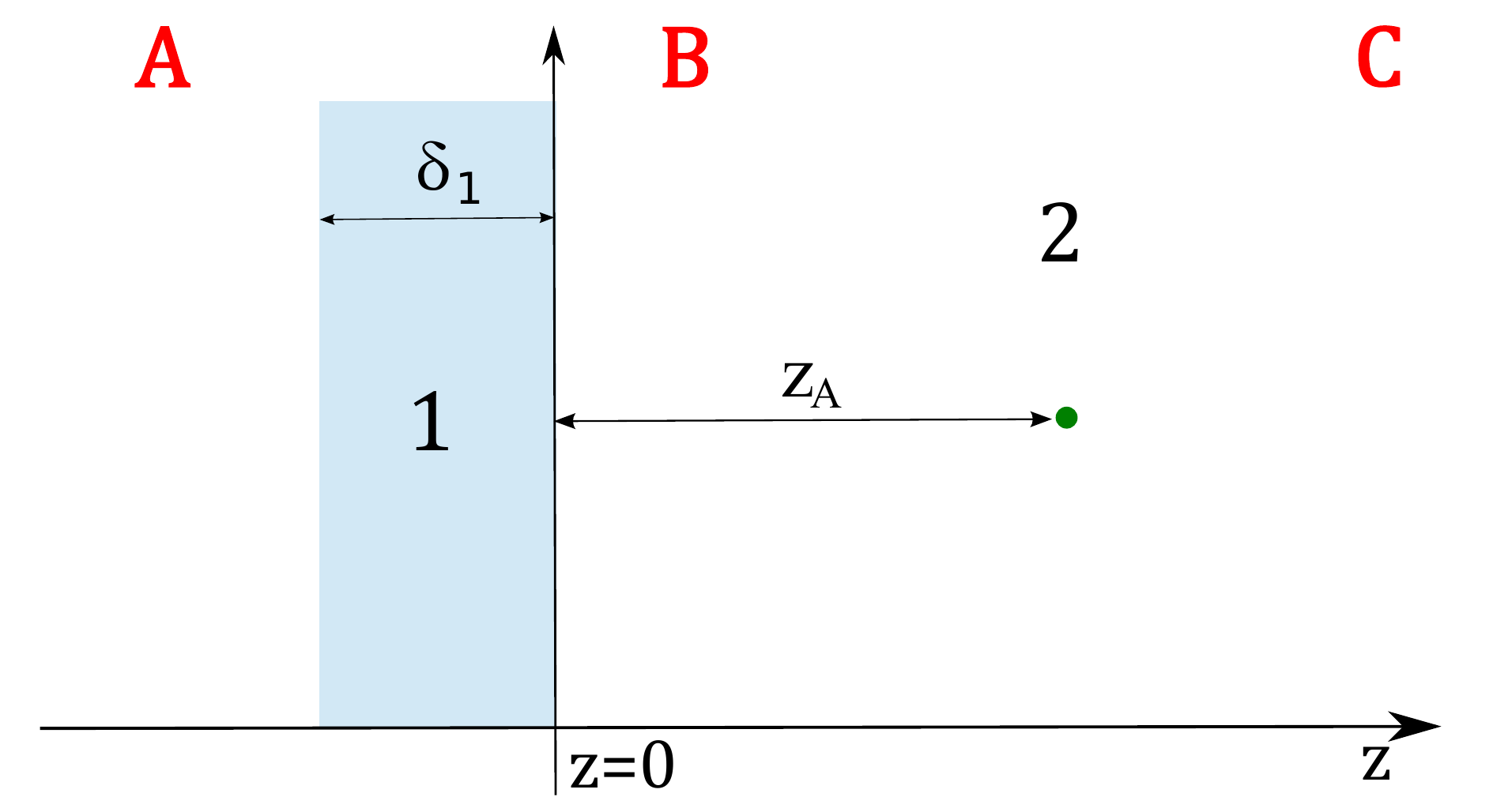}
\caption{Geometry of the atom\,-\,slab configuration.}\label{AtomSlabGeometry}\end{figure}

Let us start with the case of a neutral atom (body 2) in front of a slab (body 1) having finite thickness $\delta_1$. The atom has position $\mathbf{R}_A=(\mathbf{r}_A,z_A)=(0,0,z_A)$ (we have chosen $\mathbf{r}_A=\mathbf{0}$
in virtue of the cylindrical symmetry of the problem with respect to the axis $z=0$) with $z_A>0$, whereas the slab is defined by
the two interfaces $z=0$ and $z=-\delta_1$, as shown in figure \ref{AtomSlabGeometry}.
This configuration is interesting since it implies the presence of a body (the atom) not characterized by translational invariance and then for which the plane-wave basis is not a natural choice. Nevertheless, we will show that
the knowledge of the atomic scattering operator in this basis, chosen in our calculation for convenience, allows us to reproduce the known results in some particular limiting cases and to give the general expression in presence
of three different temperatures $T_1$, $T_2$ and $T_3$.

We now first discuss the reflection and transmission operators for the slab $\mathcal{R}^{(1)+}$ and
$\mathcal{T}^{(1)+}$. For homogeneous flat slabs, these operators are diagonal and given by
\begin{equation}\label{RT1slab}\begin{split}\bra{p,\mathbf{k}}\mathcal{R}^{(1)+}\ket{p',\mathbf{k}'}&=(2\pi)^2\delta(\mathbf{k}-\mathbf{k}')\delta_{pp'}\rho_{1p}(\mathbf{k},\omega)\\
\bra{p,\mathbf{k}}\mathcal{T}^{(1)+}\ket{p',\mathbf{k}'}&=(2\pi)^2\delta(\mathbf{k}-\mathbf{k}')\delta_{pp'}\tau_{1p}(\mathbf{k},\omega)\\\end{split}\end{equation}
and defined in terms of the Fresnel reflection and transmission coefficients modified by the finite thickness $\delta_1$
\begin{equation}\begin{split}\rho_{1p}(\mathbf{k},\omega)&=r_{1p}(\mathbf{k},\omega)\frac{1-e^{2ik_{z1}\delta_1}}{1-r_{1p}^2(\mathbf{k},\omega)e^{2ik_{z1}\delta_1}}\\
\tau_{1p}(\mathbf{k},\omega)&=\frac{t_{1p}(\mathbf{k},\omega)\bar{t}_{1p}(\mathbf{k},\omega)e^{ik_{z1}\delta_1}}{1-r_{1p}^2(\mathbf{k},\omega)e^{2ik_{z1}\delta_1}}.\\\end{split}\end{equation}
In these definitions we have introduced the $z$ component of the
$\mathbf{K}$ vector inside medium 1
\begin{equation}k_{z1}=\sqrt{\varepsilon_1(\omega)\frac{\omega^2}{c^2}-\mathbf{k}^2},\end{equation}
the ordinary vacuum-medium Fresnel reflection coefficients
\begin{equation}r_{1,\text{TE}}=\frac{k_z-k_{z1}}{k_z+k_{z1}}\qquad r_{1,\text{TM}}=\frac{\varepsilon_1(\omega)k_z-k_{z1}}{\varepsilon_1(\omega)k_z+k_{z1}}\end{equation}
as well as both the vacuum-medium (noted with $t$) and
medium-vacuum (noted with $\bar{t}$) transmission coefficients
\begin{equation}\begin{split}t_{1,\text{TE}}&=\frac{2k_z}{k_z+k_{z1}}\qquad\hspace{.3cm}t_{1,\text{TM}}=\frac{2\sqrt{\varepsilon_1(\omega)}k_z}{\varepsilon_1(\omega)k_z+k_{z1}}\\
\bar{t}_{1,\text{TE}}&=\frac{2k_{z1}}{k_z+k_{z1}}\qquad
\bar{t}_{1,\text{TM}}=\frac{2\sqrt{\varepsilon_1(\omega)}k_{z1}}{\varepsilon_1(\omega)k_z+k_{z1}}.\end{split}\end{equation}

The other ingredient of our calculation is represented by the atomic scattering operators. As discussed in \cite{MessinaPRA09}, these operators can be deduced, in dipole approximation, starting from the description
of the atom as an induced dipole  $\mathbf{d}(\omega)=\alpha(\omega)\mathbf{E}(\mathbf{R}_A,\omega)$ proportional to the component of the electric field at frequency $\omega$ calculated at the atomic position
$\mathbf{R}_A$. The proportionality factor coincides with the atomic dynamical polarizability $\alpha(\omega)$. The field radiated by the induced dipole can thus be written analytically as a function of any incoming field:
this produced field has then to be decomposed in plane waves. The expression of the outgoing amplitudes as a function of the incoming ones provides the explicit expression of the atomic reflection and transmission operators.
They read (for $\phi=+,-$)
\begin{equation}\label{AtomScatt}\begin{split}\langle\mathbf{k},p|&\mathcal{R}_A^\phi(\omega)|\mathbf{k}',p'\rangle=\frac{i\omega^2\alpha(\omega)}{2\epsilon_0c^2k_z}\Bigl(\hat{\bbm[\epsilon]}_p^{\phi}(\mathbf{k},\omega)\cdot\hat{\bbm[\epsilon]}_{p'}^{-\phi}(\mathbf{k}',\omega)\Bigr)\\
&\,\times\exp[i(\mathbf{k}'-\mathbf{k})\cdot\mathbf{r}_A]\exp[-i\phi(k_z+k'_z)z_A]\\
\langle\mathbf{k},p|&\tilde{\mathcal{T}}_A^\phi(\omega)|\mathbf{k}',p'\rangle=\frac{i\omega^2\alpha(\omega)}{2\epsilon_0c^2k_z}\Bigl(\hat{\bbm[\epsilon]}_p^{\phi}(\mathbf{k},\omega)\cdot\hat{\bbm[\epsilon]}_{p'}^{\phi}(\mathbf{k}',\omega)\Bigr)\\
&\,\times\exp[i(\mathbf{k}'-\mathbf{k})\cdot\mathbf{r}_A]\exp[-i\phi(k_z-k'_z)z_A].\end{split}\end{equation}
We remark that the we provided the \emph{modified} atomic transmission operator $\tilde{\mathcal{T}}_A^\phi$ (as a matter of fact, it clearly goes to zero in absence of the atom) and that both operators are not diagonal with respect
to the wavevector $\mathbf{k}$ and the polarization $p$, as a result of the lack of translational invariance on the $x-y$ plane. Moreover, since we have attributed a temperature $T_2$ to the atom, the atomic polarizability
$\alpha(\omega)$ must be the one associated to a thermal state at the same temperature.

We are now ready to calculate the equilibrium and non-equilibrium force on the atom. Coherently with the dipole approximation, we have to keep only the leading-order terms in these expressions with respect to the atomic
polarizability, and thus to its scattering operators \eqref{AtomScatt}. Moreover, the appropriate changes have to be made in eq. \eqref{Noneq}, considering that we are in this case calculating the force on the body 2. As
shown in \cite{MessinaPRA09}, this procedure leads to the expression of the force on the atom at thermal equilibrium deduced using several different independent approaches. Focusing on the nonequilibrium contribution, after
some simple algebraic manipulations, we obtain
\begin{equation}\label{Atom}\begin{split}&\Delta_2(T_1,T_2,T_3)=-\frac{\hbar}{4\pi^2\epsilon_0c^2}\Ima\Bigl\{\sum_p\int_0^{+\infty}d\omega\,\omega^2\alpha(\omega)\\
&\times\Bigl[n_{13}\int_0^{\frac{\omega}{c}}dk\,k\bigl(|\rho_{1p}|^2+|\tau_{1p}|^2-1\bigr)\\
&\,+\int_0^{\frac{\omega}{c}}dk\,k\bigl(\bbm[\hat{\epsilon}]_p^+\cdot\bbm[\hat{\epsilon}]_p^-\bigr)\bigl(n_{31}\rho_{1p}\,e^{2ik_zz_A}+n_{23}\rho_{1p}^*e^{-2ik_zz_A}\bigr)\\
&\,+n_{21}\int_{\frac{\omega}{c}}^{+\infty}dk\,k\bigl(\bbm[\hat{\epsilon}]_p^+\cdot\bbm[\hat{\epsilon}]_p^-\bigr)\rho_{1p}^* e^{2ik_zz_A}\Bigr]\Bigr\}\end{split}\end{equation}
where the dependence on the variables $\omega$ and $k$ of all the quantities inside the integral is kept implicit. The first term in the square bracket in eq. \eqref{Atom} does not depend on the atomic position $z_A$ and it
was already identified in \cite{AntezzaPRL05}. On the contrary, the second and the third terms do depend on the atom-slab distance $z_A$, but they come from different regions of the spectrum: the former results from propagative waves
only, the latter from the evanescent sector. As a check of coherence with previous results, we have reobtained the expression deduced in \cite{AntezzaPRL05} using a different approach. To this aim, we assumed that the atom occupies
its ground state ($T_2=0$\,K) and that the slab and environmental temperatures $T_1$ and $T_3$ are such that no atomic excitation is possible: this corresponds to the replacement of the frequency-dependent dynamical polarizability
$\alpha(\omega)$ with its static value $\alpha(0)$.

\subsection{Force between two slabs}\label{SecFSlab}

\begin{figure}[htb]
\includegraphics[height=4.5cm]{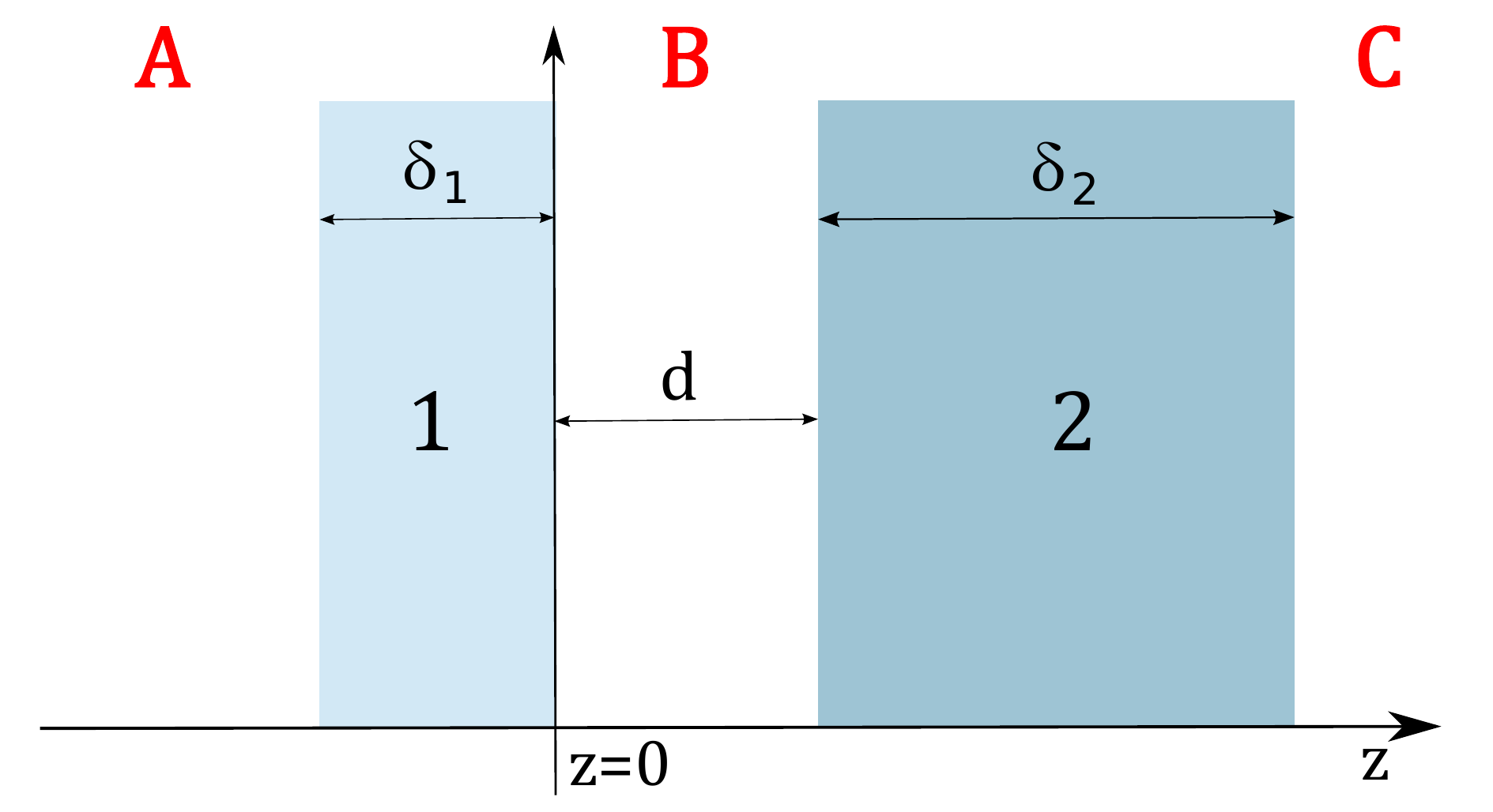}
\caption{Geometry of the slab-slab
configuration.}\label{SlabGeometry}\end{figure}

The case of two parallel homogeneous dielectric slabs of finite thickness will be now examined. This configuration, already studied in \cite{AntezzaPRL06} in the case of infinite thickness, shows the advantage of keeping
the translational symmetry, making all the scattering operators diagonal in the $(\mathbf{k},p)$ basis, allowing at the same time to study the effect of the environmental temperature, in virtue of the finite thickness of the slabs.
Let us assume that the slab $i$ ($i=1,2$) has thickness $\delta_i$ and call $d$ the distance between the two slabs: in particular, the slab 1 occupies the region $-\delta_1<z<0$ (as in the atom-slab configuration described in
sec. \ref{SecAtomSlab}) whereas the slab 2 coincides with $d<z<d+\delta_2$, as shown in figure \ref{SlabGeometry}.
The reflection and transmission operators $\mathcal{R}^{(1)+}$ and
$\mathcal{T}^{(1)+}$ for the slab 1 are still given by eq. \eqref{RT1slab}. The operators $\mathcal{R}^{(1)-}$ and $\mathcal{T}^{(1)-}$ associated to the left side of body 1 as well as all the
scattering operators of body 2 can be obtained by solving the problem of the behavior of the scattering operators with respect to changes of frame of reference, discussed in appendix
\ref{AppTransl}. The result is that the matrix elements of $\mathcal{T}^{(1)+}$ coincide with the ones of $\mathcal{T}^{(1)-}$ given by eq. \eqref{RT1slab}, while the
interchange of 1 and 2 provides directly the elements of the transmission operator $\mathcal{T}^{(2)-}$. As far as the reflection operators are concerned we have
\begin{equation}\label{RTslab2}\begin{split}\bra{p,\mathbf{k}}\mathcal{R}^{(1)-}\ket{p',\mathbf{k}'}&=(2\pi)^2\delta(\mathbf{k}-\mathbf{k}')\delta_{pp'}\rho_{1p}(\mathbf{k},\omega)e^{-2ik_z\delta_1}\\
\bra{p,\mathbf{k}}\mathcal{R}^{(2)-}\ket{p',\mathbf{k}'}&=(2\pi)^2\delta(\mathbf{k}-\mathbf{k}')\delta_{pp'}\rho_{2p}(\mathbf{k},\omega)e^{2ik_zd}.\\\end{split}\end{equation}

Before moving to the explicit calculation of the force we remark that in this geometrical configuration the matrix element of any
scattering operator between the states $\ket{\mathbf{k},p}$ and $\ket{\mathbf{k}',p'}$ is proportional to the Dirac delta
$(2\pi)^2\delta(\mathbf{k}-\mathbf{k}')$, as evident from eqs. \eqref{RT1slab} and \eqref{RTslab2}. This property reflects indeed
the translational symmetry with respect to $x$ and $y$ characterizing this system. As a consequence, the total force
\eqref{F1tot} acting on body 1 is proportional to $(2\pi)^2\delta(\mathbf{0})$ and then formally divergent. This
happens since we are calculating the total force on slab 1, which is by definition infinite, while the force density, i.e. the force
per unit of surface, is a finite quantity. Nevertheless, the
simple analysis of the symmetrized average of the $zz$ component
of the stress tensor \eqref{Tzz} shows us that this quantity is in
this case finite and independent on $\mathbf{r}$, coherently with
the translational invariance. Moreover, the result for the average
value of $T_{zz}$, which means the force per unit of area, is the
same we would get by using the formula \eqref{F1tot} for the force
and neglecting the divergent term $(2\pi)^2\delta(\mathbf{0})$.

We are thus now ready to give the explicit expression of the pressure
acting on slab 1 given by eq. \eqref{F1tot} after neglecting the
divergent term. As for the equilibrium contribution at temperature
$T$, it is given by
\begin{equation}\label{PeqSlab}\begin{split}P_{1z}^{\text{(eq)}}(T)=-4\Rea\sum_p\int_0^{+\infty}&\frac{d\omega}{2\pi}\int\frac{d^2\mathbf{k}}{(2\pi)^2}\frac{k_z}{\omega}\\
&\,\times
N(\omega,T)\frac{\rho_{1p}\rho_{2p}e^{2ik_zd}}{D_p}\end{split}\end{equation}
where
\begin{equation}D_p=1-\rho_{1p}\rho_{2p}e^{2ik_zd}\end{equation}
and the quantities $\rho_{1p}$, $\rho_{2p}$ and $D_p$ implicitly depend on $\omega$ and $\mathbf{k}$. We now turn to the the
non-equilibrium contribution $\Delta F_{1z}(T_1,T_2,T_3)$
appearing in eq. \eqref{F1tot}. This gives in this case the
non-equilibrium pressure
\begin{equation}\label{DeltaPSlab}\begin{split}\Delta P_{1z}&(T_1,T_2,T_3)=A^{\text{(ew)}}(T_1)-A^{\text{(ew)}}(T_2)\\
&\,+B_1^{\text{(pw)}}(T_1)-B_1^{\text{(pw)}}(T_2)+B_2^{\text{(pw)}}(T_3)-B_2^{\text{(pw)}}(T_1)\\
&\,+B_3^{\text{(pw)}}(T_3)-B_3^{\text{(pw)}}(T_2)\\\end{split}\end{equation}
where we have defined
\begin{widetext}
\begin{equation}\begin{split}A^{\text{(ew)}}(T)&=\frac{\hbar}{2\pi^2}\sum_p\int_0^{+\infty}d\omega\int_{\frac{\omega}{c}}^{+\infty}dk\,k\Ima k_z\,n(\omega,T)\frac{\Ima(\rho_{1p}\rho_{2p}^*)}{|D_p|^2}e^{-2d\Ima k_z}\\
B_1^{\text{(pw)}}(T)&=-\frac{\hbar}{4\pi^2}\sum_p\int_0^{+\infty}d\omega\int_0^{\frac{\omega}{c}}dk\,k\,k_z\,n(\omega,T)\frac{|\rho_{2p}|^2-|\rho_{1p}|^2+|\tau_{1p}|^2(1-|\rho_{2p}|^2)}{|D_p|^2}\\
B_2^{\text{(pw)}}(T)&=-\frac{\hbar}{4\pi^2}\sum_p\int_0^{+\infty}d\omega\int_0^{\frac{\omega}{c}}dk\,k\,k_z\,n(\omega,T)\Biggl[\frac{|\tau_{1p}|^2\bigl(1+|\rho_{2p}|^2(1-|\tau_{1p}|^2)\bigr)}{|D_p|^2}-|\rho_{1p}|^2-2\Rea\Bigl(\frac{\rho_{1p}^*\rho_{2p}\tau_{1p}^2}{D_p}e^{2ik_z(d+\delta_1)}\Bigr)\Biggr]\\
B_3^{\text{(pw)}}(T)&=-\frac{\hbar}{4\pi^2}\sum_p\int_0^{+\infty}d\omega\int_0^{\frac{\omega}{c}}dk\,k\,k_z\,n(\omega,T)\Biggl[\frac{|\tau_{2p}|^2}{|D_p|^2}(1+|\rho_{1p}|^2-|\tau_{1p}|^2)-1\Biggr].\\\end{split}\end{equation}
\end{widetext}
We start noticing that the last term of the last line can be explicitly integrated. It gives to the nonequilibrium force \eqref{DeltaPSlab} a contribution $2\sigma(T_3^4-T_2^4)/3c$ where $\sigma=\pi^2k_B^4/60c^2\hbar^3$: this term is the well-known Stefan-Boltzmann radiation pressure. We have verified that in the limit of infinite thickness,
corresponding to $\tau_{1p},\tau_{2p}\to0$ and the replacement of
$\rho_{1p}$ and $\rho_{2p}$ with the ordinary Fresnel
coefficients, we analytically reobtain the results already deduced
in \cite{AntezzaPRL06,AntezzaPRA08}.

\begin{figure}[htb]
\includegraphics[height=8.2cm]{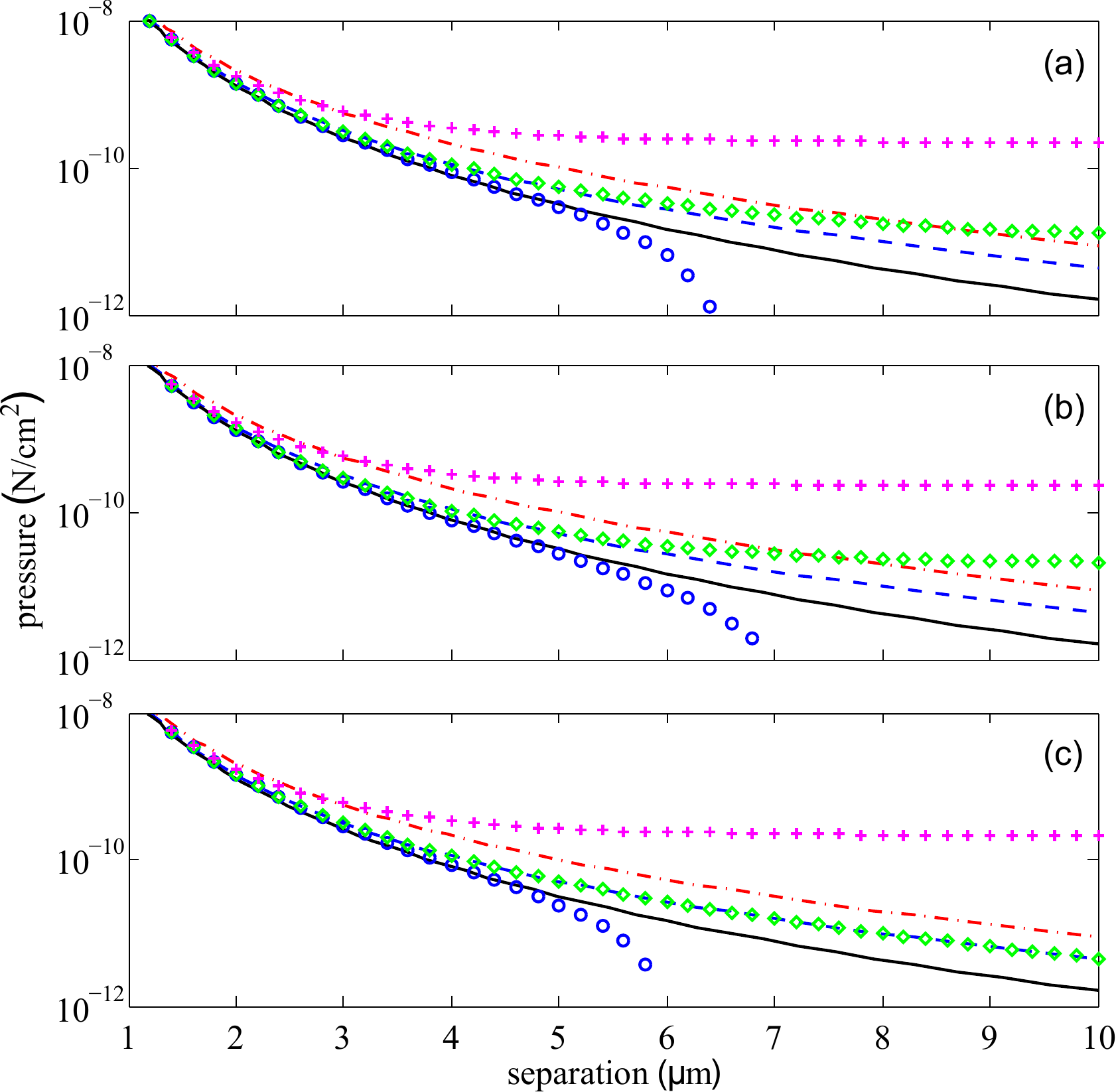}
\caption{(color online) Slab-slab geometry (see Sec. \ref{SecFSlab}). Pressure \eqref{DeltaPSlab} acting on a $\delta_1=2\,\mu$m thick slab (body 1, fused silica)
parallel to a $\delta_2=1000\,\mu$m thick slab (body 2, silicon). Lines: equilibrium
pressures at $T=0\,$K (black solid), 300\,K (blue dashed), 600\,K
(red dash-dotted). Symbols: non-equilibrium pressures, $T_3=0\,$K
(blue circles), 300\,K (green diamonds), 600\,K (magenta plus), with $T_1=300\,$K
and $T_2=0\,$K in (a), $T_1=0\,$K
and $T_2=300\,$K in (b), $T_1=T_2=300\,$K in
(c).}\label{ForceSlab1}\end{figure}

\begin{figure}[htb]
\includegraphics[height=8.5cm]{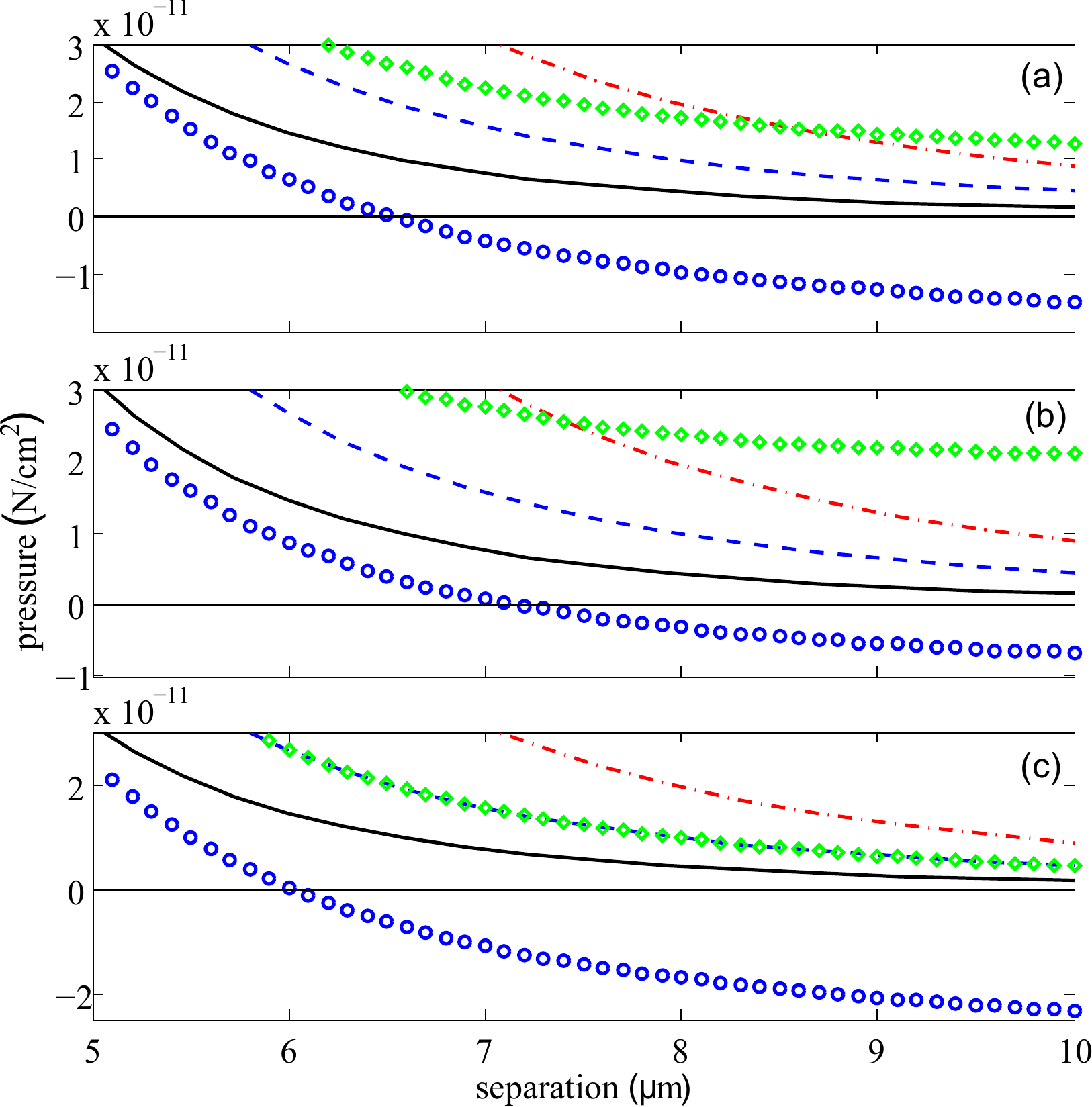}
\caption{(color online) Zoom of figure \ref{ForceSlab1} in linear scales, with the same conventions. Here the change of sign of the force clearly appears.}\label{ForceSlab2}\end{figure}

We have then numerically evaluated, using eqs. \eqref{PeqSlab} and \eqref{DeltaPSlab} arranged as in eq. \eqref{Ffinal}, the total pressure acting on a 2$\mu$m thick slab 1, made of fused silica, in front of a 1000$\mu$m thick slab 2,
made of silicon. The optical data for the two materials are taken from \cite{Palik98}. We have considered different sets of temperatures ($T_1,T_2,T_3$). The results
are shown in figure \ref{ForceSlab1} and \ref{ForceSlab2}.


In figure \ref{ForceSlab1} a wide range of distances, from 1 to 10\,$\mu$m, has been considered, for different equilibrium and nonequilibrium thermal configurations. In particular, (a), (b) and (c) correspond to three
different choices of the slab temperatures $T_1$ and $T_2$ (see caption of figure \ref{ForceSlab1} for details). For each case we have represented the nonequilibrium pressure corresponding to the values of the
environmental temperature $T_3=0,300,600\,$K, as well as the equilibrium pressure at the same three temperatures. We note that the transition from an equilibrium to a nonequilibrium configuration can dramatically change
both the qualitative and quantitative behavior of the interaction. It is worth noting that, even for fixed values of $T_1$, $T_2$ and the slab-slab distance $d$, the value of $T_3$ may significantly
affect the value of the pressure, even by orders of magnitude. As a consequence, the environmental temperature can be remarkably considered as an efficient tool to tune the interaction. This feature is equally present
in the case described in figure \ref{ForceSlab1}(c), where $T_1$ equals $T_2$. This underlines that even in experiments devoted to the measure of the force at thermal equilibrium the environmental temperature should be
carefully controlled.

All these effects prove to be even more spectacular by looking at figure \ref{ForceSlab2}, where linear scales are employed. Indeed, in the case of $T_3=0\,$K the pressure becomes exactly zero at a given distance
around 6\,$\mu$m, and repulsive for larger distances. The appearance of repulsive interactions with nonequilibrium systems has been previously showed only for microscopic bodies, and in particular for the
atom-surface interaction \cite{AntezzaJPhysA06}. Moreover, the possibility of drastically reducing the Casimir-Lifshitz force may be useful in investigations of hypothetical smaller forces of different origins
\cite{DimopoulosPRD03,CarusottoPRL05,WolfPRA07,SorrentinoPRA09,MessinaPRA11}.

\subsection{Heat transfer between two slabs}\label{SecHSlab}

\begin{figure}[htb]
\includegraphics[height=8.3cm]{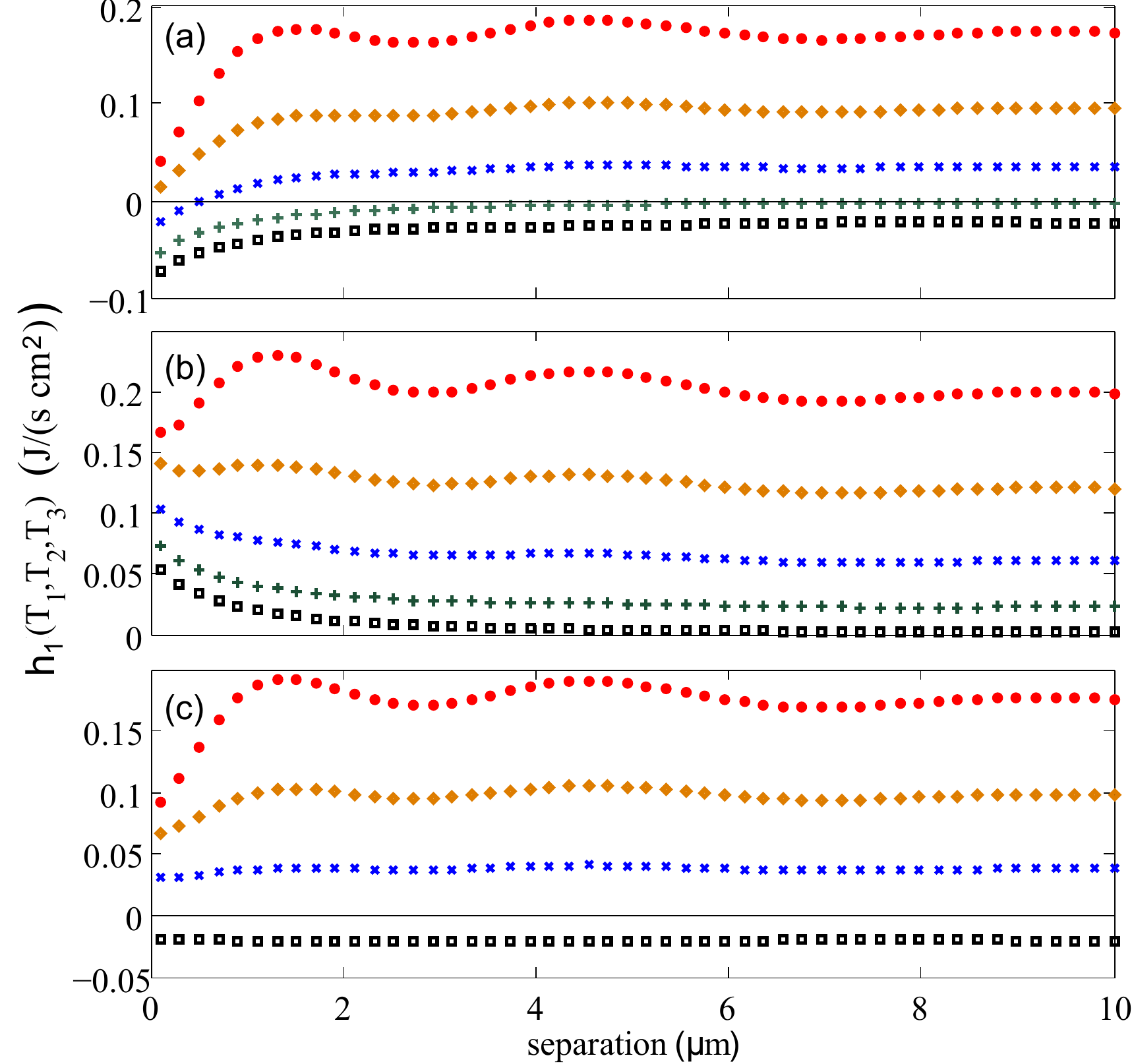}
\caption{(color online) Slab-slab geometry (see Sec.
\ref{SecHSlab}). Radiative heat transfer per unit of surface
\eqref{hSlab} on a $\delta_1=2\,\mu$m thick slab (body 1, fused
silica) parallel to a $\delta_2=1000\,\mu$m thick slab (body 2,
silicon). The temperatures of the slabs are $T_1=300\,$K and
$T_2=0\,$K in (a), $T_1=0\,$K and $T_2=300\,$K in (b),
$T_1=T_2=300\,$K in (c). Symbols: $T_3=0\,$K (black squares),
300\,K (green plus), 400\,K (blue crosses), 500\,K (brown
diamonds), 600\,K (red circles).}\label{HTSlab}\end{figure}

In analogy with the force, we have chosen the same slab-slab
configuration to provide a numerical application of the eq.
\eqref{Hfinal} giving the heat transfer on body 1 for any choice
of $T_1$, $T_2$ and $T_3$. This case was already studied in \cite{BenAbdallahJApplPhys09}, where the influence of the environmental temperature $T_3$ was not considered. Also in this case, the result is
expressed per unit of surface: we then obtain the energy $h_1$
absorbed per unit of surface and per unit of time by the slab 1.
Its analytic expression, using the same formalism of sec.
\ref{SecFSlab}, reads
\begin{equation}\label{hSlab}\begin{split}h_1&(T_1,T_2,T_3)=\mathcal{A}^{\text{(ew)}}(T_1)-\mathcal{A}^{\text{(ew)}}(T_2)\\
&\,+\mathcal{B}_1^{\text{(pw)}}(T_1)-\mathcal{B}_1^{\text{(pw)}}(T_2)+\mathcal{B}_2^{\text{(pw)}}(T_3)-\mathcal{B}_2^{\text{(pw)}}(T_1)\\
&\,+\mathcal{B}_3^{\text{(pw)}}(T_3)-\mathcal{B}_3^{\text{(pw)}}(T_2)\\\end{split}\end{equation}
where we have defined
\begin{widetext}
\begin{equation}\begin{split}\mathcal{A}^{\text{(ew)}}(T)&=\frac{\hbar}{2\pi^2}\sum_p\int_0^{+\infty}d\omega\int_{\frac{\omega}{c}}^{+\infty}dk\,k\,\omega\,n(\omega,T)\frac{e^{-2d\Ima k_z}}{|D_p|^2}\Bigl[\Rea(\rho_{1p}\rho_{2p})-\Rea(\rho_{1p}\rho_{2p}^*)\Bigr]\\
\mathcal{B}_1^{\text{(pw)}}(T)&=\frac{\hbar}{4\pi^2}\sum_p\int_0^{+\infty}d\omega\int_0^{\frac{\omega}{c}}dk\,k\,\omega\,n(\omega,T)\frac{|\rho_{1p}|^2+|\rho_{2p}|^2-1-|\rho_{1p}\rho_{2p}|^2+|\tau_{1p}|^2(1-|\rho_{2p}|^2)}{|D_p|^2}\\
\mathcal{B}_2^{\text{(pw)}}(T)&=\frac{\hbar}{4\pi^2}\sum_p\int_0^{+\infty}d\omega\int_0^{\frac{\omega}{c}}dk\,k\,\omega\,n(\omega,T)\Biggl[1-|\rho_{1p}|^2-\frac{|\tau_{1p}|^2\bigl(1-|\rho_{2p}|^2(1-|\tau_{1p}|^2)\bigr)}{|D_p|^2}\\
&\hspace{4cm}-2\Rea\Bigl(\frac{\rho_{1p}^*\rho_{2p}\tau_{1p}^2}{D_p}e^{2ik_z(d+\delta_1)}\Bigr)\Biggr]\\
\mathcal{B}_3^{\text{(pw)}}(T)&=\frac{\hbar}{4\pi^2}\sum_p\int_0^{+\infty}d\omega\int_0^{\frac{\omega}{c}}dk\,k\,\omega\,n(\omega,T)\frac{|\tau_{2p}|^2}{|D_p|^2}(1-|\rho_{1p}|^2-|\tau_{1p}|^2).\\\end{split}\end{equation}
\end{widetext}
We have numerically evaluated the heat transfer \eqref{hSlab} on body
1 for different set of temperatures ($T_1,T_2,T_3$). In particular, the cases (a), (b) and (c) of figure \ref{HTSlab} correspond to three different choices of the slab temperatures $T_1$ and $T_2$ (see caption for details).
For each case we have represented the heat transfer corresponding to the values of the environmental temperature $T_3=0,300,400,500,600\,$K.
 As in
the case of pressure, the role of the environmental radiation for
heat transfer is particularly interesting. In figure \ref{HTSlab}
the heat transfer $h_1$ shows an oscillating behavior with an
amplitude increasing with the temperature $T_3$. As far as the
positions of minima and maxima are concerned, they are almost
insensitive to the slab thicknesses $\delta_1$ and $\delta_2$ and
to the three temperatures $T_1$, $T_2$ and $T_3$, being on the
contrary connected to the dielectric properties of the two bodies.
Furthermore, these oscillations originate from the propagative
sector, as evident from the analysis of case (c): as a matter of
fact, in this configuration, where $T_1$ and $T_2$ coincide, eq.
\eqref{hSlab} contains only contributions of pure propagative
nature ($B_2$ and $B_3$). These oscillations were already
theoretically studied in \cite{PolderPRB71}. Another interesting
property emerging from figure \ref{HTSlab}(a) is the occurrence of
a change of sign in the heat transfer. Focusing on the blue
crosses, corresponding to
$(T_1,T_2,T_3)=(300\,\text{K},0\,\text{K},400\,\text{K})$, we
observe that at large separations, where the propagative waves
play a dominant role, the heat transfer is positive, i.e. the slab
1 absorbs energy. At smaller separations, and in particular for
distances of the order of 1\,$\mu$m the heat transfer changes
sign, becoming negative (i.e. the slab 1 radiates energy). This
can be understood in terms of the evanescent-wave coupling between
body 1 and body 2, which is at zero temperature. We also note that
the higher is the value of $T_3$, the smaller is the distance at
which the change of sign occurs. From the figure, we deduce that
for $T_3=500,600\,$K this transition happens at distances below
$0.5\,\mu$m.

\section{Conclusions}

We presented a systematic derivation of the radiative heat transfer and the Casimir-Lifshitz force between two arbitrary bodies. We have first expressed the correlating functions of the electromagnetic field in any region of the
system as a function of the scattering operators of each body. This result has been used to provide to closed-form unified analytic expression of the heat transfer and the force. This expression fully takes into account the
interaction between bodies of finite size, any shape and any temperature, as well as the presence of a thermal external radiation coming from the environment.

We applied this theory to two simple but instructive examples: an atom in front of a slab and a couple of parallel slabs. The former configuration is an example of non translationally-invariant system, and generalizes previous
results out of thermal equilibrium. The latter represents the simplest geometrical configuration in which the effects of finite size and external temperature can be quantitatively analyzed. As far as the force is concerned,
we observed that the environmental temperature can substantially tune the interaction, eventually producing a repulsive force. As for the heat transfer, it shows an oscillatory behavior with respect to distance connected to
the dielectric properties of the two slabs and whose amplitude increases with the temperature. Moreover, we found that some given choices of the environmental temperature are able to produce a heat flux whose sign changes as a
function of distance.

This study shows the interest of nonequilibrium configurations, mainly consisting in a strong tunability of force and heat transfer. It would be thus interesting to apply our results to other geometrical configurations
and to test them experimentally.

\begin{acknowledgments}
The authors thank D. Felbacq, B. Guizal and E. Rousseau for fruitful and stimulating discussions.
\end{acknowledgments}

\appendix

\section{Transformations of scattering operators under translations}\label{AppTransl}

We will discuss here the problem of the transformation of the
scattering (both reflection and transmission) operators with
respect to translations. Let us suppose to have a first frame of
reference $\mathcal{O}$ and a second one $\tilde{\mathcal{O}}$
obtained from the first one by a translation of $\mathbf{R}_S$.
For the purpose of this section it is useful to gather the
reflection and transmission operators $\mathcal{R}^\pm$ and
$\mathcal{T}^\pm$ in a single operator $\mathcal{S}^{\phi\phi'}$
connecting the outgoing modes propagating in direction $\phi$ to
the incoming modes propagating in direction $\phi'$. As a
consequence, we have the following identification
\begin{equation}\begin{split}\mathcal{S}^{++}&=\mathcal{T}^+\qquad\mathcal{S}^{+-}=\mathcal{R}^+\\
\mathcal{S}^{-+}&=\mathcal{R}^-\qquad\mathcal{S}^{--}=\mathcal{T}^-.\end{split}\end{equation}
From the definition of the electric field in $\mathcal{O}$
\begin{equation}\begin{split}\mathbf{E}(\mathbf{R},\omega)&=\sum_{\phi,p}\int\frac{d^2\mathbf{k}}{(2\pi)^2}
\exp(i\mathbf{K}^\phi\cdot\mathbf{R})\hat{\bbm[\epsilon]}_p^\phi(\mathbf{k},\omega)E_p^\phi(\mathbf{k},\omega)\\
&=\sum_{\phi,p}\int\frac{d^2\mathbf{k}}{(2\pi)^2}
\exp[i\mathbf{K}^\phi\cdot(\mathbf{R}-\mathbf{R}_S)]\hat{\bbm[\epsilon]}_p^\phi(\mathbf{k},\omega)\\
&\,\times\exp(i\mathbf{K}^\phi\cdot\mathbf{R}_S)E_p^\phi(\mathbf{k},\omega)\end{split}\end{equation}
we deduce that the amplitude $\tilde{E}_p^\phi(\mathbf{k},\omega)$
in the new frame of reference $\tilde{\mathcal{O}}$ equals
\begin{equation}\tilde{E}_p^\phi(\mathbf{k},\omega)=\exp(i\mathbf{K}^\phi\cdot\mathbf{R}_S)E_p^\phi(\mathbf{k},\omega).\end{equation}
From this properties we deduce
\begin{equation}\begin{split}&\tilde{E}_p^{\text{(out)}\phi}(\mathbf{k},\omega)=\exp(i\mathbf{K}^\phi\cdot\mathbf{R}_S)E_p^{\text{(out)}\phi}(\mathbf{k},\omega)\\
&=\exp(i\mathbf{K}^\phi\cdot\mathbf{R}_S)\sum_{p'}\int\frac{d^2\mathbf{k}'}{(2\pi)^2}\bra{p,\mathbf{k}}\mathcal{S}^{\phi\phi'}\ket{p',\mathbf{k}'}\\
&\,\times E_{p'}^{\text{(in)}\phi'}(\mathbf{k}',\omega)\\
&=\exp(i\mathbf{K}^\phi\cdot\mathbf{R}_S)\sum_{p'}\int\frac{d^2\mathbf{k}'}{(2\pi)^2}\bra{p,\mathbf{k}}\mathcal{S}^{\phi\phi'}\ket{p',\mathbf{k}'}\\
&\,\times\exp(-i\mathbf{K}^{'\phi'}\cdot\mathbf{R}_S)\tilde{E}_{p'}^{\text{(in)}\phi'}(\mathbf{k}',\omega)\end{split}\end{equation}
and finally the link between the matrix element of the scattering
operators in the two frames of reference
\begin{equation}\begin{split}
\bra{p,\mathbf{k}}\tilde{\mathcal{S}}^{\phi\phi'}\ket{p',\mathbf{k}'}&=\exp[i(\mathbf{K}^\phi-\mathbf{K}^{'\phi'})\cdot\mathbf{R}_S)]\\
&\,\times\bra{p,\mathbf{k}}\mathcal{S}^{\phi\phi'}\ket{p',\mathbf{k}'}.\end{split}\end{equation}
In the particular and important case of translation along the $z$
axis, assuming that the origin of $\tilde{\mathcal{O}}$ has
coordinates $(0,0,d)$ with respect to $\mathcal{O}$, we have
\begin{equation}\begin{split}\bra{p,\mathbf{k}}\tilde{\mathcal{R}}^+\ket{p',\mathbf{k}'}&=\exp[i(k_z+k'_z)d]\bra{p,\mathbf{k}}\mathcal{R}^+\ket{p',\mathbf{k}'}\\
\bra{p,\mathbf{k}}\tilde{\mathcal{R}}^-\ket{p',\mathbf{k}'}&=\exp[-i(k_z+k'_z)d]\bra{p,\mathbf{k}}\mathcal{R}^-\ket{p',\mathbf{k}'}\\
\bra{p,\mathbf{k}}\tilde{\mathcal{T}}^+\ket{p',\mathbf{k}'}&=\exp[i(k_z-k'_z)d]\bra{p,\mathbf{k}}\mathcal{T}^+\ket{p',\mathbf{k}'}\\
\bra{p,\mathbf{k}}\tilde{\mathcal{T}}^-\ket{p',\mathbf{k}'}&=\exp[-i(k_z-k'_z)d]\bra{p,\mathbf{k}}\mathcal{T}^-\ket{p',\mathbf{k}'}.\end{split}\end{equation}
Apart from their general theoretical interest, these relations are used in Sec. \ref{NumAppl} in order to deduce any reflection and transmission operator associated to a planar slab
as a function of the ordinary Fresnel coefficient (modified to take into account the finite thickness), usually calculated assuming that the interface coincides with the surface $z=0$.

\section{Reciprocity relations of scattering operators}\label{AppRec}

The matrix elements of the scattering operators are not all
mutually independent. By exploiting some properties of the
electromagnetic field, it is in fact possible to deduce some
relations connecting these elements. This is the case, for
example, of the reciprocity relations presented in
\cite{CarminatiJOptSocAmA98}. In this appendix, we derive and
express these relations using our field decomposition and
notation. In \cite{CarminatiJOptSocAmA98}, the authors start their
derivation of the reciprocity relations from Lorentz's reciprocity
theorem in presence of sources. To formulate this theorem, we
start by supposing to have, in presence of a given body, a dipole
$\mathbf{p}_1$ ($\mathbf{p}_2$) in position $\mathbf{R}_1$
($\mathbf{R}_2$) and oscillating at frequency $\omega$. Each of
these dipoles produces an electromagnetic field which is then
scattered by the body and reaches the other dipole. Lorentz's
reciprocity theorem can then formulated by imposing that
\begin{equation}\label{Lorentz}\mathbf{p}_1\cdot\mathbf{E}_2(\mathbf{R}_1)=\mathbf{p}_2\cdot\mathbf{E}_1(\mathbf{R}_2)\end{equation}
where $\mathbf{E}_1(\mathbf{R}_2)$ ($\mathbf{E}_2(\mathbf{R}_1)$)
is the result of the scattering on the body of the field produced
by the dipole $\mathbf{p}_1$ ($\mathbf{p}_2$), then calculated at
the position of dipole $\mathbf{p}_2$ ($\mathbf{p}_1$).

In order to fix the notation, let us suppose that the dipole
$\mathbf{p}_1$ is on the side $\phi$ of the body, while
$\mathbf{p}_2$ is on the side $\phi'$. In order to calculate the
field $\mathbf{E}_1(\mathbf{R}_2)$, we need to take the component
of the field emitted by $\mathbf{p}_1$ propagating in direction
$-\phi$, then apply the operator $\mathcal{S}^{\phi',-\phi}$ and
calculate the resulting field in position $\mathbf{R}_2$. The
component at frequency $\omega$ of the field produced by a dipole
$\mathbf{p}$ in position $\mathbf{R}_p$ reads
\begin{equation}\mathbf{E}(\mathbf{R},\omega)=\frac{1}{4\pi\epsilon_0}\nabla_\mathbf{R}\times\nabla_\mathbf{R}\times\Biggl[\mathbf{p}\frac{e^{i\frac{\omega}{c}R_d}}{R_d}\Biggr]\end{equation}
where $R_d=|\mathbf{R}_d|=|\mathbf{R}-\mathbf{R}_p|$ and
$\nabla_\mathbf{R}$ represents the gradient with respect to
$\mathbf{R}$. As discussed in \cite{MessinaPRA09}, the passage
from the spherical wave emitted by the dipole and our angular
spectrum representation can be performed by using the Weyl
representation \cite{NietoVesperinas91}: the resulting field
propagating in the $\phi$ direction from position $\mathbf{R}_p$
reads, after straightforward algebraic manipulations,
\begin{equation}\begin{split}\mathbf{E}^{\phi}(\mathbf{R},&\omega)=\frac{i\omega^2}{2\epsilon_0c^2}\sum_p\int\frac{d^2\mathbf{k}}{(2\pi)^2}\frac{1}{k_z}\hat{\bbm[\epsilon]}^\phi_p(\mathbf{k},\omega)\\
&\times\Bigl(\hat{\bbm[\epsilon]}^\phi_p(\mathbf{k},\omega)\cdot\mathbf{p}\Bigr)\exp\bigl[i\mathbf{K}^\phi\cdot\bigl(\mathbf{R}-\mathbf{R}_p\bigr)\bigr].\end{split}\end{equation}
We can then simply deduce the expression
\begin{equation}\begin{split}&\mathbf{E}_1(\mathbf{R}_2)=\frac{i\omega^2}{2\epsilon_0c^2}\sum_{pp'}\int\frac{d^2\mathbf{k}}{(2\pi)^2}\int\frac{d^2\mathbf{k}'}{(2\pi)^2}\hat{\bbm[\epsilon]}^{\phi'}_p(\mathbf{k},\omega)\\
&\times\frac{1}{k'_z}\Bigl(\mathbf{p}_1\cdot\hat{\bbm[\epsilon]}^{-\phi}_{p'}(\mathbf{k}',\omega)\Bigr)\bra{\mathbf{k},p}\mathcal{S}^{\phi',-\phi}\ket{\mathbf{k}',p'}\\
&\times\exp\bigl[i\bigl(\mathbf{K}^{\phi'}\cdot\mathbf{R}_2-\mathbf{K}^{'-\phi}\cdot\mathbf{R}_1\bigr)\bigr].\end{split}\end{equation}
and analogously
\begin{equation}\begin{split}&\mathbf{E}_2(\mathbf{R}_1)=\frac{i\omega^2}{2\epsilon_0c^2}\sum_{pp'}\int\frac{d^2\mathbf{k}}{(2\pi)^2}\int\frac{d^2\mathbf{k}'}{(2\pi)^2}\hat{\bbm[\epsilon]}^\phi_{p'}(-\mathbf{k}',\omega)\\
&\times\frac{1}{k_z}\Bigl(\mathbf{p}_2\cdot\hat{\bbm[\epsilon]}^{-\phi'}_p(-\mathbf{k},\omega)\Bigr)\bra{-\mathbf{k}',p'}\mathcal{S}^{\phi,-\phi'}\ket{-\mathbf{k},p}\\
&\times\exp\bigl[i\bigl(\mathbf{K}^{\phi'}\cdot\mathbf{R}_2-\mathbf{K}^{'-\phi}\cdot\mathbf{R}_1\bigr)\bigr].\end{split}\end{equation}
where we have performed the change of variables
\begin{equation}(\mathbf{k},\mathbf{k}',p,p')\longrightarrow(-\mathbf{k}',-\mathbf{k},p',p).\end{equation}
By imposing the condition \eqref{Lorentz} we get
\begin{equation}\begin{split}&k_z\Bigl(\mathbf{p}_1\cdot\hat{\bbm[\epsilon]}^{-\phi}_{p'}(\mathbf{k}',\omega)\Bigr)\Bigl(\mathbf{p}_2\cdot\hat{\bbm[\epsilon]}^{\phi'}_p(\mathbf{k},\omega)\Bigr)\bra{\mathbf{k},p}\mathcal{S}^{\phi',-\phi}\ket{\mathbf{k}',p'}\\
&=k'_z\Bigl(\mathbf{p}_1\cdot\hat{\bbm[\epsilon]}^\phi_{p'}(-\mathbf{k}',\omega)\Bigr)\Bigl(\mathbf{p}_2\cdot\hat{\bbm[\epsilon]}^{-\phi'}_p(-\mathbf{k},\omega)\Bigr)\\
&\,\times\bra{-\mathbf{k}',p'}\mathcal{S}^{\phi,-\phi'}\ket{-\mathbf{k},p}.\end{split}\end{equation}
from which we deduce, by using the properties of the polarization
unit vectors \eqref{PropEps}, the final relation
\begin{equation}k_z\bra{\mathbf{k},p}\mathcal{S}^{\phi',-\phi}\ket{\mathbf{k}',p'}=k'_z(-1)^{p+p'}\bra{-\mathbf{k}',p'}\mathcal{S}^{\phi,-\phi'}\ket{-\mathbf{k},p}.\end{equation}
By choosing the four possible values of the couple $(\phi,\phi')$
we obtain the relations
\begin{equation}\label{ReciprocityRel}\begin{split}k_z\bra{\mathbf{k},p}\mathcal{T}^\pm\ket{\mathbf{k}',p'}&=k'_z(-1)^{p+p'}\bra{-\mathbf{k}',p'}\mathcal{T}^\mp\ket{-\mathbf{k},p}\\
k_z\bra{\mathbf{k},p}\mathcal{R}^\pm\ket{\mathbf{k}',p'}&=k'_z(-1)^{p+p'}\bra{-\mathbf{k}',p'}\mathcal{R}^\pm\ket{-\mathbf{k},p}.\end{split}\end{equation}
These equations clearly show that the matrix elements of each
transmission operator are connected to elements of the specular
operator, while the matrix elements of each reflection operator
are not independent. The relations \eqref{ReciprocityRel} contribute to the derivation presented in appendix \ref{AppCorrObj}.

\section{Green function and scattering operators}\label{AppGreen}

In this appendix we are going to derive the relation between the Green function in the presence of a single body and its reflection
and transmission operators. This feature is a main point in the calculation given in appendix \ref{AppCorrObj}. In order to derive this connection we
will start from the definition of the Green function. Suppose to
have a dipole electric moment $\mathbf{p}$ located at $\mathbf{R}'$ oscillating at
frequency $\omega$ and thus producing an electric field
oscillating at the same frequency proportional to the components
of the dipole moment itself. The component
$G_{ij}(\mathbf{R},\mathbf{R}')$ at frequency $\omega$ of the
Green function can be interpreted as the part of component $i$ of
the total electric field at the point $\mathbf{R}$, namely
$E^{\text{tot}}_i(\mathbf{R})$, proportional to the component $j$
of the dipole moment, divided by $p_j$. Of course, in our
description of the electromagnetic field in the presence of
scatterers, the field directly produced by the dipole $\mathbf{p}$
will result in reflection and transmission: as a consequence, the
Green function will prove to be linked to the scattering operators
$\mathcal{R}^\pm$ and $\mathcal{T}^\pm$.

It is important at this point to remind that our choice of mode
decomposition of the field naturally introduces a left and a right
side for a given body. Thus, we will separately discuss the cases
in which the arguments $\mathbf{R}$ and $\mathbf{R}'$ appearing in
the Green function are either on the same side or on opposite
sides of the body. Let us suppose first that the two points
$\mathbf{R}$ and $\mathbf{R}'$ are located on the same side $\phi$
of the body. In this case, the field directly produced by the
dipole at the point $\mathbf{R}'$ will be directly observed in
$\mathbf{R}$. Moreover, this field will produce a reflected field
defined in the same region $\phi$, and consequently observed in
$\mathbf{R}$ as well. We argue then that for $\mathbf{R}$ and
$\mathbf{R}'$ on the same side $\phi$ of the body the Green
function can be expressed as a sum of two terms, a \emph{free} one
independent on the scattering operators, and a \emph{reflected}
one proportional to $\mathcal{R}^\phi$. Analogously, if the first
argument $\mathbf{R}$ of the Green function is located on the side
$\phi$, while $\mathbf{R}'$ is located in the $-\phi$ region, the
Green function will be made up of a unique \emph{transmitted}
term, proportional to $\mathcal{T}^\phi$.

In order to make this description analytic we need the explicit
expression of the dipole field propagating in direction $\phi$
given in appendix \ref{AppRec}. As far as the free contribution is
concerned (existing if $\mathbf{R}$ and $\mathbf{R}'$ are located
on the same side of the body), if $\mathbf{R}$ is on the right
(left) side of $\mathbf{R}'$, the Green function will contain the
component of the field emitted by the dipole propagating toward
the right (left). Let us now suppose that both $\mathbf{R}$ and
$\mathbf{R}'$ are on the same side $\phi$ of the body. In this
case, apart from the direct contribution we have just discussed,
the field contains the component propagating in direction $\phi$,
resulting from the reflection by means of the operator
$\mathcal{R}^\phi$ of the dipole field propagating in the opposite
direction $-\phi$. Then, the $ij$ component of the Green function
for $\mathbf{R}$ and $\mathbf{R}'$ on the side $\phi$ of the body
reads
\begin{equation}\label{GS1}\begin{split}&G_{ij}(\mathbf{R},\mathbf{R}',\omega)=G^{(0)}_{ij}(\mathbf{R},\mathbf{R}',\omega)+G^{\text{(R)}}_{ij}(\mathbf{R},\mathbf{R}',\omega),\\
&G^{(0)}_{ij}(\mathbf{R},\mathbf{R}',\omega)=\frac{i\omega^2}{2\epsilon_0c^2}\sum_p\int\frac{d^2\mathbf{k}}{(2\pi)^2}\exp\bigl[i\mathbf{k}\cdot\bigl(\mathbf{r}-\mathbf{r}'\bigr)\bigr]\\
&\times\frac{1}{k_z}\Bigl[\theta(z-z')\Bigl(\hat{\bbm[\epsilon]}^+_p(\mathbf{k},\omega)\Bigr)_i\Bigl(\hat{\bbm[\epsilon]}^+_p(\mathbf{k},\omega)\Bigr)_j\exp\bigl[ik_z\bigl(z-z'\bigr)\bigr]\\
&+\theta(z'-z)\Bigl(\hat{\bbm[\epsilon]}^-_p(\mathbf{k},\omega)\Bigr)_i\Bigl(\hat{\bbm[\epsilon]}^-_p(\mathbf{k},\omega)\Bigr)_j\exp\bigl[ik_z\bigl(z'-z\bigr)\bigr]\Bigr],\\
&G^{\text{(R)}}_{ij}(\mathbf{R},\mathbf{R}',\omega)=\frac{i\omega^2}{2\epsilon_0c^2}\sum_{pp'}\int\frac{d^2\mathbf{k}}{(2\pi)^2}\int\frac{d^2\mathbf{k}'}{(2\pi)^2}\\
&\times\exp\bigl[i\bigl(\mathbf{k}\cdot\mathbf{r}-\mathbf{k}'\cdot\mathbf{r}'\bigr)\bigr]\frac{1}{k'_z}\Bigl(\hat{\bbm[\epsilon]}^\phi_p(\mathbf{k},\omega)\Bigr)_i\Bigl(\hat{\bbm[\epsilon]}^{-\phi}_{p'}(\mathbf{k}',\omega)\Bigr)_j\\
&\times\exp\bigl[i\phi(k_zz+k'_zz'\bigr)\bigr]\bra{\mathbf{k},p}\mathcal{R}^\phi\ket{\mathbf{k}',p'}.\end{split}\end{equation}
On the contrary, if $\mathbf{R}$ ($\mathbf{R}'$) is located on the
side $\phi$ ($-\phi$) of the body, we observe a field, propagating
in the $\phi$ direction, resulting from the transmission described
by the operator $\mathcal{T}^\phi$ of the component of the field
emitted by the dipole propagating in direction $\phi$ as well. As
a consequence, in this case we conclude
\begin{equation}\label{GS2}\begin{split}&G_{ij}(\mathbf{R},\mathbf{R}',\omega)=\frac{i\omega^2}{2\epsilon_0c^2}\sum_p\int\frac{d^2\mathbf{k}}{(2\pi)^2}\exp\bigl[i\bigl(\mathbf{k}\cdot\mathbf{r}-\mathbf{k}'\cdot\mathbf{r}'\bigr)\bigr]\\
&\times\sum_{p'}\int\frac{d^2\mathbf{k}'}{(2\pi)^2}\frac{1}{k'_z}\Bigl(\hat{\bbm[\epsilon]}^\phi_p(\mathbf{k},\omega)\Bigr)_i\Bigl(\hat{\bbm[\epsilon]}^\phi_{p'}(\mathbf{k}',\omega)\Bigr)_j\\
&\times\exp\bigl[i\phi(k_zz-k'_zz'\bigr)\bigr]\bra{\mathbf{k},p}\mathcal{T}^\phi\ket{\mathbf{k}',p'}.\end{split}\end{equation}
The expressions \eqref{GS1} and \eqref{GS2} describe, for any
position of the points $\mathbf{R}$ and $\mathbf{R}'$ with respect
to the body, the connection between the Green function and the
scattering operators.

\section{Correlators of the field emitted by the bodies}\label{AppCorrObj}

As discussed in sec. \ref{FieldCorr}, when a body is at thermal
equilibrium with the environment at temperature $T$, the
fluctuation-dissipation theorem gives complete knowledge of the
correlation functions of the total field, resulting from the
environmental field, the one emitted by the body, and the result
of scattering processes. Nevertheless, even at thermal equilibrium
the correlation functions describing the field emitted by the body
are not straightforward. They were given in eq. \eqref{Cois} for
two components of the field propagating in the same direction,
whilst \eqref{Coio} gives the corresponding quantity for
counterpropagating components. The derivation of eqs. \eqref{Cois}
and \eqref{Coio} is the main scope of this appendix.

Let us consider a body at thermal equilibrium at temperature $T$ and described by
the scattering operator $\mathcal{R}^\pm$ and $\mathcal{T}^\pm$.
The first step of our calculation is to write the expression of
the total field both on the left and on the right side of the
body. Their components at frequency $\omega$ are given by
\begin{equation}\label{Etotphi}\begin{split}\mathbf{E}^{\text{(tot)}\phi}(\mathbf{R},\omega)&=\mathbf{E}^{\text{(env)}-\phi}(\mathbf{R},\omega)+\mathbf{E}^{\text{(b)}\phi}(\mathbf{R},\omega)\\
&\,+\mathbf{E}^{\text{(re)}\phi}(\mathbf{R},\omega)+\mathbf{E}^{\text{(tr)}\phi}(\mathbf{R},\omega).\end{split}\end{equation}
In this expression the superscript $\phi$ for the total field in
the l.h.s. refers to the region we are considering, while for the
four fields in the r.h.s. it corresponds, as usual in the rest of
the paper, to the direction of propagation. Eq. \eqref{Etotphi}
tells us that the total field in region $\phi$ contains first the
environmental field propagating in direction $-\phi$ (i.e. toward
the body) and of course the field $\mathbf{E}^{\text{(b)}\phi}$
emitted by the body itself and propagating in direction $\phi$.
Moreover, the environmental field $\mathbf{E}^{\text{(env)}-\phi}$
produces a reflected field $\mathbf{E}^{\text{(re)}\phi}$
connected to $\mathbf{E}^{\text{(env)}-\phi}$ by the operator
$\mathcal{R}^\phi$. Finally, in region $\phi$ there is a field
$\mathbf{E}^{\text{(tr)}\phi}$ resulting from the transmission of
the environmental field $\mathbf{E}^{\text{(env)}\phi}$ (existing
in the other region $-\phi$) by the operator $\mathcal{T}^\phi$.

We are looking for correlation functions such as
\begin{equation}\langle E^{\text{(b)}\phi}_p(\mathbf{k},\omega)E^{\text{(b)}\phi'\dag}_{p'}(\mathbf{k}',\omega')\rangle_\text{sym}\end{equation}
of the field emitted by the body, both for $\phi=\phi'$ and for
$\phi\neq\phi'$. Let us start by the case of copropagating
components of the field, namely by the case $\phi=\phi'$. It is
first useful to calculate the correlation function of the $i$ and
$j$ components of eq. \eqref{Etotphi}, for two couples of
coordinates $(\mathbf{R},\omega)$ and $(\mathbf{R}',\omega')$. As
for the l.h.s., the result is directly given by the
fluctuation-dissipation theorem stated in eq. \eqref{FluDiss}. The
discussion of the r.h.s. requires a more accurate analysis. We
first point out that we already know the correlation function
characterizing the environmental field: it is given by eq.
\eqref{Co3} and it also tells us that two counterpropagating
components of this field are uncorrelated. As a consequence,
$\mathbf{E}^{\text{(env)}-\phi}$ is correlated (apart from with
itself) with the reflected field $\mathbf{E}^{\text{(re)}\phi}$,
but not with the transmitted field $\mathbf{E}^{\text{(tr)}\phi}$,
coming from the counterpropagating component of the environmental
field. As a consequence $\mathbf{E}^{\text{(tr)}\phi}$ is
correlated only with itself: the same holds for the field
$\mathbf{E}^{\text{(b)}\phi}$, since the field produced by the
body alone is of course uncorrelated with everything which results
from the environment. We now observe that, since the reflected and
transmitted field are connected by means of scattering operators
to the environmental field, whose correlation functions are known,
the only unknown remaining is the correlation function
\begin{equation}\langle E_i^{\text{(b)}\phi}(\mathbf{R},\omega)E_j^{\text{(b)}\phi\dag}(\mathbf{R}',\omega')\rangle_\text{sym}\end{equation}
strictly connected to the one we are looking for. In particular we
obtain the expression
\begin{widetext}
\begin{equation}\label{LUR}\begin{split}L&=R,\\
L&=\langle E_i^{\text{(tot)}\phi}(\mathbf{R},\omega)E_j^{\text{(tot)}\phi\dag}(\mathbf{R}',\omega')\rangle,\\
R&=\langle E_i^{\text{(env)}-\phi}(\mathbf{R},\omega)E_j^{\text{(env)}-\phi\dag}(\mathbf{R}',\omega')\rangle
+\langle E_i^{\text{(re)}\phi}(\mathbf{R},\omega)E_j^{\text{(re)}\phi\dag}(\mathbf{R}',\omega')\rangle+\langle E_i^{\text{(env)}-\phi}(\mathbf{R},\omega)E_j^{\text{(re)}\phi\dag}(\mathbf{R}',\omega')\rangle\\
&\,+\langle
E_i^{\text{(re)}\phi}(\mathbf{R},\omega)E_j^{\text{(env)}-\phi\dag}(\mathbf{R}',\omega')\rangle+\langle
E_i^{\text{(tr)}\phi}(\mathbf{R},\omega)E_j^{\text{(tr)}\phi\dag}(\mathbf{R}',\omega')\rangle
+\langle
E_i^{\text{(b)}\phi}(\mathbf{R},\omega)E_j^{\text{(b)}\phi\dag}(\mathbf{R}',\omega')\rangle\end{split}\end{equation}
where the subscript (sym) has been dropped for simplicity. We will
start by working on the r.h.s. term $R$. In this case, expression
all the fields by their Fourier decomposition \eqref{DefE} (all
for a given value of $\phi$), it is easy to see that all the
correlation functions will contain a double integral on
$\mathbf{k}$ and $\mathbf{k}'$ as well as the factor
$\exp[i(\mathbf{k}\cdot\mathbf{r}-\mathbf{k}'\cdot\mathbf{r}')]$,
which we will drop here, in virtue of the fact that it will also
appear in the l.h.s. $L$. Moreover we will also drop the factor
$2\pi\delta(\omega-\omega')$, appearing in both $L$ and $R$. With
these conventions and using the definition of the scattering
operators given in Sec. \ref{SRT} we have
\begin{equation}\label{RHS}\begin{split}\langle E_i^{\text{(env)}-\phi}(\mathbf{R},\omega)E_j^{\text{(env)}-\phi\dag}(\mathbf{R}',\omega')\rangle&=\sum_{pp'}\exp[-i\phi(k_zz-k_z^{'*}z')]
\Bigl(\hat{\bbm[\epsilon]}^{-\phi}_p(\mathbf{k},\omega)\Bigr)_i\Bigl(\hat{\bbm[\epsilon]}^{-\phi}_{p'}(\mathbf{k}',\omega)\Bigr)^*_j\bra{\mathbf{k},p}C^{(3)}\ket{\mathbf{k}',p'},\\
\langle E_i^{\text{(re)}\phi}(\mathbf{R},\omega)E_j^{\text{(re)}\phi\dag}(\mathbf{R}',\omega')\rangle&=\sum_{pp'}\exp[i\phi(k_zz-k_z^{'*}z')]
\Bigl(\hat{\bbm[\epsilon]}^\phi_p(\mathbf{k},\omega)\Bigr)_i\Bigl(\hat{\bbm[\epsilon]}^\phi_{p'}(\mathbf{k}',\omega)\Bigr)^*_j\bra{\mathbf{k},p}\mathcal{R}^\phi C^{(3)}\mathcal{R}^{\phi\dag}\ket{\mathbf{k}',p'},\\
\langle E_i^{\text{(env)}-\phi}(\mathbf{R},\omega)E_j^{\text{(re)}\phi\dag}(\mathbf{R}',\omega')\rangle&=\sum_{pp'}\exp[-i\phi(k_zz+k_z^{'*}z')]
\Bigl(\hat{\bbm[\epsilon]}^{-\phi}_p(\mathbf{k},\omega)\Bigr)_i\Bigl(\hat{\bbm[\epsilon]}^\phi_{p'}(\mathbf{k}',\omega)\Bigr)^*_j\bra{\mathbf{k},p}C^{(3)}\mathcal{R}^{\phi\dag}\ket{\mathbf{k}',p'},\\
\langle E_i^{\text{(re)}\phi}(\mathbf{R},\omega)E_j^{\text{(env)}-\phi\dag}(\mathbf{R}',\omega')\rangle&=\sum_{pp'}\exp[i\phi(k_zz+k_z^{'*}z')]
\Bigl(\hat{\bbm[\epsilon]}^\phi_p(\mathbf{k},\omega)\Bigr)_i\Bigl(\hat{\bbm[\epsilon]}^{-\phi}_{p'}(\mathbf{k}',\omega)\Bigr)^*_j\bra{\mathbf{k},p}\mathcal{R}^\phi C^{(3)}\ket{\mathbf{k}',p'},\\
\langle E_i^{\text{(tr)}\phi}(\mathbf{R},\omega)E_j^{\text{(tr)}\phi\dag}(\mathbf{R}',\omega')\rangle&=\sum_{pp'}\exp[i\phi(k_zz-k_z^{'*}z')]
\Bigl(\hat{\bbm[\epsilon]}^\phi_p(\mathbf{k},\omega)\Bigr)_i\Bigl(\hat{\bbm[\epsilon]}^\phi_{p'}(\mathbf{k}',\omega)\Bigr)^*_j\bra{\mathbf{k},p}\mathcal{T}^\phi C^{(3)}\mathcal{T}^{\phi\dag}\ket{\mathbf{k}',p'},\\
\langle
E_i^{\text{(b)}\phi}(\mathbf{R},\omega)E_j^{\text{(b)}\phi\dag}(\mathbf{R}',\omega')\rangle&=\sum_{pp'}\exp[i\phi(k_zz-k_z^{'*}z')]
\Bigl(\hat{\bbm[\epsilon]}^\phi_p(\mathbf{k},\omega)\Bigr)_i\Bigl(\hat{\bbm[\epsilon]}^\phi_{p'}(\mathbf{k}',\omega)\Bigr)^*_j\bra{\mathbf{k},p}C^{\phi\phi}\ket{\mathbf{k}',p'}.\\\end{split}\end{equation}
\end{widetext}
We now need to calculate the l.h.s by using the fluctuation-dissipation theorem \eqref{FluDiss} and inserting the explicit
expression found in appendix \ref{AppGreen} of the Green function
as a function of the scattering operators. Since in this case we
are calculating the correlator of $E_i^{\text{(tot)}\phi}$ with
$E_j^{\text{(tot)}\phi}$, we are using the Green function
calculated at positions $\mathbf{R}$ and $\mathbf{R}'$ being on
the same side $\phi$ of the body. We will thus make use of eq.
\eqref{GS1} to connect Green function to scattering operators. In
particular, the fluctuation-dissipation theorem tells us that we
need to calculate the imaginary part of the Green function
\eqref{GS1}. Starting from the free term $G^{(0)}_{ij}$, and
inserting a term $(2\pi)^2\delta(\mathbf{k}-\mathbf{k}')$
integrated over $\mathbf{k}'$ as well as a $\delta_{pp'}$ in order
to have the desired factor
$\exp[i(\mathbf{k}\cdot\mathbf{r}-\mathbf{k}'\cdot\mathbf{r}')]$
and the same structure of the terms in \eqref{RHS} (we remark here
that the factor $2\pi\delta(\omega-\omega')$ is already in the
fluctuation-dissipation theorem \eqref{FluDiss}), we obtain
\begin{equation}\begin{split}&\Ima G^{(0)}_{ij}(\mathbf{R},\mathbf{R}',\omega)=\frac{\omega^2}{4\epsilon_0c^2}\sum_{pp'}\delta_{pp'}(2\pi)^2\delta(\mathbf{k}-\mathbf{k}')\\
&\times\Bigl\{\theta(z-z')\Bigl[\frac{1}{k_z}\Bigl(\hat{\bbm[\epsilon]}^+_p(\mathbf{k},\omega)\Bigr)_i\Bigl(\hat{\bbm[\epsilon]}^+_{p'}(\mathbf{k}',\omega)\Bigr)_j\exp\bigl[i\bigl(k_zz-k'_zz'\bigr)\bigr]\\
&+\frac{1}{k^*_z}\Bigl(\hat{\bbm[\epsilon]}^+_p(-\mathbf{k},\omega)\Bigr)^*_i\Bigl(\hat{\bbm[\epsilon]}^+_{p'}(-\mathbf{k}',\omega)\Bigr)^*_j\exp\bigl[-i\bigl(k^*_zz-k^{'*}_zz'\bigr)\bigr]\Bigr]\\
&+\theta(z'-z)\Bigl[\frac{1}{k_z}\Bigl(\hat{\bbm[\epsilon]}^-_p(\mathbf{k},\omega)\Bigr)_i\Bigl(\hat{\bbm[\epsilon]}^-_{p'}(\mathbf{k}',\omega)\Bigr)_j\exp\bigl[i\bigl(k'_zz'-k_zz\bigr)\bigr]\\
&+\frac{1}{k^*_z}\Bigl(\hat{\bbm[\epsilon]}^-_p(-\mathbf{k},\omega)\Bigr)^*_i\Bigl(\hat{\bbm[\epsilon]}^-_{p'}(-\mathbf{k}',\omega)\Bigr)^*_j\exp\bigl[-i\bigl(k^{'*}_zz'-k^*_zz\bigr)\bigr]\Bigr]\Bigr\}\end{split}\end{equation}
where for the terms obtained by complex conjugation we made a
change of variable from $\mathbf{k}$ to $-\mathbf{k}$. Making use
of the properties of the polarization unit vectors, it is easy to
show that each term multiplying a $\theta$ step function is zero
in the evanescent sector of $\mathbf{k}$ (and then $\mathbf{k}'$).
After simple algebraic manipulations the imaginary part of
$G_{ij}^{(0)}$ finally takes the form
\begin{equation}\label{G0fin}\begin{split}&\Ima G^{(0)}_{ij}(\mathbf{R},\mathbf{R}',\omega)=\frac{\omega^2}{4\epsilon_0c^2}\sum_{pp'}\delta_{pp'}(2\pi)^2\delta(\mathbf{k}-\mathbf{k}')\\
&\times\theta(\omega-ck)\frac{1}{k_z}\\
&\times\Bigl[\Bigl(\hat{\bbm[\epsilon]}^+_p(\mathbf{k},\omega)\Bigr)_i\Bigl(\hat{\bbm[\epsilon]}^+_{p'}(\mathbf{k}',\omega)\Bigr)_j\exp\bigl[i\bigl(k_zz-k'_zz'\bigr)\bigr]\\
&+\Bigl(\hat{\bbm[\epsilon]}^-_p(\mathbf{k},\omega)\Bigr)_i\Bigl(\hat{\bbm[\epsilon]}^-_{p'}(\mathbf{k}',\omega)\Bigr)_j\exp\bigl[i\bigl(k'_zz'-k_zz\bigr)\bigr]\Bigr].\end{split}\end{equation}
Observing from eq. \eqref{Co3} that
\begin{equation}\begin{split}\bra{\mathbf{k},p}C^{(3)}\ket{\mathbf{k}',p'}&=\frac{\omega}{2\epsilon_0c^2}N(\omega,T)\delta_{pp'}(2\pi)^2\delta(\mathbf{k}-\mathbf{k}')\\
&\,\times\theta(\omega-ck)\frac{1}{k_z}\end{split}\end{equation}
and taking back the factor $\frac{2}{\omega}N(\omega,T)$ in eq.
\eqref{FluDiss} we conclude that the term \eqref{G0fin} cancels
the first term in \eqref{RHS} and gives in the l.h.s. a
contribution
\begin{equation}\begin{split}\sum_{pp'}\exp[i\phi(k_zz-k_z^{'*}z')]&\Bigl(\hat{\bbm[\epsilon]}^\phi_p(\mathbf{k},\omega)\Bigr)_i\Bigl(\hat{\bbm[\epsilon]}^\phi_{p'}(\mathbf{k}',\omega)\Bigr)_j\\\
&\times\bra{\mathbf{k},p}C^{(3)}\ket{\mathbf{k}',p'}.\end{split}\end{equation}
We are now left with the calculation of the imaginary part of
$G_{ij}^{\text{(R)}}$. We have
\begin{equation}\begin{split}&\Ima G^{\text{(R)}}_{ij}(\mathbf{R},\mathbf{R}',\omega)=\frac{\omega^2}{4\epsilon_0c^2}\sum_{pp'}\\
&\times\Bigl[\frac{1}{k'_z}\Bigl(\hat{\bbm[\epsilon]}^\phi_p(\mathbf{k},\omega)\Bigr)_i\Bigl(\hat{\bbm[\epsilon]}^{-\phi}_{p'}(\mathbf{k}',\omega)\Bigr)_j\exp\bigl[i\phi(k_zz+k'_zz'\bigr)\bigr]\\
&\qquad\times\bra{\mathbf{k},p}\mathcal{R}^\phi\ket{\mathbf{k}',p'}\\
&+\frac{1}{k^{'*}_z}\Bigl(\hat{\bbm[\epsilon]}^{-\phi}_p(\mathbf{k},\omega)\Bigr)^*_i\Bigl(\hat{\bbm[\epsilon]}^\phi_{p'}(\mathbf{k}',\omega)\Bigr)^*_j\exp\bigl[-i\phi(k^*_zz+k^{'*}_zz'\bigr)\bigr]\\
&\qquad\times(-1)^{p+p'}\bra{-\mathbf{k}',p'}\mathcal{R}^{\phi\dag}\ket{-\mathbf{k},p}\Bigr].\end{split}\end{equation}
where the second term was obtained by the change of variables
$(\mathbf{k},\mathbf{k}')\longrightarrow(-\mathbf{k},-\mathbf{k}')$
and using the properties of the polarization unit vectors
\eqref{PropEps}. Starting from the first term we note that its
part which is propagative in $\mathbf{k}'$ exactly cancels the
fourth term in eq. \eqref{RHS}. Observing that in the evanescent
sector we have $k^{'*}_z=-k'_z$ and using again \eqref{PropEps} we
are left from the first term with a contribution
\begin{equation}\begin{split}&\frac{\omega^2}{4\epsilon_0c^2}\sum_{pp'}\Bigl(\hat{\bbm[\epsilon]}^\phi_p(\mathbf{k},\omega)\Bigr)_i\Bigl(\hat{\bbm[\epsilon]}^\phi_{p'}(\mathbf{k}',\omega)\Bigr)^*_j\\
&\times\exp\bigl[i\phi(k_zz-k^{'*}_zz'\bigr)\bigr]\bra{\mathbf{k},p}\mathcal{R}^\phi\mathcal{P}_{-1}^{\text{(ew)}}\ket{\mathbf{k}',p'}.\end{split}\end{equation}
As far as the second term is concerned, we exploit the reciprocity
relations of scattering operators presented in appendix
\ref{AppRec} to conclude that
\begin{equation}\frac{1}{k^{'*}_z}(-1)^{p+p'}\bra{-\mathbf{k}',p'}\mathcal{R}^{\phi\dag}\ket{-\mathbf{k},p}=\frac{1}{k^*_z}\bra{\mathbf{k},p}\mathcal{R}^{\phi\dag}\ket{\mathbf{k}',p'}.\end{equation}
As a consequence the propagative part with respect to $\mathbf{k}$
of this second term analogously cancels the third term in
\eqref{RHS} and we are left with
\begin{equation}\begin{split}&-\frac{\omega^2}{4\epsilon_0c^2}\sum_{pp'}\Bigl(\hat{\bbm[\epsilon]}^\phi_p(\mathbf{k},\omega)\Bigr)_i\Bigl(\hat{\bbm[\epsilon]}^\phi_{p'}(\mathbf{k}',\omega)\Bigr)^*_j\\
&\times\exp\bigl[i\phi(k_zz-k^{'*}_zz'\bigr)\bigr]\bra{\mathbf{k},p}\mathcal{P}_{-1}^{\text{(ew)}}\mathcal{R}^{\phi\dag}\ket{\mathbf{k}',p'}.\end{split}\end{equation}
Having considered all the terms in the equality \eqref{LUR}, we
remain now with quantities which have all exactly the same
structure, namely the same sum over $p$ and $p'$, the same
$z$-dependent exponential and the same polarization unit vectors.
As a consequence we can identify the matrices whose elements are
calculated between $(\mathbf{k},p)$ and $(\mathbf{k}',p')$ and
obtain immediately the equality \eqref{Cois}.

In the case of the correlation function between two components of
the field emitted by the body propagating in two opposite
directions the structure of the calculation is the same. Now one
has to use eq. \eqref{GS2} instead of \eqref{GS1} in order to
connect the Green function to the scattering operators. The
analytic expression \eqref{Coio} of the correlation function for
counterpropagating fields has been already given in the paper.

\end{document}